\documentclass[twocolumn, pra, superscriptaddress]{revtex4-1}
\usepackage{upgreek}
\usepackage{tikz}
\usepackage{dcolumn}
\usepackage{color,tcolorbox}
\usepackage{amsmath,mathrsfs,amssymb,esint}
\usepackage{dcolumn}
\usepackage{bm}
\usepackage{blindtext}
\usepackage{natbib}
\usepackage{graphicx}
\usepackage{rotating}
\usepackage{titlesec}
\usepackage{multirow}
\usepackage{mathtools}
\usepackage{soul}
\usepackage{amsmath}
\usepackage[pdfborder={0 0 0},colorlinks=true,linkcolor=blue,urlcolor=blue,citecolor=blue]{hyperref}
\colorlet{orange1}{green!10!orange!90!}
\usepackage{enumitem}

\def\be{\begin{equation}}
\def\ee{\end{equation}}
\def\bea{\begin{eqnarray}}
\def\eea{\end{eqnarray}}
\def\bi{\begin{itemize}}
\def\ei{\end{itemize}}

\newcommand{\npp}[1]{{\leavevmode\color{black}#1}}

\usepackage[T1]{fontenc}

\definecolor{ao(english)}{rgb}{0.0, 0.5, 0.0}
\begin{document}
\title{Universality of Bose-Einstein Condensation and Quenched Formation Dynamics}
\author{Nick P. Proukakis}
\email{nikolaos.proukakis@newcastle.ac.uk}
\affiliation{Joint Quantum Centre (JQC) Durham--Newcastle, School of Mathematics, Statistics and Physics, Newcastle University,Newcastle upon Tyne, NE1 7RU, United Kingdom}

\begin{abstract}

%Under appropriate conditions (e.g.~combination of high density and low temperature), a system of bosonic (constituent) particles  acquires significant macroscopic coherence, with all particles behaving collectively: in a typical three-dimensional system at equilibrium this manifests itself as a so-called Bose-Einstein Condensate.
%Nonetheless, the nature of the underlying properties and the dynamical transition to a fully coherent state are both heavily dependent on system details, most notably the system dimensionality. 
The emergence of macroscopic coherence in a many-body quantum system is a ubiquitous phenomenon across different physical systems and scales.
This Chapter reviews key concepts characterizing such systems (correlation functions, condensation, quasi-condensation) and applies them to the study of emerging non-equilibrium features in
 the dynamical path towards such a highly-coherent state:
 particular emphasis is placed on emerging universal features in the dynamics of conservative and open quantum systems, their equilibrium or non-equilibrium nature,
  and the extent that these can be observed in current experiments with quantum gases.  
 Characteristic examples include symmetry-breaking in the Kibble-Zurek mechanism, coarsening and phase-ordering kinetics, and universal spatiotemporal scalings around non-thermal fixed points and in the context of the Kardar-Parisi-Zhang equation; the Chapter concludes with a brief review of the potential relevance of some of these concepts in modelling the large-scale distribution of dark matter in the universe. 
  %Although the presentation is set in the specific context of an ultracold trapped atomic gas of bosons -- addressing also effects of system geometry, dimensionality, inhomogeneity and other relevant considerations -- much of the underlying discussion is generic, with appropriate connections made to other related physical systems exhibiting similar effects.

\end{abstract}

\maketitle

\vspace{-2.0cm}
\begin{center}
{\bf Keywords:}
\vspace{-0.2cm}
\end{center}
{\footnotesize
{\em 
Berezinskii-Kosterlitz-Thouless; Bose-Einstein condensation; 
 classical field; coarsening dynamics; correlation function; 
 critical exponents; 
 dynamical criticality; exciton-polaritons; fuzzy dark matter; Gross-Pitaevskii equation; Kardar-Parisi-Zhang; Kibble-Zurek mechanism; Penrose-Onsager mode; non-equilibrium; non-thermal fixed point; order parameter; persistent current; 
 phase-ordering; phase transition; quantum fluid; quasi-condensation; quenched dynamics; Schr\"{o}dinger-Poisson equation; scaling hypothesis; self-similarity; stochastic modelling; symmetry-breaking; turbulence; ultracold atoms; universality; vortex; winding number.
}
}

%\vspace{-0.5cm}
\begin{center}
{\bf Objectives:}
\end{center}
\vspace{-0.5cm}
\begin{itemize}
\small{
    \item Demonstration of ubiquitous nature of Bose-Einstein condensation and key observables characterizing it.
    \item Unified overview of the dynamical phase transition crossing from an incoherent initial condition to a (highly/fully) coherent final state, introducing key universal scaling laws applicable during such process. %and key observables used for their analysis.
    \item Detailed explanation of the Kibble-Zurek mechanism for a driven dynamical phase transition crossing through an external time-dependent parameter.
    \item Discussion of phase-ordering process and universal scaling at late evolution times following an instantaneous quench from an incoherent state.
    \item Applications of above frameworks to the dynamics of ultracold atoms and exciton-polaritons, with concrete examples and experimental observations.
    \item Discussion of other highly-non-equilibrium features of quantum gases originally stemming from cosmology (non-thermal fixed points) and classical systems (Kardar-Parisi-Zhang).
    \item %Highlighting to condensed matter physicists the 
    Review of cosmological condensation model for dark matter, and its analogies to laboratory condensates.
    }
\end{itemize}

\vspace{-0.9cm}
{\small
\tableofcontents
}

%-------------------------------------
\begin{figure*}[t!]
\centering
\includegraphics[width=0.9\linewidth]{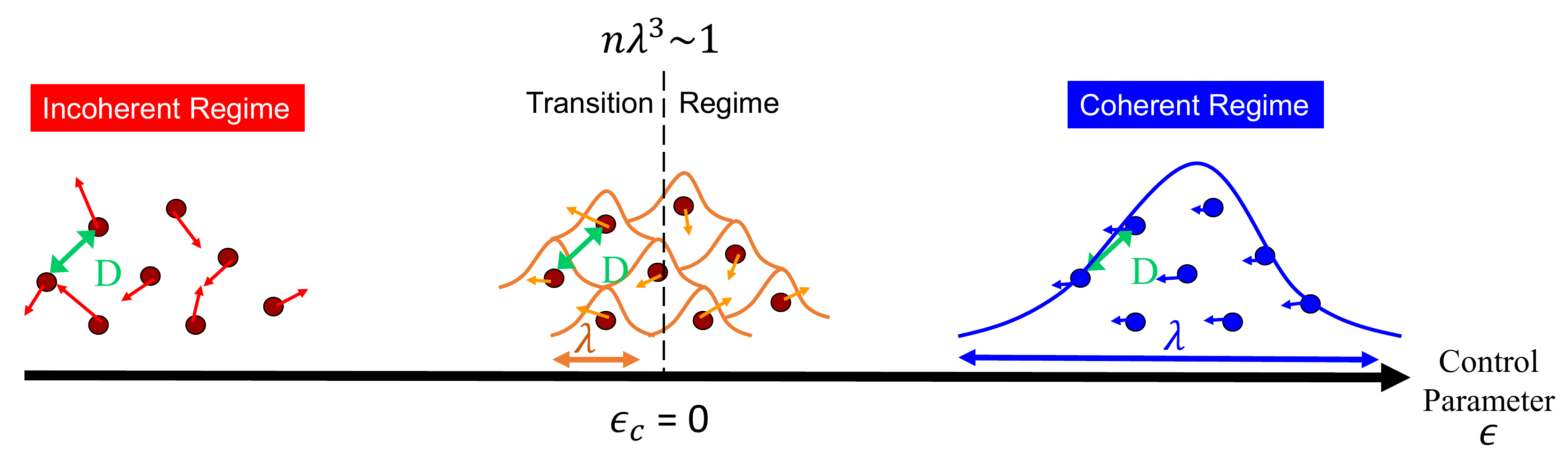}%{Fig-1.png}
\caption{
Schematic of the transition from a system of incoherent, distinguishable (potentially composite) bosonic particles having a mean interparticle distance $D \sim n^{-1/3}$ (left), through the critical region in which the de Broglie wavelength, $\lambda$, becomes comparable to $D$ (when $n \lambda^3 \sim 1$) (middle), to the quantum-degenerate (highly coherent) regime
%and ultimately exceeds, such the mean inter as a function 
in which the typical wavelength much exceeds the mean interparticle distance (right)
in terms of an external control parameter $\epsilon$ (e.g.~corresponding to decreasing system temperature). Depicted arrows indicate particle velocities, which typically start aligning in the transition region (when the external control parameter, $\epsilon$ acquires a critical value $\epsilon_c$, which is typically set to zero for convenience) eventually forming a coherent matter wave, known as a Bose-Einstein condensate, with a unique (randomly chosen) direction; the reduced vector magnitudes shown here indicate cooling.
}
\label{fig:1}
\end{figure*}
%-------------------------------------

\section{Introduction} \label{sec:intro}

Bose-Einstein condensation (BEC), first  predicted nearly 100 years ago, is a phenomenon associated with the emergence of a highly-coherent state which manifests itself across vastly different physical systems: Interestingly, not only the existence of the effect, but also the manner in which it emerges dynamically across such systems can exhibit the same characteristic features~\cite{ProukakisUBEC}. This chapter reviews such universality of the existence and the dynamical emergence of a Bose-Einstein condensate starting from an incoherent (or thermal) state: in particular, attention is paid to universal scaling laws and (dynamical) self-similarity  associated with the initial growth stages when the system breaks the Hamiltonian symmetry to randomly choose phase domains in a potentially `turbulent' manner, and with the subsequent relaxation of such a non-equilibrium state to the corresponding final equilibrium state (or steady state).
%in particular I note that although the final state is defined by the particular system properties and dimensionality, the dynamical relaxation path to such a state can still exhibit a range of universal dynamics.
The main aim of this chapter is to give an overview of key concepts of (selected) universal scaling laws which 
have developed over decades across distinct branches of physics, and focus on those with recent applications to experimentally-controlled quantum gases; as such, this Chapter is only intended to give a `flavour' and not a thorough overview of such topics across other systems, for which the reader is referred to suitable review articles.
Nonetheless, before concluding, I also briefly discuss a less widely known emerging new example characterizing the interplay between condensed matter physics, non-equilibrium statistical physics and cosmology/high-energy physics, in the context of how quasi-condensation, Bose-Einstein condensation and turbulence, characterized through relevant correlation functions,
could potentially relate to our understanding of large-scale cosmological structures.
%as this is an emerging new example characterizing the interplay between condensed matter physics, non-equilibrium statistical physics and cosmology/high-energy physics.

\subsection{`Universality' of Bose-Einstein Condensation} \label{sec:ubec}

Bose-Einstein condensation (BEC) is a fundamental manifestation of many-particle coherence in a quantum system, occurring when a system of many particles acquires wave-like properties. Such behaviour emerges  when the de Broglie wavelength $\lambda \sim h/p$ (where $h$ is Planck's constant and $p$ is momentum) of the underlying particles becomes comparable to, or exceeds, the typical inter-particle separation in the medium (Fig.~1). 
The textbook discussion of this effect~\cite{Huang:1987,Dalfovo1999,Pitaevskii:2003,Pethick:2002,ueda:2010}
 {\em (... see also Chapters by Stringari and Smith ...)} typically focusses on the case of a three-dimensional (3D) system of bosonic particles (i.e.~particles of integer spin obeying Bose-Einstein statistics), for which such condensation emerges when the dimensionless phase-space density $n \lambda^3 \gtrsim O(1)$ (where $n=N/V$ is the number density of particles in the system) exceeds a value of order 1 (in a 3D non-interacting system, the exact value is $\zeta(3/2) \approx 2.612$). This occurs at a characteristic, or critical, value ($\epsilon_c$) of the relevant external parameter, $\epsilon$, which varies across, and thus induces, the transition. 
For example, in the well-studied case of particles being cooled (and noting that for such systems $\lambda \sim (mT)^{-1/2}$),  BEC emerges at a critical temperature, $T_c$ -- a well-known characteristic in superfluidity/condensation phase diagrams of liquid helium and ultracold quantum gases.

However, in general, the role of the critical external control parameter could be played by many different physical variables~\cite{UBEC-Intro}: for example, for the systems considered in this Chapter, beyond a critical temperature or a critical density, such a phenomenon could also arise in terms of a a critical interaction -- whose magnitude can also be controlled by the system confinement/dimensionality -- or a critical pumping rate, although more broadly a system could be induced to criticality through a critical magnetic field or pressure. 
%(e.g.~in exciton-polariton condensates~\cite{deng2010exciton,carusotto2013quantum}).

The bosonic particle undergoing such change of state can be fundamental (e.g.~photons~\cite{Klaers:2010}), or composite -- noting that an even number of fermions also acquires integer spin, thus behaving bosonically.
The simplest example of a composite bosonic particle is that of a Hydrogen $^{1}$H atom, which comprises of two fermions: a proton in the core, surrounded by a single electron.
As it turns out, significant success in laser cooling and trapping~\cite{Chu-RevModPhys.70.685,Tannoudji-RevModPhys.70.707,Phillips-RevModPhys.70.721} has facilitated BEC to be achieved in a range of dilute (and thus typically weakly-interacting) alkali gases~\cite{Cornell-RevModPhys.74.875,Ketterle-RevModPhys.74.1131}: in fact the first observations of a {\em weakly-interacting} BEC across any physical system were made in ultracold atomic gases of $^{87}$Rb and $^{23}$Na (prior to its observation in spin-polarized $^{1}$H which followed a few years later~\cite{Fried:1998,UBEC-Kleppner}).
Helium is the first known superfluid (and the only one known to exist in thermodynamic equilibrium). Although the well-studied bosonic $^{4}$He, which is also a composite boson, exhibits a clear transition at some characteristic temperature and reaches 100\% superfluidity at $T=0$, the fraction of condensed particles in such a system nonetheless remains under 10\% due to the relatively strong atomic interactions.
Composite bosonic quasiparticles can also be formed from an even number of fermions: the most striking example is that of electron-hole Cooper pairs in superconductors or pairs of fermionic $^{3}$He atoms exhibiting superfluidity. Although the detailed microscopic mechanism here is different (the well-known Bardeen--Cooper--Schrieffer, or BCS, mechanism~\cite{cooper2010bcs}) {\em (... see Chapter ...)}, ultimately the composite bosonic quasiparticle still effectively undergoes a BEC-type transition to a state of macroscopic coherence~\cite{Leggett:2006}. In fact, the exquisite controlled setting of ultracold atomic gases enables one to probe, even experimentally, the transition from a BEC to a BCS coherent system~\cite{zwerger2011bcs} {\em (... see  Chapter ....)}.
Another example receiving significant current attention over the past 15 years is the condensation of a composite half-matter-half-light quasiparticle known as an exciton-polariton condensate, which manifests itself in pumped semiconductor microcavities~\cite{sanvitto2012exciton,deng2010exciton,carusotto2013quantum} {\em (... see also chapter on quantum fluids of light ...)}.

%-------------------------------------
\begin{figure}[b!]
\centering
\includegraphics[width=1.0\linewidth]{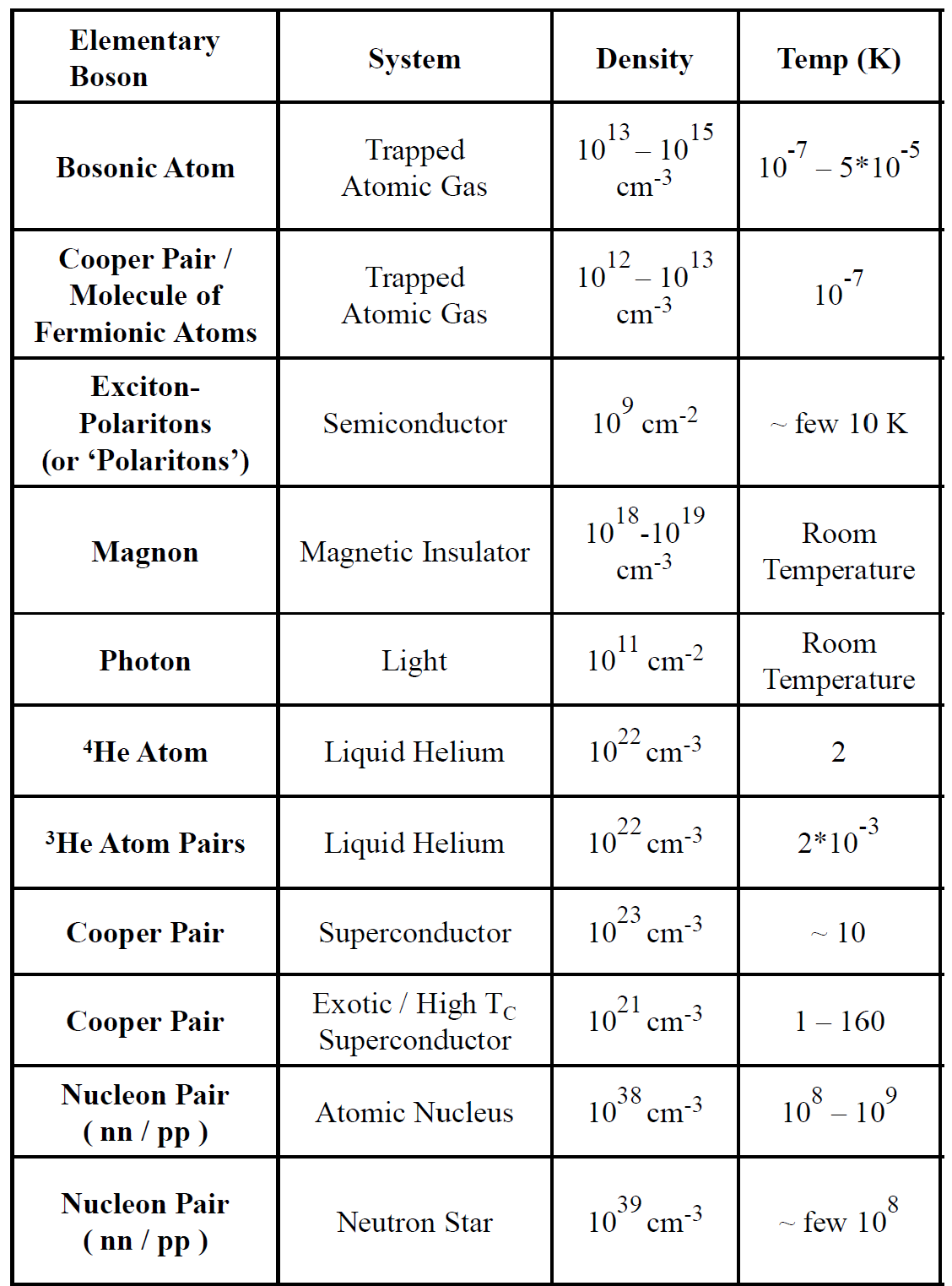}
\caption{
Examples of very different systems exhibiting macroscopic quantum coherence and Bose-Einstein condensation, facilitated through a high dimensionless phase-space density $n \lambda^d \sim 1$ (where $d$ the system dimensionality). Adapted with permission from N.~P.~Proukakis, Quantum Gases: Finite temperature and non-equilibrium dynamics, N.~P.~Proukakis {\em et al.} (Eds.), Imperial College Press (London, 2013)~\cite{FINESS-Proukakis}. For a more extended discussion of typical control parameters, probes, observables and related characteristic properties and dimensionalities across such diverse systems see Ref.~\cite{UBEC-Intro}.
}
\label{fig:2-table}
\end{figure}
%-------------------------------------

Signatures of BEC (and macroscopic coherence in general) in fact arise across an immensely broad range of length and energy/temperature scales \npp{(see Fig.~\ref{fig:2-table})}, spanning from nuclear and atomic physics, through optical and condensed matter systems to astrophysical (neutron star cores) and even cosmological (fuzzy dark matter) settings -- a more extended discussion on such diverse manifestations can be found in Ref.~\cite{ProukakisUBEC}.

Although BEC was initially discussed in the context of a homogeneous, non-interacting 3D system, realistic systems are in fact of finite size and do exhibit some interactions -- which can mask the effects of BEC, as in the case of superfluid $^{4}$He. This led to a many-decade-long search for experimentally-controllable weakly-interacting systems which could exhibit BEC: typical ultracold atomic experiments have provided a natural setting for achieving and observing nearly pure condensates, despite their (typically) inhomogeneous nature and finite system size.
Current experiments in such systems in fact facilitate a detailed understanding of the modifications to both equilibrium coherence properties and dynamical formation induced by inhomogeneous/trapped and finite-size configurations, the influence of (realistic) weak interactions, the transition to more strongly interacting settings, and the confinement to lower effective dimensionalities, alongside studies of mixtures, superfluid turbulence, quantum phase transitions and even long-range interactions and supersolidity. 
Moreover, the last 2 decades have facilitated another well-controlled system, namely that of exciton-polaritons, which additionally offer glimpses to non-equilibrium phase transition features.

%where appropriate system-specific modifications to such standard picture have been adopted.

Beyond a classification of equilibrium properties of such systems, a fascinating generic underlying question with a history of significant contributions spanning many decades and across diverse physical settings, concerns  the fundamental question of how such systems form, and what common -- or universal -- features may exist during such a process.
%
%This is an extremely rich topic, with a history of significant contributions spanning many decades, with notable recent progress made in the context of controlled experiments with ultracold atomic gases.

In this Chapter, I discuss in a broad sense the main aspects of the dynamical path towards such a highly-coherent state, focussing primarily on the context of trapped ultracold (bosonic) atomic gases, with some emphasis on realistic inhomogeneous settings which facilitate direct links to controlled experiments:
 applicability and extension of such concepts to other physical settings and systems -- most notably exciton-polariton condensates -- are also discussed at relevant points within this Chapter.
 %most notably, explicit discussion is made on the modifications to macroscopic coherence phenomena in a system of reduced dimensionality, noting that strictly speaking 2D systems exhibit a distinct phase transition, known as the Berezinskii-Kosterlitz-Thouless (BKT) phase transition rather than BEC.
The presentation is kept at a broad introductory level, geared to a non-expert audience, with the intention to give a flavour of such rich questions -- for a more complete picture, and historical overview, the reader is referred to a number of excellent reviews on such topics~\cite{ProukakisUBEC,del_campo_universality_2014,delcampo-kibble-zurek,bray_theory_1994,rutenberg_energy-scaling_1995,schmied_non-thermal_2019,chantesana_kinetic_2019,40YearsBKT,PolaritonReview-Nature2022}. %[UBEC,KZM reviews,NTFP,BKT,QFL].

This Chapter is structured as follows:
The rest of Sec.~\ref{sec:intro} gives a unified qualitative overview of the dynamical stages in the transition process from an initial incoherent state to the emerging final equilibrium or steady-state of the system (Sec.~\ref{sec:stages}), discussing the key physical quantities used to characterize such a system (Sec.~\ref{sec:coherence}), with a brief introduction to the basic phenomenological description of a BEC (Sec.~\ref{sec:modelling}).
Sec.~\ref{sec:kz-main}, which forms the core of this Chapter, addresses how coherence grows dynamically during an external parameter quench through spontaneous symmetry breaking and defect formation in the context of the so-called Kibble-Zurek mechanism giving rise to universal scaling laws:
after giving a brief historical introduction (Sec.~\ref{sec:earlyideas}) and presenting a minimal numerical model capturing such effects (Sec.~\ref{sec:sgpe}), the key underlying principles and scaling law of Kibble-Zurek are described mathematically by extending the ideas of equilibrium criticality (which are thus also briefly reviewed here in Sec.~\ref{sec:eqm-criticality}) to dynamical transition crossing (Sec.~\ref{sec:dyn-criticality}-\ref{sec:kz-hom}) and then applying them to recent experiments (Sec.~\ref{sec:kz-spgpe}) which highlight the need for a more extended framework to account for inhomogeneous systems and their interplay with causality, which is also presented here (Sec.~\ref{sec:kz-sonic}).
Sec.~\ref{sec:phaseordering-main} discusses universal phase-ordering dynamics following an instantaneous quench, demonstrating the emergence of universal scaling of the spatial correlation function in characteristic late-time evolution stages of a relaxing quantum gas (Sec.~\ref{sec:scalinghypothesis}), with particular emphasis on highly-non-equilibrium quenches giving rise to `turbulent' states evolving in the vicinity of non-thermal fixed points (Sec.~\ref{sec:ntfp}).
Sec.~\ref{sec:other} presents a highly-relevant (but heuristic) selection of applications of such concepts to other systems, briefly mentioning for completeness other effects and manifestations in ultracold quantum gases not discussed here (Sec.~\ref{sec:other-atomic}). Sec.~\ref{sec:excitonpolariton} focuses on the emergence of dynamical features discussed in Secs.~\ref{sec:kz-main}-\ref{sec:phaseordering-main} and the related spatiotemporal scaling according to a (quantum) mapping 
of phase evolution in an open, driven-dissipative, quantum gas of exciton-polaritons
to the well-known (classical) Kardar-Parisi-Zhang equation characterizing interfacial growth.
Sec.~\ref{sec:cosmo-bec} discusses the potential relevance of some of the discussed characterizations to the context of a cosmological-scale BEC, which is shown to exhibit many characteristics similar to harmonically-confined ultracold quantum gases.
General concluding remarks and outlook are given in Sec.~\ref{sec:conclusions}.

%-------------------------------------
\begin{figure*}[t!]
\centering
\includegraphics[width=0.8\linewidth]{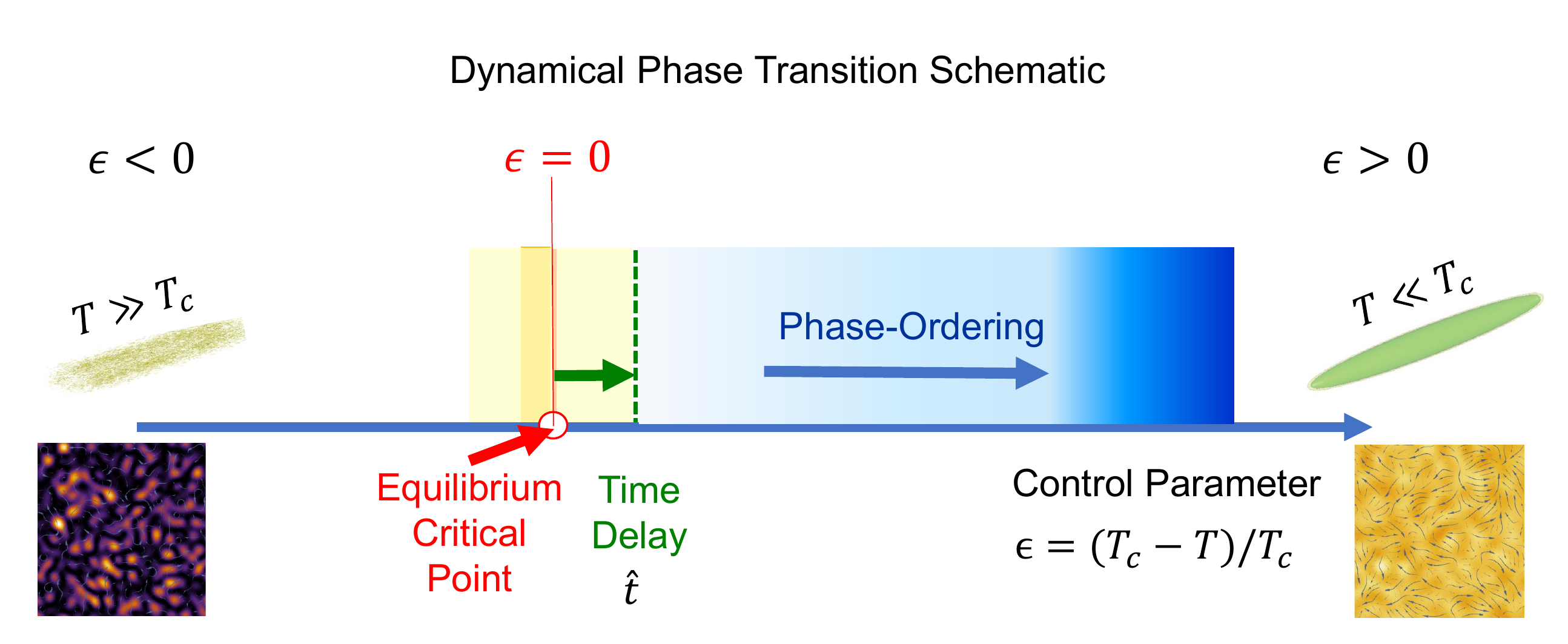}%{Fig-2a.png}
\caption{
Dynamical Phase Transition Schematic: As the external control parameter, $\epsilon$, is dynamically varied across the transition (i.e.~$\epsilon\rightarrow\epsilon(t)$), the system exhibits a delayed response compared to the corresponding equilibrium critical point at $\epsilon=\epsilon_c$ (corresponding, e.g.~to $T=T_c$), with $\epsilon_c$ typically set to $0$, thus separating the incoherent ($\epsilon <0$) from the coherent ($\epsilon >0$) side of the phase transition. As soon as $\epsilon(t)$ dynamically crosses to a positive value, and still within the critical regime (i.e.~$\epsilon(t)\rightarrow 0^+$), it acquires (in general) a chaotic, turbulent non-equilibrium configuration containing  phase defects: such physics is dominated by a universal Kibble-Zurek scaling law governing the dynamical phase transition. The subsequent gradual relaxation of such defects leads to the emergence of macroscopic coherence  over much longer timescales: this includes a self-similar `phase ordering' process obeying universal scaling laws. Other universal features discussed in this Chapter (such as spatio-temporal evolution around non-thermal fixed points and Kardar-Parisi-Zhang scaling) may also appear at appropriate scales and evolutionary stages during the extended process, depending on the particular system, initial state preparation and quenching protocol.
Specific equilibrium density snapshots shown here sufficiently before/after the transition are for the specific cases of an elongated inhomogeneous 3D, and a homogeneous 2D, ultracold atomic gas -- both analyzed within this Chapter.
}
\label{fig:2}
\end{figure*}
%-------------------------------------

\subsection{Dynamical Phase Transition Stages} \label{sec:stages}

Condensate formation is a fundamental non-equilibrium problem with a very long history dating over many decades (see, e.g.~selected influential early works~\cite{kagan-greenbook,stoof-greenbook,Stoof-JLTP-long}, discussions~\cite{davis-growth-98,zaremba-stoof-growth-PhysRevA.62.063609} of the first quantitative ultracold atomic growth experiments~\cite{ketterle-bec-growth,davis-essliner-growth-PhysRevLett.88.080402} and the review~\cite{UBEC-Proukakis} which contains many more highly-relevant references).
The main question to be addressed here is how a system, originally in a disordered/incoherent state transitions, through the breaking of a continuous symmetry, to an equilibrium state (or steady-state)  exhibiting a high degree of macroscopic coherence, the extent of which is set by the particular system under consideration (e.g. system geometry and dimensionality, nature and strength of interactions between constituent particles, mass of constituent boson).
A schematic of the main steps during such a process is shown in Fig.~\ref{fig:2} in terms of the external control parameter $\epsilon$.%(for the specific case study of an elongated 3D atomic gas undergoing cooling at a constant rate -- details of which will be reviewed later).

From an equilibrium perspective, such a transition can be visualized as a gradual passage through different equilibrium states, with a well-identified critical region (around $\epsilon \sim \epsilon_c \sim 0$) separating purely incoherent from emerging coherent states. But how is such picture affected when the external control parameter is dynamically varied, i.e.~$\epsilon \rightarrow \epsilon(t)$?
Let us consider here %for concreteness 
by means of an example the specific case of a temperature quench, for which $\epsilon = (T_c-T)/T_c$.

During the first stage of cooling, and while the system is still (sufficiently) above the critical temperature, particle dynamics can be described in the usual (classical) phase-space picture, based on a Boltzmann equation with collisions~\cite{lifshitz1995physical}. Under continuous cooling towards the critical point, particles are gradually redistributed into lower-energy states (with elastic collisions typically occuring on a much faster timescale than that governing the variation of the external control parameter, thus providing the required thermalization).

The second, and arguably most interesting, stage in the evolution addresses the question of how the system dynamically crosses the phase transition, and what the nature of the non-equilibrium state generated in this vicinity of the critical point is. %How is such process affected by dimensionality? 
This is discussed in detail in Sec.~\ref{sec:kz-main} below: for now, it suffices to say that this leads to a delay in the crossing of the phase transition, and the emergence of interesting (potentially `chaotic', or `turbulent-like') non-equilibrium states displaying a (potential) abundance of phase defects and relevant universal properties:
%
%After dynamically crossing the phase transition, the system is left in a rather non-equilibrium (potentially `chaotic', or `turbulent-like') state, which exhibits interesting properties and a (potential) abundance of phase defects -- details of which 
such emerging transient states depend, in general, on both the dimensionality/geometry of the system and the rate of crossing of the phase transition (e.g.~adiabatically vs.~instantaneously).

The final stage in the equilibration/thermalization of the system is intricately related to the subsequent relaxation of such phase defects, and the related emergence of (a certain degree of) macroscopic coherence across the system: universal behaviour is again expected to emerge here in some dynamical scenarios/regimes (e.g.~for homogeneous systems, or very rapidly quenched, highly non-equilibrium states), but the specific system details can potentially also leave their imprint on this process, thus masking (or even potentially modifying the details of) such universality.

%Such a transition can of course proceed adiabatically (or effectively adiabatically), by transitioning gradually through similar equilibrium states, in each of which the system particles are coherent over the entire spatial extent of the system under consideration. While such distinct equilibria are of course interesting in their own right, as they highlight the exact location and nature of the underlying phase transition in the region of critical fluctuations, here we focus on the more interesting case of the dynamical evolution from incoherent to coherent properties. In such a scenario, a system originally in a purely incoherent/chaotic state dynamically transitions, via the breaking of an underlying symmetry, to a coherent phase-ordered (or condensed) state. (Fig.~2)

%This is a fundamental physical question valid across a broad range of systems, and has a long history on its own. 
Such discussion can be formulated in terms of two commonly considered types of quenched regimes: instantaneous quenches, where the system is prepared, or induced, into a highly-non-equilibrium state, and finite-duration (i.e.~non-instantaneous) quenches starting from an incoherent equilibrium configuration, in which the system is typically modelled on the basis of a smoothly varying external parameter inducing such transition.
Both such cases lead to interesting universal physical laws, associated with the dynamical phase transition crossing and the subsequent phase-ordering process, as discussed in more detail below (Secs.~\ref{sec:kz-main}-\ref{sec:phaseordering-main}), starting from the finite-duration phase transitions.

%%%%%%%%%%% HERE %%%%%%%%%%%%%%%%%%%%%%%%%%%%%

\subsection{Relevant Coherence Measures} \label{sec:coherence}

Classification of the system equilibrium and dynamical properties is intricately related to appropriate characterization of the degree of system coherence.

\subsubsection{Correlation Functions}

For a quantum system described by Bose field operators $\hat{\Psi}({\bf r},t)$ and $\hat{\Psi}^{\dag}({\bf r},t)$, a classification of the coherence in both the spatial and temporal domains can be obtained through the $n^{\rm th}$-order correlation function
\begin{eqnarray}
G^{(n)}({\bf r}_1,{\bf r}_1',{\bf r}_2,{\bf r}_2',\cdots ,t_1,t_1',t_2,t_2',\cdots) &=& \nonumber \\ && \hspace{-5.0cm} = \langle\,\, \hat{\Psi}^{\dag}({\bf r}_1,t_1) \hat{\Psi}^{\dag}({\bf r}_2,t_2) \cdots \hat{\Psi}^{\dag}({\bf r}_n,t_n)
\nonumber \\ && \hspace{-4.3cm}
\hat{\Psi}({\bf r}_n',t_n') \cdots  \hat{\Psi}({\bf r}_2',t_2') \hat{\Psi}({\bf r}_1',t_1') \,\,\,\, \rangle
\label{eq:gn}
\end{eqnarray}
measured across $n$ different spatial locations and at $n$ different times (with $\cdots$ intended to reflect remaining terms such that there are $n$ pairs of $\hat{\Psi}^*({\bf r}_n,t_n) \, \hat{\Psi}({\bf r}_n',t_n')$ contained within the above expectation value expression).

In practice, the most elementary -- and thus most commonly discussed -- correlation functions are those associated with spatial and temporal one-body correlations, respectively defined as
\bea
 {\rm Fixed}\,\, t:\,\,\, C({\bf r},{\bf r'}) = C({\bf r},{\bf r'},t,t)  = \langle \hat{\Psi}^{\dag}({\bf r},t) \hat{\Psi}({\bf r'},t) \rangle \\ %\hspace{1.0cm} {\rm and} \hspace{1.0cm}
 {\rm Fixed}\,\, {\bf r}:\,\,\,\, C(t,t') = C({\bf r},{\bf r},t,t') = \langle \hat{\Psi}^{\dag}({\bf r},t) \hat{\Psi}({\bf r},t') \rangle \;,
\eea
%\be
%C(r,r') = C(r,r';t,t) \hspace{1.0cm} {\rm and} \hspace{1.0cm} C(t,t') = C(r,r;t,t') \;,
%\ee
as these contain the primary information about spatial and temporal coherence.
%The first of these (spatial $C(r,r')$) is also known as the one-body density matrix $\rho(r,r';t)=\langle \hat{\Psi}^{\dag}(r,t) \, \hat{\Psi}(r',t) \rangle$.

The typical definition of equilibrium Bose-Einstein condensation is in the form of so-called Off-Diagonal-Long-Range Order (ODLRO), associated with the non-vanishing asymptotic (spatial) behaviour of  the one-body density matrix $\rho({\bf r},{\bf r'},t)=\langle \hat{\Psi}^{\dag}({\bf r},t) \, \hat{\Psi}({\bf r'},t) \rangle$~\cite{Pitaevskii:2003,ueda:2010,Leggett:2006}.
In particular, ODLRO is defined by
%$\langle \hat{\Psi}^{\dag}({\bf r};t) \hat{\Psi}({\bf r}',t) \rangle$, as
\be
%{\rm Lim}_{|{\bf r}-{\bf r}'| \to \infty} \rho({\bf r},{\bf r}',t) = 
{\rm Lim}_{|{\bf r}-{\bf r'}| \to \infty} \langle \hat{\Psi}^{\dag}({\bf r},t) \hat{\Psi}({\bf r'},t) \rangle \neq 0 \;,
\ee
with its value tending to some non-zero constant limit 
$\psi^*(r,t)\,\psi(r',t)$,
%$\langle \hat{\Psi}^{\dag}(r,t) \rangle \langle \hat{\Psi}(r',t) \rangle$, 
consistent with the existence of off-diagonal correlations in the one-body density matrix.
This introduces the quantity $\psi(r,t)$
%$\langle \hat{\Psi} \rangle$ 
as the so-called order parameter of the system.
To highlight the existence of a non-zero value, this is usually expressed as $\psi=|\psi|e^{i \theta}$, a form which facilitates an analogy to classical fluids through the mass density ${\cal \rho} = m |\psi|^2$ and superfluid velocity $v = (\hbar/m) \nabla \theta$~\cite{Pitaevskii:2003}. 

Moreover, one often connects the BEC order parameter with the expectation value of the Bose field operator through
$\psi({\bf r},t) = \langle \hat{\Psi} \rangle$. Such an argument is motivated by the analogous classical description of the electromagnetic field, in which
one introduces a complex `classical' electric field as the expectation value of the corresponding (quantum-mechanical) electric field operator. While somewhat useful, there are some subtleties in following such procedure for massive particles, like atoms, which (unlike photons) cannot be created/destroyed from the vacuum by single-particle creation/annihilation operators: an insightful discussion of this issue, and the definition of BEC can be found in the book by Leggett~\cite{Leggett:2006}.
While models explicitly maintaining the quantum-mechanical nature of the operator also exist, it is not essential to discuss such an extension here, with the reader instead referred to recent reviews~\cite{proukakis_finite-temperature_2008,Proukakis13Quantum}.
For the purposes of this discussion, we approximate the actual field operator by a fluctuating classical field (obeying a classical probability distribution in phase space), thus accommodating both density and phase fluctuations into the subsequent description
%
%At this point, we make some important remarks which will facilitate the subsequent analysis:
%Firstly, 
%
%Although the quantum description (and typical Hamiltonian) of a system is cast in terms of the Bose field operators $\hat{\Psi}^{(\dag)}(r,t)$, it is often convenient to re-express these in terms of complex classical fields $\Phi^{(*)}(r,t)$, as this 
in a manner which significantly simplifies the direct numerical modelling (see later). 
%This is equivalent to making a semiclassical approximation to the Bose field operator %in the form $\langle \hat{\Psi}^{(\dag)}(r,t) \rangle = \Phi^{(*)}(r,t)$.
%and is notionally analogous to the classical description of the electromagnetic field. 
%Performing the replacement $\langle \hat{\Psi}^{(\dag)}(r,t) \rangle \rightarrow \Phi^{(*)}(r,t)$ introduces
%the order parameter of the system no longer has the same symmetries (in terms of a global phase rotation) as the original system hamiltonian -- i.e.~the system chooses a configuration with a definite phase.
%
%This is often discussed in terms of $U(1)$ symmetry breaking, in the sense that the arising complex field giving the order parameter of the system no longer has the same symmetries (in terms of a global phase rotation) as the original system hamiltonian -- i.e.~the system chooses a configuration with a definite phase.
%While models explicitly maintaining the quantum-mechanical nature of the operator also exist, it is not essential to discuss such an extension here, with the reader instead referred to~\cite{proukakis_finite-temperature_2008,Proukakis13Quantum}.
%
All discussion from this point onwards is thus expressed in terms of the complex/multimode classical field, which in this work will be denoted by $\Phi({\bf r},t)$ and also referred to occasionally as the order parameter.

When studying the correlation in a given system, it is
often useful to deal with appropriately normalized correlation functions, denoted by $g^{(n)}$: for example,  the above general expression [Eq.~(\ref{eq:gn})] (written now in terms of the classical field $\Phi({\bf r},t)$) can be normalized by the product $\Pi_{i} \sqrt{ \langle |\Phi({\bf r}_i,t_i)|^2 \rangle \langle |\Phi({\bf r}_i',t_i')|^2 \rangle }$. This is useful as it implies that  the same-location (all ${\bf r}_i,\,{\bf r}_i'$ the same), equal-time (all $t_i,\,t_i'$ the same) normalized correlation function $g^{(n)}(0)$ 
has the elegant limiting cases~\cite{Cohen-Tannoudji:2001}
\bea
g^{(n)}({\bf r}_i\cdots {\bf r}_i;t_i\cdots t_i) = \left\{ \begin{array}{ll} 1 \,\,\,\,\, {\rm (Purely \, Coherent)}\\ n! \,\,\,\,\,\,\,\,\,\,\,\, {\rm (Incoherent)} \end{array} \right\} \;.
\label{eq:gn-normalized}
\eea
%In this context we also note that a (somewhat idealized) fully-coherent system will exhibit a value of the normalized correlation function $g^{(n)}(r_1,r_2;t_1,t_2,\cdots) = 1$ for all $r_i$ and $t_i$.
%While a corresponding general expression is hard to obtain in the chaotic, thermal, or incoherent regime, we note that the special case of all $r_i$ and $t_i$ being the same leads to the very useful result (used later) that
%\be
%g^{(n)}(r_1,r_1,\cdots r_1;t_1,t_1,\cdots t_1) = n! \;.
%\ee

In what follows, I will use interchangeably the (unnormalized) correlation functions
$C({\bf r},{\bf r'})$ and $C(t,t')$,
or their normalized versions respectively defined by
%In practice, the most commonly considered correlation functions are those which investigate the dependence on the spatial, or temporal, variation of the spatial coherence, or phase correlation function, keeping the other parameter fixed, i.e., 
\be
g^{(1)}({\bf r},{\bf r'}) = \frac{\langle \Phi^*({\bf r},t)\Phi({\bf r'},t) \rangle}{\sqrt{\langle |\Phi({\bf r},t)|^2\rangle}\sqrt{\langle |\Phi({\bf r'},t)|^2\rangle}}\;, \label{eq:g1r}
\ee
and
\be
g^{(1)}(t,t') = \frac{\langle \Phi^*({\bf r},t)\Phi({\bf r},t') \rangle}{\sqrt{\langle |\Phi({\bf r},t)|^2\rangle}\sqrt{\langle |\Phi({\bf r},t')|^2\rangle}}\;. \label{eq:g1t}
\ee
%The former contains crucial information on spatial correlations, such as the extent of spatial coherence, while the latter does the same for temporal correlations, giving a timescale over which such coherence is maintained: in fact, 
Note that although Eqs.~(\ref{eq:g1r})-(\ref{eq:g1t}) are often considered independently, a combined analysis of these two quantities can provide further crucial information about the extent of the `equilibrium' (or `non-equilibrium')  nature of the system, or the emerging spatiotemporal scaling relations (see subsequent discussion on exciton-polaritons in Sec.~\ref{sec:excitonpolariton}).

As previously mentioned, the hallmark of BEC in a 3D system is the striking emergence of ODLRO, signifying that (some degree of) coherence is maintained across all space, i.e.~$g^{(1)}({\bf r},{\bf r'};t)$ tends to a non-zero-valued plateau (having a value between 0 and 1) as $|{\bf r}-{\bf r'}| \rightarrow \infty$: for a finite-size system, it is of course only meaningful to study the dependence of $g^{(1)}$ on $|{\bf r}-{\bf r'}|$ over the spatial extent of the (coherent part of the) system. In fact, the extent of deviation of the plateau value from the starting value of 1 (taken at ${\bf r}={\bf r'}$) provides information about the fraction of incoherent particles co-existing with the condensed part~\cite{Cohen-Tannoudji:2001}.

Key emerging concepts utilized throughout this Chapter, and described in more detail later, are those of self-similarity and scaling hypothesis. % and the concept of self-similarity.
In a nutshell, the underlying idea is that under appropriate conditions (e.g.~in a restricted spatial and/or temporal window) the microscopic details of the system are not relevant, with the physical description of key system observables acquiring a universal description in terms of appropriately rescaled variables and a specific functional form.
Elimination of non-essential details at short scales (times) in favour of emphasizing large-scale (long-time) features -- a key step in the universal description of a physical system associated with the analysis of critical phenomena and the concept of renormalization group theory (see e.g.~\cite{nishimori-phasetransition-book,binney1992theory,RG-newphysics} {\em ... or the Chapter by Fradkin  in this volume ....}  ) -- is possible when there exists a characteristic lengthscale (timescale) which is much larger (longer) than all other relevant system scales (but smaller than the system size/evolution time). %which allows us to identify universal scaling functions for the system correlations through our ability to 
This in turn allows us to express relevant correlation functions over a limited spatial and/or temporal range in terms of a unique (but not {\em a priori} known) functional form whose arguments are appropriately rescaled spatial (temporal) variables, combined with a small number of critical exponents characterizing such behaviour~\cite{nishimori-phasetransition-book,binney1992theory}.
Details of such scalings will become more apparent at relevant points throughout this Chapter.

%these will depend in a universal way on a general functional form (yet to be determined) which can be written during relevant spatial/temporal windows in terms of the appropriately rescaled spatial/temporal variables and a set of characteristic critical exponents characterizing such behaviour.

Given the existence of both density and phase fluctuations in a realistic system, it is natural to examine their interplay and relative importance. In general, as a system approaches quantum degeneracy from the incoherent regime, density fluctuations start becoming suppressed at an earlier point (larger value of $|\epsilon|$) than phase fluctuations.
In the specific context of cooling (thermal phase transition), this implies the emergence of two distinct characteristic temperatures for suppression of density and phase fluctuations.
%with their corresponding characteristic temperatures being thus decoupled (somewhat analogously to conventional condensed matter phase diagrams [ ]): 
This is particularly pronounced in lower effective dimensionalities (i.e.~in quasi-1D and quasi-2D geometries)~\cite{Petrov:Low-D-Review}, but such effect is still present in 3D geometries, even if largely obscured (and so, sub-dominant).

In order to also investigate the emergence of density fluctuations, 
we thus introduce here the second-order, or density-density correlation function (taken here for simplicity at equal times)
\begin{eqnarray}
g^{(2)}({\bf r},{\bf r'},t) &=& \frac{\langle \Phi^*({\bf r},t)\Phi^*({\bf r'},t)\Phi({\bf r'},t)\Phi({\bf r},t) \rangle}{\langle |\Phi({\bf r},t)|^2\rangle \langle |\Phi({\bf r'},t)|^2\rangle} \nonumber \\ 
 &=& \frac{\langle n({\bf r},t)\,n({\bf r'},t) \rangle}{\langle n({\bf r},t) \rangle\, \langle n({\bf r'},t) \rangle}
\;,
\end{eqnarray}
%Such off-diagonal quantity is formally related to $g^{(1)}(r,r';t)$. 
%In fact one typically only probes
with local density-density correlations accessible by evaluating such quantity at $r'=r$.

As a direct consequence of Eq.~(\ref{eq:gn-normalized}), a pure BEC can be identified through $g^{(1)}({\bf r},{\bf r},t)=g^{(2)}({\bf r},{\bf r},t)=1$, whereas an incoherent (thermal) field has $g^{(2)}({\bf r},{\bf r},t)=2$: this simple identification provides a further probe for testing the degree of coherence of a system at a given time $t$.

\subsubsection{Penrose-Onsager Condensation vs.~Quasi-Condensation} \label{sec:PO}

The above discussion has introduced the concept of ODLRO for characterizing a condensate at equilibrium: However, one is typically dealing with a finite-size system, and so such discussion has to be necessarily limited to within the spatial extent of the system (which should much exceed all other characteristic system lengthscales to make this non-trivial), and as limiting behaviour in the thermodynamic limit.

In practice, for a finite-sized interacting system one can instead proceed with a different characterization, following a seminal contribution by Penrose and Onsager~\cite{Penrose-Onsager}.
In particular, one can diagonalize the off-diagonal one-body density matrix $\rho({\bf r},{\bf r'},t)$ to calculate its eigenvalues $N_n$ through $\rho({\bf r},{\bf r'},t) = \sum_{n} N_n(t) \phi_n^*({\bf r},t) \phi_n({\bf r'},t)$, where $\phi_n({\bf r},t)$ denotes the corresponding $n$-th eigenfunction. 
If in doing so one finds a macroscopically dominating eigenvalue which is on the {\em same order} as the total particle number, then one can identify the mode corresponding to such eigenvalue with the Bose-Einstein condensate mode~\cite{Leggett:2006}.
Mathematically, a dominant eigenvalue (if one exists) and corresponding eigenfunction of the density matrix can be identified from the solution of
\be
\int {\rm d}{\bf  r'} \varrho ({\bf r}, {\bf r'}) \phi_n({\bf r'}) =  N_{n} \phi_n({\bf r}) \;.
\ee
%where the index $n$ labels the $n$-th eigenfunction $\phi_n$ of $\varrho(r,r')$. 
%
In practice one often uses a slightly relaxed criterion associated with the existence of a single eigenvalue which is much larger than all other eigenvalues of the system, but does not necessarily need to be on the same order as the total number of particles in the system.
Such a mode is nowadays routinely referred to as the Penrose-Onsager mode of the system (and the above condition known as the Penrose-Onsager criterion)~\cite{Blakie-AdvPhys-2008}.

In the context of numerical simulations of a fluctuating classical field,  it is often numerically prohibitive to perform such a calculation by means of an ensemble/trajectory averaging.
Thus, for purely practical reasons, one often resorts to time-averaging of a single trajectory as a numerical trick to achieve the same effect: the idea here is to compute the off-diagonal density matrix by averaging over a large number of samples spread over a timescale which is both longer than that governing the evolution of intrinsic fluctuations, and also  shorter than any typical evolution of interest in the system (e.g.~collective mode, large-scale motion of defects): in such a scenario, fluctuations in the field quickly randomise a single trajectory, thus making an average over different time evolutions of a single trajectory statistically analogous to averaging over many different independent trajectories: in other words, with careful choice of parameters, averaging over different time samples can also mimic the effect of averaging over different trajectories not only in a dynamical steady-state in which the order parameter fluctuates about some mean value, but even in a dynamical setting~\cite{Blakie-AdvPhys-2008}. 
%In that case, the averagging timescale should be
Note that global observables depending critically on the dynamical path additionally require averaging over (at least) a small number of trajectories to ensure the result is indeed statistically independent (and does not contain information about the system history): this is for example the case when probing correlation functions and defect numbers later in the context of quenched (Kibble-Zurek) dynamics in an inhomogeneous geometry [Sec.~\ref{sec:kz-spgpe}].

%Such identification only makes interpretative sense if one, or few, eigenvalues are dominant: in the former case the system is said to exhibit a well-defined Bose-Einstein condensate, with $\psi_0$ the condensate mode, and $N_\mathrm{PO}=N_0$ the condensate particle number.
%A system with no (significantly) prevailing eigenvalue, but exhibiting co-existence of few macroscopically occupied eigenmodes is characteristic of a quasi-condensate state, in which there is no overall phase coherence (except locally) due to the competition between dominant eigenmodes (exhibiting different random phases).

At this stage it is important to also comment on the role of system dimensionality:
As well-known, ODLRO -- and so BEC -- is strictly speaking restricted to 3D (or higher) configurations, with corresponding homogeneous ideal 2D systems (for all $T \neq 0$) only exhibiting a reduced degree of coherence, known as quasi-long-range (or topological) order; such distinction arises despite the fact that both systems exhibit superfluidity~\cite{Pitaevskii:2003,ueda:2010}.
Reduced coherence in 2D arises as a result of the dominant role of long-wavelength excitations destroying the overall phase coherence of the system, according to the well-known Hohenberg-Mermin-Wagner theorem. In such a case, the system exhibits a distinct phase transition, known as the Berezinskii-Kosterlitz-Thouless (BKT) transition~\cite{Kosterlitz-RevModPhys.89.040501,40YearsBKT}: this is associated for $\epsilon \gtrsim 0$ with the binding of the free vortices which typically proliferate above the critical temperature ($\epsilon >0$) -- and whose presence randomizes the system phase -- into vortex-antivortex pairs; such binding effectively screens their existence, which can be visualized as annihilation. 
%Since thermal fluctuations in 2D lead to the random creation and breaking of vortex-antivortex pairs, the
Thermal fluctuations in 2D at any non-zero temperature reduce the coherence of the system 
%becomes reduced for all $T \neq 0$ 
compared to the corresponding 3D case, with the corresponding the spatial profile of the correlation %function in a homogeneous 2D system is different from the 3D case, 
exhibiting instead in 2D a power-law decay and quasi-long-range order (as opposed to ODLRO).

Given that the dominant factor in the behaviour of the spatial correlation function $g^{(1)}({\bf r},{\bf r}')$ is the distance between the two points, i.e.~$|{\bf r}-{\bf r}'|$, and to simplify the expressions, here I arbitrarily choose to label one of these points as ${\bf 0}$, and the  distance between them as ${\bf r}$, such that
the relevant normalized spatial correlation function takes the form $g^{(1)}({\bf r})=g^{(1)}({\bf 0},{\bf r})$.
Below the critical temperature ($\epsilon\rightarrow 0^+$) -- and in the specific context of an infinite homogeneous system -- this takes in the $r \rightarrow \infty$ limit the dimensionality-dependent form
%This distinction is evident in the behaviour of $g^{(1)}(0,r)$ below the critical temperature:
\begin{eqnarray}
g^{(1)}(0,{\bf r})|_{0<T < T_c} =
\left\{ \begin{array}{cc}
    {\rm constant} & {\rm (3D)}   \\
     r^{-\alpha(T)} & {\rm (2D)} 
\end{array} \right\} \,, \label{eq:g1D}
\end{eqnarray}
with the value of the constant being temperature-dependent and lying between 0 and 1.
In the above expression $\alpha(T) = 1/n_s \lambda^2$ is a temperature-dependent exponent characterizing the rate at which the correlation function decays to zero in a 2D system, which is inversely proportional to the superfluid density $n_s$. In an {\em equilibrium} 2D system, $\alpha(T) \lesssim 1/4$, with equality reached at the critical point $T=T_c$.
Since no ODLRO can emerge in an infinite homogeneous 2D system (except for the specific case of $T=0$), such a system cannot possess a condensate in the infinite limit~\footnote{Of course experimental systems are finite in nature, and so in confined atomic condensates one still often speaks of a `condensate' in the sense of significant coherence over the system size.}. In that case one instead introduces a quasi-condensate,  $n_{qc}$, which corresponds to that part of the density $n$ which exhibits (relatively) suppressed density fluctuations, but whose phase is still largely fluctuating -- this is possible due to the previously-mentioned decoupling of the characteristic temperatures for the onset of phase and density fluctuations. 

The quasi-condensate is defined as~\cite{Svistunov-QC-PhysRevA.66.043608,Proukakis-2006}
\be
n_{qc}({\bf r}) = n({\bf r}) \,\, \sqrt{2-g^{(2)}({\bf r})} \;,
\ee
where here $g^{(2)}({\bf r}) = g^{(2)}({\bf r},{\bf r};\,t,t)$,
%and is particularly important in low-dimensional geometries.
%For example, we note that -- except at $T=0$ -- a 2D homogeneous system does not exhibit Bose-Einstein condensation; nonetheless a quasi-condensate fraction can be useful in calculating certain quantities free of divergences [ ].
From a density matrix perspective,
a quasi-condensate corresponds to a state with many relatively large eigenvalues of (broadly) comparable values. Put differently, a quasi-condensate arises when the system has multiple modes exhibiting relatively enhanced macroscopic occupation, but there is no single mode which can be identified as dominant. 
%This is because of the existence of long-range fluctuations prohibiting the manifestation of true long-range order -- for example due to the presence of phase defects (such as vortices). 
Although a quasi-condensate is thus a more natural state to consider in a lower-dimensional setting (both in 2D and, for weakly-interacting systems, also in 1D), it is important to note that such a state also plays a crucial (transient) role even in 3D systems during the formation stage of a true Bose-Einstein condensate: in a nutshell, the transition from an incoherent 3D system to a coherent 3D BEC precedes through states of quasi-condensation\npp{~\cite{kagan-qc-growth,ProukakisUBEC,Berloff2002a}}, which simply implies that, as the system occupation is redistributed towards lower momenta, there is initially some competition of various low-lying modes before one ultimately dominates and forms the Bose-Einstein condensate (in the sense of the dominant Penrose-Onsager mode).
%The latter can thus be identified, in the regimes where it exists, as the mode corresponding to a single dominant, macroscopically-populated, eigenvalue of the off-diagonal correlation function. 
 %Such a state emerges when the mode competition between a number of highly-populated modes ends with one state becoming increasingly more dominant.

%The relevant quantity to probe, and diagonalize, is the off-diagonal one-body density matrix, $g^{(1)}(x,x';t)$. Following the procedure of Penrose-Onsager [ ] , we identify the condensate as the mode with the largest eigenvalue. This mode exhibits both suppressed density and suppressed phase fluctuation. As such, this 

The Penrose-Onsager mode exhibits both suppressed density and suppressed phase fluctuations. As such, this mode can also be (approximately numerically) reconstructed from the two lowest-order correlation functions. For example, for the specific case of a harmonically-trapped condensate which provides a natural peak in the condensate density at the trap centre ${\bf r}=0$, one can approximate~\cite{AlKhawaja:2002,Proukakis-Henkel-2011} (for a fixed time $t$, suppressed here)
\bea
n_{c}({\bf r}) &=& g^{(1)}({\bf r}) \,\, n_{qc}({\bf r}) \nonumber \\ &=& g^{(1)}({\bf r}) \,\, \sqrt{2 - g^{(2)}({\bf r})} \,\,\, n({\bf r}) \,\,\;.
\eea

Notwithstanding the above, I note that for finite systems trapped in inhomogeneous potentials, such as harmonic traps, the system maintains sizeable coherence over its spatial extent, 
as the trap sets a lower limit for the momenta of the elementary excitations, thus suppressing the extent of phase fluctuations.
As a result, in the limit of relatively low temperatures (much below $T_c$), one can still identify the equilibrium state of the system as a `true' BEC~\cite{Petrov:Low-D-Review,AlKhawaja:2002,ueda:2010}.

%and is therefore ofter referred to in the literature simply as a condensate: for further details we refer the reader to [ , ].

%In the context of classical field simulations -- based on the models introduced earlier -- although one can typically perform an ensemble average over different trajectories (numerical realizations, different noise), such a procedure is extremely numerically demanding, and so is not generally possible in 3D. Instead, on short-enough timescales compared to typical system evolution (or in the final equilibrium setting), one often performs an average over a number of different closely-located time snapshots, which is intended to convey the same information as an ensemble average over different realizations: this approximate method -- which becomes technically exact in an ergodic system at equilibrium -- works generally remarkably well in ultracold quantum gases (and has even been successfully applied to other systems such as exciton-polariton condensates [ ] and fuzzy dark matter [ ]), and allows us to use classical field simulations and still extract information about the condensate and coherent vs.~incoherent modes of the system during its dynamical evolution [Stoof JLTP].

\subsection{Workhorse Model of Bose-Einstein Condensation} \label{sec:modelling}

The usual (and simplest)  textbook description of a 
continuous phase transition is in terms of a phenomenological Ginzburg-Landau  model, whose free energy functional takes the general form~\cite{Leggett:2006}
\be
F[\Phi({\bf r})] = \int d^3{\bf r} \left[ F_0 + \alpha |\Phi|^2 + \frac{\beta}{2} |\Phi|^4  + \gamma |\nabla \Phi|^2 \right] \;,
\label{eq:GL}
\ee
for an appropriate (system-specific) choice of parameters $\alpha, \beta$ and $\gamma$.
The usefulness of such a model is that it supports a phase transition from a symmetric state with a global minimum of the free energy when $\Phi=0$, to a Mexican-hat style potential exhibiting a minimum at a $\Phi \neq 0$ value [see illustration in subsequent Fig.~\ref{fig:5}]: in other words, it facilitates -- under appropriate conditions -- the emergence of a non-zero order parameter of the system.

In the context of BEC, such a model is cast in terms of a
%To shed more light into such process let us consider a commonly used and highly effective model of such quenched dynamics in the context of a weakly-interacting Bose gas, for which we only need to consider elastic two-body collisions (scattering processes) with an effective strength $g$. In this case (consistent with the usual $\lambda \phi^4$ scalar field theory and the generic Ginzburg-Landau model), the Bose field, $\Phi(x,t)$, can be described by 
 nonlinear Schr\"{o}dinger equation -- known as the Gross-Pitaevskii (GP) equation -- which takes the form~\cite{Pethick:2002,Pitaevskii:2003,ueda:2010,Dalfovo1999}
\be
i \hbar \frac{\partial \Phi}{\partial t} = - \frac{\hbar^2}{2m}\nabla^2 \Phi +V \Phi + g|\Phi|^2 \Phi \;.
\label{gpe}
\ee
%sometimes also explicitly written with a $-\mu \Phi$ term on the right-hand-side, where $\mu$ is the system chemical potential.
Here $m$ is the mass of the (composite) boson, $V$ is the external potential and $g$ is the effective interaction strength between constituent bosons, assumed here for simplicity to be local [i.e.~$V({\bf r}-{\bf r'})=g \delta({\bf r}-{\bf r'})$], based on the lowest order (or s-wave) scattering process.
This equation is often explicitly written with an additional $-\mu \Phi$ term on the right-hand-side, where $\mu$ is the system chemical potential.

This equation arises both as (trivially) the lowest-order equation for the symmetry-broken field $\Phi = \langle \hat{\Psi} \rangle$, 
%(where $\hat{\Psi}$ is the usual Bose field operator), 
and as the effective (low-energy) field equation for a semi-classical field (or order parameter) $\Phi$ -- the latter applicable for all highly-occupied modes in the system (i.e.~those with occupation exceeding one particle per mode): for more information see recent reviews~\cite{proukakis_finite-temperature_2008,Proukakis13Quantum,berloff_brachet_14,Blakie-AdvPhys-2008}.

Beyond ultracold atomic gases, an appropriately modified form of this equation which takes account of specific system details (e.g.~non-local interactions, pumping, dissipation, rotation, gravitational effects) is applicable (with some limitations) to a range of diverse systems including liquid helium~\cite{Berloff-Nonlocal}, photon~\cite{Klaers:2010} and exciton-polariton condensates~\cite{carusotto2013quantum}, neutron star superfluidity~\cite{melatos-berloff-neutronstars,melatos-neutronstars,brachet-pulsars-PhysRevResearch.4.013026} and fuzzy dark matter~\cite{Marsh:2016,Hui:2021,Ferreira} -- many of which are more explicitly discussed in subsequent sections.
%and neutron star superfluidity [ ]. 
%To describe such settings one generally adds a pumping and/or decay term to the equation, or couples the GPE to a Poisson equation.
Specifically to account for a coupling to a heat bath (e.g.~providing pumping, and/or dissipation), and the associated dynamical stochasticity, one often further supplements the above picture with further pumping/dissipation terms and an associated (typically additive) noise contribution, thus generalizing the `basic' Ginzburg-Landau model of Eq.~(\ref{eq:GL}) to a stochastic complex Ginzburg-Landau equation
[see also subsequent Eqs.~(\ref{spgpe}) and (\ref{eq:pol})].

Having identified the relevant physical systems, key considerations and quantities useful in their description, I proceed to discuss features of universal dynamics emerging during the phase transition crossing (Sec.~\ref{sec:kz-main}) and the subsequent system phase-ordering and relaxation to the final (equilibrium) state (Sec.~\ref{sec:phaseordering-main}) -- with some emphasis on ultracold atomic gases as the physical system.
Further remarks in the context of driven-dissipative quantum gases, and other related features are then discussed in Sec.~\ref{sec:other}, with some overall concluding remarks in Sec.~\ref{sec:conclusions}.

Key to the subsequent discussion is the idea of a scaling hypothesis in the spatial -- and, by extension, also the temporal -- domain, i.e.~the idea that correlation functions acquire in some limited `universal' regime the same functional form, in terms of a limited set of parameters -- with the obvious example being that of a time-dependent lengthscale -- and a specific set of exponents describing such dependence.
Interestingly, the concepts discussed below are interdisciplinary, benefiting from cross-fertilization of ideas between (among other fields) condensed-matter theory, statistical physics, elementary particle theory and cosmology.

\section{Universality during Phase Transition Crossing: The Kibble-Zurek Scaling Law} \label{sec:kz-main}

\subsection{Early Ideas} \label{sec:earlyideas}

The concepts of spontaneous symmetry breaking and order formation at a critical temperature in ferromagnets and superconductors have found analogues in the context of the hot big bang model, with the universe transiting from a symmetric, to a symmetry-broken phase. Building on such (mathematical/notional) analogies, Tom Kibble wrote a seminal paper discussing the formation, topology and evolution of domains emerging during a dynamical (second-order) continuous phase transition  in the context of early-universe cosmology~\cite{kibble1976topology}. % in a pioneering 1976 work [ ], on the topology of cosmic domains and strings (in the context of the hot big-bang model), Tom 
In such work, Kibble noted the initial chaotic formation of small quasi-ordered localized regions, or `protodomains', between which the system expectation value would vary randomly from region to region as the system breaks its $U(1)$ symmetry, and starts choosing a preferred phase. In particular, Kibble raised the question of whether any residue of such protodomains may remain detectable  -- in the form of a long-lived topologically stable structure such as monopole, string, or domain wall -- as the overall system coherence grows: such a structure would form naturally at the interface of two spatially-separated regions of different phase. This question has since been appropriately reformulated, finding significant applications in diverse lab-based condensed-matter systems (including liquid helium, liquid crystals, superconductors, and trapped ions), owing to the intuition of Wojciech Zurek who later discussed the analogy between such cosmological strings and vortex lines in a superfluid in 1985~\cite{zurek1985cosmological}, and thus proposed schemes for its experimental study. 

For completeness, I briefly note here that 
analogies between superfluids and the early universe are abundant, with
liquid helium having been proposed as a system in which to study a range of different phenomena of relevance to the physics of the universe~\cite{zurek1996cosmological,volovik2009universe,UBEC-Pickett}, considerations which have since been extended to other controlled quantum gases (ultracold atomic systems and exciton-polariton condensates) in which aspects such as analogue gravity~\cite{analogue-black-hole-review}, false vacuum decay~\cite{Fialko_2015-false-vacuum-decay,Moss-falsevacuumdecay} and hydrodynamic QCD plasma analogues~\cite{qcd-coldatoms}
are currently under investigation.

%-------------------------------------
\begin{figure*}[t!]
\centering
\includegraphics[width=0.95\linewidth]{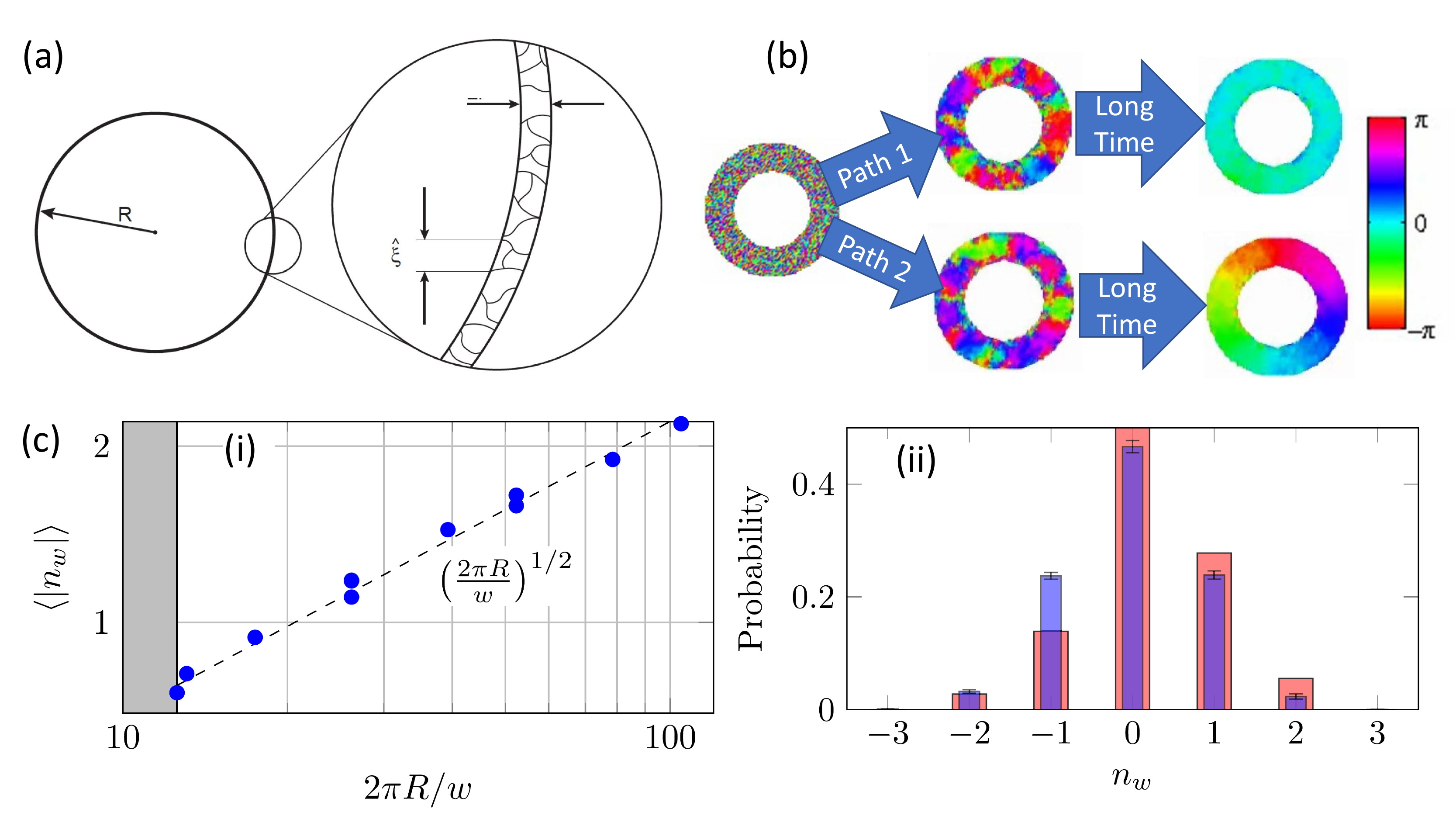}%{Fig-4-wide.png}
\caption{
Phase Evolution in an annular geometry (ring trap).
(a) Original thought experiment of Zurek about an ultra-thin, effectively 1D ring in which the crossing of the phase transition induces typical domains of mean size $\hat{\xi} \ll R$, where $R$ is the ring radius.
Reprinted with permission from W.~H.~Zurek, Cosmological experiments in superfluid Helium?, Nature {\bf 317}, 505~\cite{zurek1985cosmological}. \npp{Copyright (1985) by the Nature Publishing Group.}
(b) Typical phase profile evolution in a 2D ultracold atomic gas based on  SPGPE simulations of Eq.~(\ref{spgpe}): a truly random, incoherent, initial state whose phase varies wildly within the ring (left) can follow under external parameter quenching different random paths of establishing larger regions of constant, but random, phase (middle), with this ultimately leading -- for example -- to the establishment of essentially constant phase across the ring (right, top), or a variation of $2 \pi$ (right, bottom), i.e.~respectively a winding number $n_W$, or `charge' of 0, or 1.
(c) (i) Dependence of mean winding number modulus $\langle |n_W| \rangle$ on mean ring radius $R$ and ring width $w$ through their ratio $(R/w)$ (blue points) and corresponding prediction based on Zurek's random region formation scenario (dashed line). (ii) Probability of occurrence of different winding number outcomes, $n_W$, in the quenched experiment of Ref.~\cite{dalibard-ring-KZ-2014} (red bars) and corresponding SPGPE predictions (blue bars)~\cite{bland_marolleau_20}: the asymmetry in the experimental measurements is likely due to a smaller number of trajectory averaging over numerical simulations.
Reprinted with permission from T.~Bland {\em et al.}, Persistent current formation in double-ring geometries, J.~Phys.~B:~At.~Mol.~Opt.~Phys. 53 {\bf 2}, 033183. \npp{Copyright (2020) by the Institute of Physics.}
}
\label{fig:3}
\end{figure*}
%-------------------------------------

A schematic from the pioneering work of Zurek is shown in Fig.~4(a), depicting a quench in a very thin (so effectively one-dimensional) doughnut-shaped (annular) helium superfluid of radius $R$ (Fig.4(a))~\cite{zurek1985cosmological,zurek1996cosmological}:
Building on the idea of `protodomains', let us consider that -- at the time the system crosses the phase transition dynamically -- the typical correlation length of the system takes a characteristic value $\hat{\xi}$; in other words, the system density around the ring circumference is divided into a number of regions, of effective mean length $\hat{\xi}$, each of which exhibits a constant phase, but such phase is chosen independently, and thus randomly, across different locations around the ring with no causal connection between them.
Assuming a random walk of the phase between regions, a net (non-zero) phase gradient can emerge from the phase mismatch $\Delta \theta$ around the ring, which is in turn translated into
%considering the typical (average) phase mismatch $\Delta \theta$ around the loop arising from a given structure size, i.e.~the arising phase gradient around the ring, after a transition quench (in this case pressure quench) and translating this to 
a superfluid velocity, via $v_s = \hbar /m \nabla (\Delta \theta)$.
Zurek thus concluded that a rapid phase transition quench (pressure quench in the case of liquid helium) could leave the superfluid rotating around the annulus, i.e.~a (random) non-zero circulation. 
By the structure of the order parameter $\Phi = |\Phi| e ^{i \theta}$, such circulation would then be in the form of integer multiples of $2 \pi$, corresponding to persistent currents around the ring -- with the value of the integer multiple characterizing the so-called winding number.
The emergence of such persistent currents was predicted to occur stochastically % with the distribution function of persistent currents being Gaussian: naturally, t
with the most likely outcome being the absence of a  persistent current.

Stochastic formation of persistent currents through a quenched phase transition of a condensed matter system in the specific context of such a (single) annulus has been experimentally studied in a single superconducting ring~\cite{KZ-superconductor-ring} (see also related earlier references therein), and, more recently, in an ultracold atomic gas ring~\cite{dalibard-ring-KZ-2014} -- the latter briefly discussed in the next section.
However, it is important to note at this stage -- as will become clearer later -- that the proposed Kibble-Zurek mechanism is very general and, as such, has been observed in a range of controlled quench experiments across different systems, such as liquid helium~\cite{KZ-He4,KZ-He3}, superconductors~\cite{KZ-superconductor-multiring}, liquid crystals~\cite{KZ-liquidcrystals}, multiferroics~\cite{KZ-Multiferroics}, trapped ions~\cite{KZ-ioncrystals} and single-component trapped ultracold atoms~\cite{Weiler:2008,lamporesi2013spontaneous,dalibard-ring-KZ-2014,Dalibard-KZ-2D,Hadzibabic-KZ,donadello2016creation,dalibard-multiring-merging,shin-KZ-Fermi-2019}-- %such persistent currents have also been found to arise spontaneously during rapid cooling in a ring-trapped atomic Bose-Einstein condensate~\cite{} -- 
see recent reviews~\cite{del_campo_universality_2014,delcampo-kibble-zurek,beugnon_navon_KZ_review} for more details.

\subsection{Numerical Phase Transition Model} \label{sec:sgpe}

Although interesting non-equilibrium properties of the system can be studied by the Gross-Pitaevskii equation [Eq.~(\ref{gpe})] for the classical field subject to appropriate initial choice of non-equilibrium states [see discussion in subsequent Sec.~\ref{sec:ntfp}], a commonly used alternative, which further introduces the element of (dynamical) stochasticity to the system, is a slightly generalized version in the form of the so-called Stochastic Gross-Pitaevskii equation~\cite{stoof_dynamics_2001,spgpe-davis,proukakis_finite-temperature_2008,Blakie-AdvPhys-2008,Proukakis13Quantum,berloff_brachet_14}
\begin{eqnarray}
 && i \hbar \frac{\partial \Phi}{\partial t}   =  \label{spgpe} \\ 
 & {\cal P} &\hspace{-0.25cm}  \left\{ (1-i \gamma)\left[ - \frac{\hbar^2}{2m}\nabla^2 \Phi +V \Phi + g|\Phi|^2 \Phi -\mu \Phi \right]  
+ \eta \right\}  \nonumber \;.
\end{eqnarray}
This is a nonlinear Langevin equation corresponding to an effective field theory for the low-energy modes of a system: implicit in this model is that such modes are driven by their coupling to higher-lying modes (representing a bath of purely thermal particles having an effective chemical potential $\mu$ and temperature $T$) through the dissipation strength $\gamma$, which -- together with $T$ -- sets the mean strength of the fluctuations;
the latter are treated
as Gaussian noise with $\delta$-correlations  in both space and time, i.e.~$\langle \eta^\star({\bf r},t) \eta({\bf r'},t') \rangle = \bar{\eta}^2 \delta({\bf r}-{\bf r'}) \delta(t-t')$, where
$\bar{\eta} = \sqrt{2 \hbar \gamma k_B T}$.
%which are in turn treated as Gaussian noise with $\delta$-correlations  in both space and time, i.e.~$\langle \eta^\star(r,t) \eta(r',t') \rangle = \langle |\eta|^2 \rangle \delta(r-r') \delta(t-t')$.
A projector ${\cal P}$ has also been included in Eq.~(\ref{spgpe}) to ensure numerical consistency, by restricting dynamics within the same low-energy band of modes~\cite{Blakie-AdvPhys-2008,berloff_brachet_14}.

%-------------------------------------
\begin{figure*}[t!]
\centering
\includegraphics[width=0.9\linewidth]{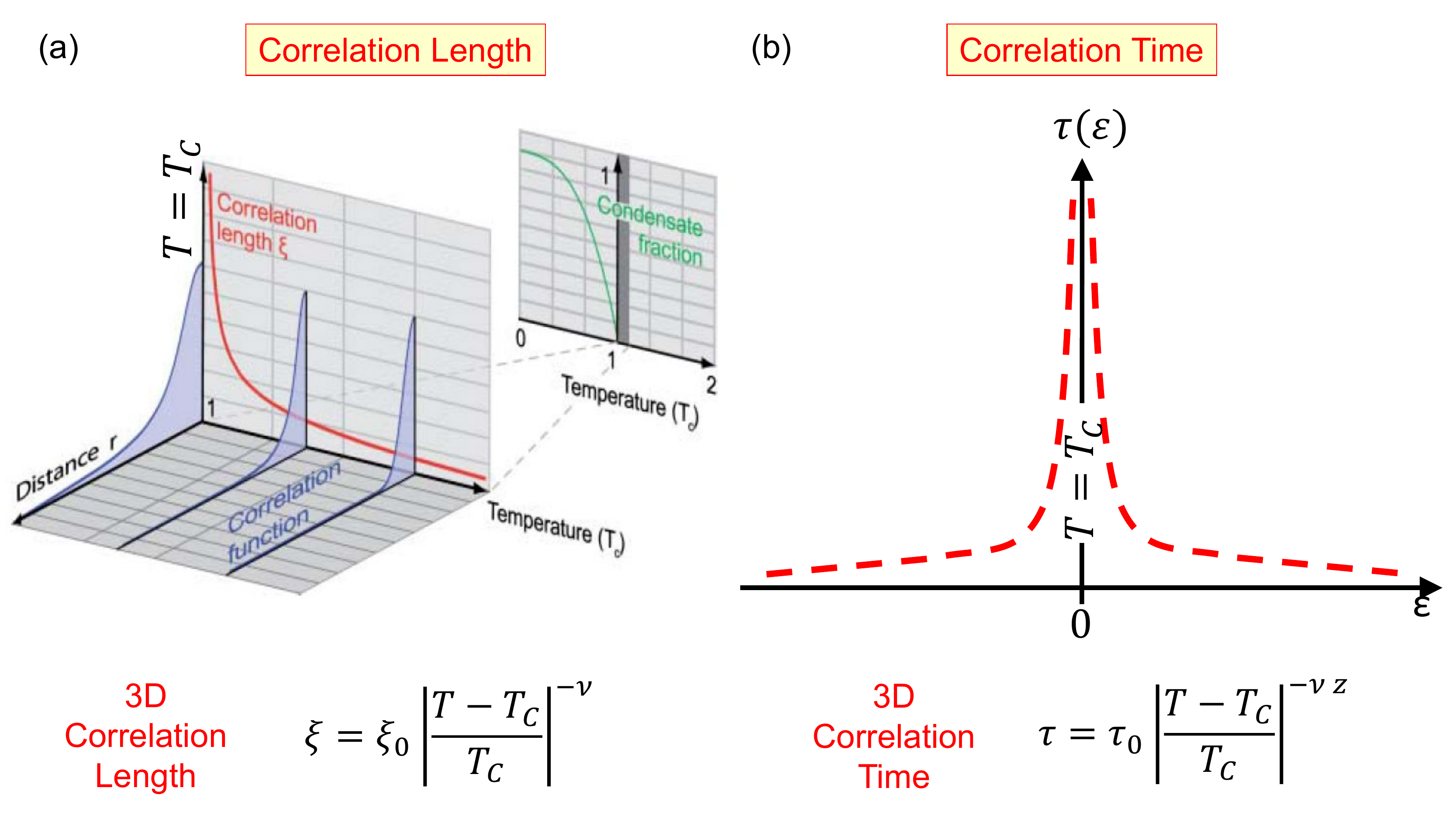}
\caption{
Divergence of (a) equilibrium correlation length and (b) correlation time in the vicinity of a phase transition occurring at $T=T_c$ (or $\epsilon=0$):
(a) Main plot shows the diverging correlation length $\xi$ (red line) as the system approaches criticality from higher temperatures (i.e.~from the right); characteristic snapshots of the unnormalized first-order spatial correlation function $G^{(1)}({\bf r})$, which decays as $1/r$ %due to the spatially decaying condensate fraction in the incoherent harmonically-trapped system 
with the distance from the trap centre, are shown at 3 different temperatures, i.e.~at different values of $\epsilon$ (filled blue curves). Inset: Temperature dependence of the overall system condensate fraction (green curve), 
with the grey band in the immediate vicinity of, and specifically above, the critical point depicting the temperature region probed in the main plot.
Figure reprinted with permission from T.~Donner {\em et al.,}, Critical behaviour of a trapped interacting Bose gas, Science {\bf 315}, 1556 (2007)~\cite{Donner:2007}. Reprinted with permission from the Science Publishing Group.
(b) Divergence of correlation time assumed symmetric on both sides of the transition, and plotted here in terms of $\epsilon = (T_c-T)/T$ for ease of comparison with subsequent time-dependent discussion. Expressions quoted in each case summarize our understanding of equilibrium criticality as a function of temperature $T$, demonstrating the relation $\tau \sim \xi^z$.
}
\label{fig:4}
\end{figure*}
%-------------------------------------

Based on such simple techniques, one can numerically visualize phase evolution within Zurek's original annular ring-trap setting in the context of ultracold quantum gases: this was beautifully demonstrated for the (effectively 1D) case of a very narrow annulus in Ref.~\cite{sabbatini-ring-KZ}. Corresponding results, generalized to a 2D setting such that the annulus width, $w$ is also explicitly taken into consideration, thus facilitating also radial phase gradient and dynamics are shown in Fig.~\ref{fig:3}(b). 
%while its generalization to the more experimentally relevant 2D regime is shown in Fig.~\ref{fig:3}(c): 
 During such evolution, the dynamical noise $\eta$ can drive an initial random phase into distinct regimes, having either zero net circulation (the most likely outcome, top right plot in Fig.~\ref{fig:3}(b)), or a non-zero phase winding of an integer multiple of $\pm 2 \pi$ around the loop: in this particular example, the non-zero winding number generated is $n_w = 1$ (bottom right plot in Fig.~\ref{fig:3}(b)).

%Following earlier work on Helium [ ], 
Such features were explicitly verified experimentally in ultracold atomic gases in Ref.~\cite{dalibard-ring-KZ-2014}, finding a spread of winding numbers around a zero mean value. The detection was done by removing the ring potential and allowing the annular condensate to  expand and interfere with an independent  (reference) condensate formed in the centre of the ring: a non-zero circulation could then be inferred from spiral distributions in the density, detected destructively through absorption imaging.
A numerical confirmation (based on Eq.~(\ref{spgpe})) of the anticipated mean winding number scaling in terms of the ring radius $R$ and width $w$, namely $\langle |n_w| \rangle = \sqrt{2 \pi R/ w}$, and a 
comparison demonstrating excellent agreement of numerically generated and observed winding numbers in the experimental set-up are respectively shown in Figs.~\ref{fig:3}(c)(i)-(ii)~\cite{bland_marolleau_20}.

Zurek's original analysis and extensive follow-on work has led to the establishment of what is now known as the universal Kibble-Zurek scaling law of defect formation in a quenched quantum liquid.
In order to be  in a position to discuss such dynamical crossing of a phase transition more explicitly, I should first comment on equilibrium features around a critical point, ideas which will subsequently be appropriately extended to provide a universal description of non-equilibrium phase transition dynamics.
%ideas from equilibrium features around a critical point.
%superfluid [ , ], which has been observed -- beyond superfluid $^{3}$He [ ] and $^{4}$He -- in superconducting Josephson junctions [ ] liquid crystals [ ], ions [ ] and ultracold atomic gases [ ].

\subsection{Criticality at Equilibrium}
\label{sec:eqm-criticality}

As the system is probed on the incoherent side through different equilibrium states of decreasing temperature, the coherence length of a system, characterizing the spatial variation of %$g^{(1)}(r)$, 
the first-order correlation function, increases.  
Standard statistical physics considerations and critical phenomena theory 
%for a continuous second-order phase transition 
enable us to describe the system correlations in terms of universal scaling relations, independent from the microscopic details of the system, and specified in terms of the distance to criticality, in a manner specified by a set of critical exponents and a scaling function~\cite{binney1992theory,nishimori-phasetransition-book}: such considerations are a key aspect of the present discussion.

A striking example of the universal approach to criticality is the divergence of the system coherence length $L$ at the actual system critical point: on the incoherent side of the phase transition, such approach is characterized via $g^{(1)}({\bf r}) \sim e^{-r/L}$; this is in fact a very useful observable in order to identify the precise location of a critical point (either experimentally, or numerically, within a given model); such points will in general be slightly shifted due to quantum fluctuations, interaction and finite-size effects~\cite{Dalfovo1999,holzmann-Tc-shift}.
One generally quantifies this by defining a `distance to criticality' parameter $\epsilon$, which is measured as the difference of the value of $\epsilon$ in a given configuration from the critical value $\epsilon_c$ which signifies the transition point (the latter typically taken as $\epsilon_c=0$).

Although the existence of a divergence at criticality is broadly applicable, the way that such a divergence manifests itself depends sensitively
on the system dimensionality and  symmetry and interaction properties of the underlying system Hamiltonian (which, in turn, define the corresponding order parameter for the particular phase transition being considered)~\cite{binney1992theory}.
Systems whose approach to criticality (for a continuous phase transition) obey the same scaling functions and the same diverging behaviours in terms of $|\epsilon|$ are said to belong to the same universality class~\cite{nishimori-phasetransition-book}: 
the Bose gas (ultracold bosonic atoms, $\lambda$ transition in $^{4}$He) belongs to the XY universality class, and thus its critical properties are expected to exhibit the same behaviour as the XY model.
%for example, BEC and the 3D XY model belong to the same universality class. 
However, as should be anticipated from the preceding discussion, 3D and 2D systems do not exhibit the same behaviour near criticality (they have distinct critical behaviour/exponents), and this is also related to the
%do not fall within the same universality class: this manifests itself in the 
very different nature of the phase transition: BEC vs.~BKT.

Once the approach to criticality has been identified, one can specify the relevant information for the transition in terms of the values of the relevant critical exponents characterizing such divergence as a function of $|\epsilon|$: systems having the same dimensionality and the same critical exponents are said to belong to the same universality class .
To make this more specific, let us consider the correlation length $\xi$ in an infinite 3D system undergoing Bose-Einstein condensation through cooling, for which the distance to criticality, $\epsilon$, is defined according to $\epsilon = (T_c - T)/T$. In the $\epsilon \lesssim 0$ limit (corresponding to $T>T_c$) this is expected to diverge in the vicinity of the critical point, i.e.~at $\epsilon \rightarrow 0^{-}$, as
\be
\xi = \xi_0 |\epsilon|^{-\nu} = \xi_0 \left| \frac{T-T_c}{T_c} \right|^{-\nu},
\label{eq:xi}
\ee
where $\xi_0$ is a non-universal parameter depending on the specific system microphysics (which is thus not considered further here) and $\nu$ is the static critical exponent.
In the case of ultracold atoms, $\nu$ has been measured as $\nu \approx 0.67$~\cite{Donner:2007} (consistent with expectations of the 3D XY model). 
An example of such behaviour is shown in Fig.~\ref{fig:4}(a).
%Note that the location of the equilibrium critical point is shifted both by mean-field and critical fluctuations, as well as by finite-size effects [ ].
%
Note that in a 2D system, the behaviour of the coherence length near criticality takes a more stringent form $\xi \sim {\rm exp} \left( b |\epsilon|^{-\nu} \right)$ (where $b \sim O(1)$ is a constant)~\cite{jelic2011quench}, with the exact form still under discussion.

\subsection{Criticality in a Dynamical Setting}\label{sec:dyn-criticality}

Generalizing to a dynamical setting, it should be noted that the relaxation dynamics of a system, i.e.~the process by which the system equilibrates in a given configuration (i.e.~for a fixed value of $\epsilon$) from a non-equilibrium initial condition, is dramatically slowed down near the transition point; this is intricately related to the fact that the correlation length of the system, i.e.~the distance over which the system has to achieve phase coherence, is -- as previously discussed -- rapidly increasing in the vicinity of the critical point. The associated increase (divergence) in time required for the relaxation process is
known as critical slowing down~\cite{nishimori-phasetransition-book}. In fact, the relaxation time (in the 3D BEC system under discussion), also known as correlation, time, diverges according to [see Fig.~\ref{fig:4}(b)]
\be
\tau=\tau_0 \, |\epsilon|^{-\nu z} \sim \xi^z \;.
\label{tau}
\ee
This additionally introduces the dynamical critical exponent $z$ governing intrinsic system evolution at a fixed value of $|\epsilon|$: note that $\tau_0$ appearing here is a non-universal time containing details of microscopic properties.

%-------------------------------------
\begin{figure*}[t!]
\centering
\includegraphics[width=1.0\linewidth]{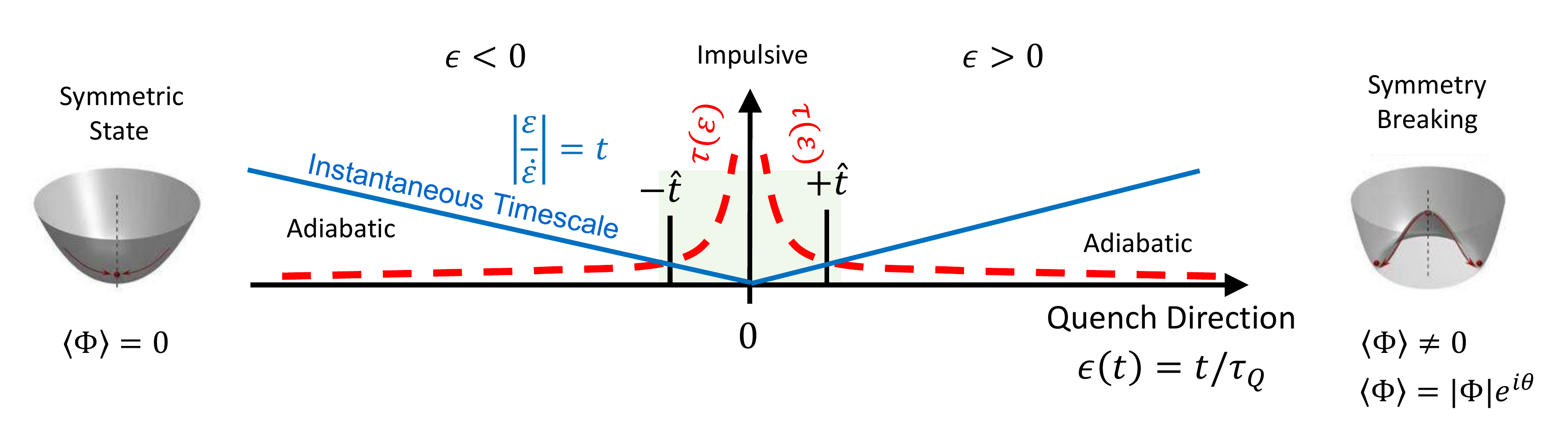}%{Fig-6-NEW.png}
\caption{
Schematic of the `cartoon' Kibble-Zurek scenario, describing the phase transition from an incoherent to a coherent state through a dynamically-induced linear crossing of the distance to criticality $\epsilon(t)$ in time. Shown in terms of $\epsilon(t)$ are the correlation time which diverges from both sides as $\epsilon(t) \rightarrow 0$ [dashed red line] and an instantaneous timescale $\epsilon/\dot{\epsilon}$ set by the rate at which the quench is induced.
Regarding the latter, I note here that
when plotting things in terms of $\epsilon(t)$, slower quenches have a steeper slope of $|\epsilon / \dot{\epsilon}|$, thus allowing the system to reach, in principle, arbitrarily close to the critical point before it can no longer adiabatically follow equilibrium states in its crossing through the phase transition.
During its crossing, the system can only adiabatically follow corresponding equilibrium states for as long as $|\tau| < |\epsilon/\dot{\epsilon}|$; otherwise, the characteristic correlation length setting the size of the domains becomes frozen to its value, $\hat{\xi}=\xi(\hat{\epsilon})$ evaluated at $\hat{\epsilon} = \epsilon(-\hat{t})$ and the system behaves impulsively between $-\hat{t}$ and $\hat{t}$.
The symmetry of the pre-quench (symmetric, $\langle \Phi \rangle =0$) and post-quench (symmetry-broken, $\langle \Phi \rangle \neq 0$) equilibria of the system are depicted in terms of the thermodynamic potential with the radial direction depicting the absence (left) or existence (right) of a non-zero order parameter: in the latter case, and while the azimuthal angle in this `Mexican hat' potential can in principle take any value with equal probability, the merging dynamics of domains of spontaneously-generated random phases ultimately leave the system at
 sufficiently large values of $\epsilon>0$ (and in this case at sufficiently late, or asymptotically-long, times such that all critical transient dynamics have died out)  with a particular spontaneously-chosen direction, or phase $\theta$ (whose gradient sets the superfluid velocity).
}
\label{fig:5}
\end{figure*}
%-------------------------------------

Having discussed the path to equilibration for a given value of $\epsilon$, one can now proceed to consider the dynamical quench scenario, and thus address the question of the system dynamics when the distance to criticality is itself tuned in a dynamical manner, i.e.~$\epsilon \rightarrow \epsilon(t)$. The simplest, and standard, scenario considered is that of a linear variation of the distance to criticality with evolution time, which is assumed to occur over a characteristic quench timescale $\tau_Q$, i.e.
\be
\epsilon(t) = \frac{t}{\tau_Q} \;.
\label{epsilon}
\ee
As such, $\tau_Q$ controls the effective quench rate:
small $\tau_Q$ implies the system is moving rapidly through different regimes along the $\epsilon$ axis, i.e.~small $\tau_Q$ corresponds to rapid quenches; conversely, the adiabatic crossing of the phase transition (through equilibrium states) is achieved in the limit $\tau_Q \rightarrow \infty$.

The evolution of the system from an (incoherent) state without symmetry-breaking (left) to a broken-symmetry one [which supports the formation of a phase-coherent state] (right) is shown schematically in Fig.~\ref{fig:5} in terms of the time-dependent distance to criticality, with $\epsilon(t)=0$ indicating the equilibrium critical point~\cite{del_campo_universality_2014}. Shown on this figure are two curves: the diverging correlation time (dashed black lines) and an instantaneous system-adjustment timescale constructed from the rate of quenching across the phase transition (i.e.~the rate of change of the distance to criticality $\dot{\epsilon}$) through the relation $t = \epsilon / \dot{\epsilon}$. By convention, during the dynamical crossing of the phase transition, the time of the transition corresponds to $t=0$, with $t<0$ (corresponding to $\epsilon<0$) labelling the incoherent side of the transition, and $t>0$ the coherent side.
Our understanding of the dynamics of the phase transition can be cast in terms of the relative magnitude of these two times (correlation time and system-adjustment timescale). 

As the system moves towards the critical point (starting from an incoherent equilibrium state), the time taken for the system to adjust to its new instantaneous equilibrium solution grows, and starts doing so very rapidly as it approaches closer to the (equilibrium) critical point (where it diverges). Thus, necessarily, as the system is dynamically induced through the phase transition, the situation will arise, in which the rate at which the system is driven through the phase transition parameter space is so fast that its corresponding inverse timescale is shorter than the time required for the system to equilibrate over an increasing lengthscale: this happens when the  correlation time $\tau$ becomes (in the vicinity of the critical point) larger than the magnitude of the instantaneous timescale $|\epsilon/\dot{\epsilon}|$, associated with the rate of quenching. 
%
%(To avoid any confusion, we note here  that -- when plotting things in terms of $\epsilon(t)$ -- slower quenches have a steeper slope of $|\epsilon / \dot{\epsilon}|$, thus allowing the system to reach, in principle, arbitrarily close to the critical point before it can no longer adiabatically follow equilibrium states in its crossing through the phase transition.) 
%
Beyond such time, the system no longer has enough time to adjust as a whole to the external quench, and so different parts of the system necessarily evolve independently of each other, no longer controlled by the external drive (e.g.~cooling ramp): in other words the system is allowed to evolve independently on a local scale, and so the global symmetry is broken.
This implies that the thermodynamic potential of the system transitions suddenly from a single minimum at the origin corresponding to the absence of any identified order parameter, i.e.~$\langle \Phi \rangle=0$,
to a typical `Mexican-hat' potential with $\langle \Phi \rangle  \neq 0$ (see Fig.~\ref{fig:5}): in such a potential, the radial location of the minima indicates a non-zero expectation value of the system order parameter $\Phi$ which thus acquires a meaning on this side of the phase transition, while the azimuthal direction contains information about the phase of $\Phi$. Note that, at this stage, all minima in such a potential, corresponding to different azimuthal orientation (at fixed radial distance) become equally probable, i.e.~all local phase choices are equally likely; this in turn implies that the phase varies randomly (following a random walk) between different domains, the size of which is (typically much) smaller than the system size. In fact, the typical size of such domains is set by the equilibrium correlation length of the system evaluated at the distance to criticality $\epsilon$ corresponding to the time when the symmetry is broken. %evaluated at the corresponding distance to criticality $\epsilon$. %Given that the distance to criticality where symmetry is broken during a quenched phase transition depends on the quench rate, this also has important implications about the size of the independently-formed domains on the symmetry-broken side.

\subsection{The Kibble-Zurek Scaling Law} \label{sec:kz-hom}

Having introduced the key concepts, one can now formulate the fundamental equations governing the Kibble-Zurek mechanism of quenched phase transitions.
The underlying idea is to use the equilibrium behaviour of the system near criticality to make specific predictions regarding the non-equilibrium consequences of the dynamics of symmetry breaking.
Mathematically, this relies on identifying the characteristic time at which the instantaneous quench timescale $|\epsilon / \dot{\epsilon}|$ [diagonal lines in Fig.~\ref{fig:5}] becomes equal to the diverging relaxation timescale $\tau$: such time, typically denoted by $\mp \hat{t}$, signifies the so-called freeze-out time, i.e.~the time during the quench when the system behaviour seizes being identified by parameters adiabatically following the equilibrium properties of the system as a whole (at a rate set by the quench rate). The typical size of subsequent independent domains -- which have (in this simplistic scenario) no causal connection between them -- is then set by the instantaneous value of the system correlation length at the freeze-out time.
The discussion below follows closely the
 review articles~\cite{delcampo-kibble-zurek,del_campo_universality_2014} based on
the considerations by Kibble and Zurek.
%as presented in the review articles~\cite{delcampo-kibble-zurek,del_campo_universality_2014}.

Mathematically, one equates the diverging relaxation time $\tau$, evaluated at the freeze-out time $- \hat{t}$, to the time $|-\hat{t}\,|$ set by $|\epsilon(-\hat{t})/\dot{\epsilon}(-\hat{t})|$.
The fundamental Kibble-Zurek relation can thus be written as
\be
\tau(-\hat{t}) = 
|-\hat{t}\,| \;,
\label{eq:kz}
\ee
As the above equation simply labels the two points of intersection of the solid and dashed lines in Fig.~\ref{fig:5}, which are the same in this (presumed symmetric) schematic across both sides of the phase transition, one often quotes this as the more `memorable' and algebraically-equivalent form $\tau(\hat{t}) = \hat{t}$.
Since the quench rate $\epsilon(t)$ depends on the quench timescale, $\tau_Q$, then so do the above-mentioned intersection points, thus characterizing the dependence of correlation time -- and, by extension, of correlation length (during the quenched evolution) -- on $\tau_Q$.

In its most basic (and most commonly presented) form )discussed here), the Kibble-Zurek scenario considers the system dynamics as being `frozen' between $-\hat{t}$ and $+\hat{t}$ (i.e.~when $\tau > |\hat{t}|$), with such temporal window -- during which the system remains fragmented into a number of smaller domains, exhibiting no correlation between them --  labelled as the `impulsive' regime.

To understand the implications of such relation further, let us now consider a specific case -- namely the example of a 3D homogeneous system, known to obey the equilibrium critical scalings of Eqs.~(\ref{eq:xi})-(\ref{tau}).
%near the critical point.
In that case, one can re-express Eq.~(\ref{eq:kz}) [or its equivalent relation $\tau(\hat{t}) = \hat{t}$] as
\be
\tau(\hat{t}) = \tau_0 |\epsilon(\hat{t})|^{- \nu z} = \tau_0 \left| \left( \frac{\hat{t}}{\tau_Q} \right)  \right|^{-\nu z} %\left| \frac{\epsilon(\hat{t})}{\dot{\epsilon}(\hat{t})} \right| 
= \hat{t} \;.
\label{eq:kz2}
\ee

Such a relation allows us to obtain a characteristic
%draw a number of interesting conclusions regarding the 
scaling of the freeze-out time, $\hat{t}$, in terms of $\tau_Q$,
with a corresponding relation for $\hat{\xi}$ obtained through $\hat{t} = \hat{\xi}^z$.
%and corresponding length scale $\hat{\xi}$ in terms of $\tau_Q$. 
Ignoring microscopic prefactors, which are not relevant for the universal system dynamics, one thus deduces that (for a homogeneous 3D system)
\be
\hat{t} \sim \tau_Q^{\nu z/(1+ \nu z)},
\label{t-hat}
\ee
and correspondingly
\be
\hat{\xi} =\xi\left(\epsilon(\hat{t})\right)=\xi_0\left| \epsilon(\hat{t}) \right|^{-\nu} \sim \tau_Q^{\nu/(1+ \nu z)} \;.
\label{xi-hat}
\ee

Although the Kibble-Zurek relation [Eq.~(\ref{eq:kz})] will always  specify 
%
%It is important to stress that while Eqs.~(\ref{eq:kz})-(\ref{eq:kz2}) and the above considerations will always lead to specific 
a characteristic scaling for $\hat{t}$ and $\hat{\xi}$ in terms of $\tau_Q$ around criticality, the exact form (and arising exponents) of such emerging relations will depend on the dimensionality and universality class of the system 
%
%of the emerging relations (and the arising exponents) depend critically on a number of factors, including the dimensionality (and universality class), and the homogeneity of the system 
-- with these results requiring revisiting in an inhomogeneous context (see later).

%Moreover, even once the dependence on the critical exponents $\nu$ and $z$ is known, their precise value depends on whether they are obtained e.g. within a mean-field model, or other models (specifically there is a classification ......).

We have thus identified the timescale $\hat{t}$ and length scale $\hat{\xi}$, as the
dominant relevant scales in the system in the critical region.
%, namely the timescale $\hat{t}$ and length scale $\hat{\xi}$,
%length and time scales of the system, 
%we are now able to make some more general remarks:
As such, within such limited regime, once can formulate the temporal and spatial dependence of physical variables in the system in a universal manner by a simple rescaling of the form
%
%\begin{center}
\begin{equation}
\begin{matrix}
{\rm Time} &  t  & \rightarrow & t/ \hat{t}  \\
{\rm Distance} & {\bf r} & \rightarrow & {\bf r}/\hat{\xi}  \\
{\rm Wavevector} & {\bf k} & \rightarrow & {\bf k} \hat{\xi} 
\end{matrix} \nonumber
\end{equation}
%\end{center}
%
%\begin{center}
%{\rm Time} & & $t$ & \rightarrow & t/ \hat{t}  \\
%{\rm Distance}& & ${\bf r}$ & \rightarrow & {\bf r}/\hat{\xi}  \\
%{\rm Wavevector}& & ${\bf k}$ & \rightarrow & {\bf k} \hat{\xi} \\
%\end{center}
%
%Irrespective of the precise nature of such relations, and based  on the Kibble-Zurek description, and the notion that the system dynamics remains frozen in the critical region (see later for a more critical view on this), we can thus note a number of important characteristics for the system (assuming of course -- as has been done throughout this section -- that quenches are sufficiently slow, i.e.~non-instantaneous).
%
%Firstly, %under such assumptions, 
%we note that the only relevant scales in the system become the timescale $\hat{t}$ and length scale $\hat{\xi}$. 
Having identified the relevant scales in the system allows us to formulate the corresponding scaling hypothesis for the system correlations (applicable in the long-wavelength, low-frequency limit): in the discussion below I also ignore for simplicity the anomalous critical exponent (denoted by $\eta$), as its value is typically zero, or very close to it~\cite{nishimori-phasetransition-book,binney1992theory}.
%We start by formulating this in the most general manner
%giving the general) Kibble-Zurek scaling hypothesis (applicable in the long-wavelength, low-frequency limit) %based on which all physical variables {\em in the critical region} become universal when appropriately scaled in the form\\
%\begin{center}
%\begin{array}{ccccc}
%{\rm Time} & & $t$ & \rightarrow & $t/ \hat{t}$  \\
%{\rm Distance}& & $r$ & \rightarrow & $r/\hat{\xi}$  \\
%{\rm Wavevector}& & $k$ & \rightarrow & $k \hat{\xi}$ 
%\end{array}
%\end{center}
%
%which  in the critical region $[-\hat{t}\,,\, \hat{t}]$ can be mathematically written (upon ignoring the role of the anomalous critical exponent which is typically zero, or very close to it~\cite{nishimori-phasetransition-book}) 
%this takes in the critical region $[-\hat{t}\,,\, \hat{t}]$ 
The Kibble-Zurek scaling hypothesis thus states that in the critical region $[-\hat{t}\,,\, \hat{t}]$, the correlation function can be cast in the
 form
\be
C({\bf r},t) \sim  \frac{1}{\hat{\xi}^{d-2}} F \left( \frac{t}{\hat{t}}\,,\,\, \frac{{\bf r}}{\hat{\xi}} \right) \;,
\label{eq:kz-scaling-hyp}
\ee
where $F$ is some system-specific function, and $d$ is the system dimensionality.
%and the scaling parameters $\hat{t}$ and $\hat{\xi}$.
%From Eq.~(\ref{eq:kz-scaling-hyp}), 
In general, this implies that the correlation function is dynamically evolving in a manner which changes both its shape, and its range.
%
%Nonetheless,
%For completeness,  we note here that although
However, the discussion so far has been focussed on the
 (adiabatic-impulsive-adiabatic) `cartoon' Kibble-Zurek scenario of Fig.~\ref{fig:5}: it is thus pertinent to note here that while such a `cartoon' picture is indeed consistent with the scaling function of Eq.~(\ref{eq:kz-scaling-hyp}), it nonetheless further constraints it by implying a particular (time-independent) form for $F$.
%moreover, for a homogeneous 3D system, the scaling parameters $\hat{t}$ and $\hat{\xi}$ depend on the quench rate $\tau_Q$ via the scalings of Eqs.~(\ref{t-hat})-(\ref{xi-hat}).

%For example, the spectral function in the critical region can be cast in a universal form 
%\be
%n(k,t) = \hat{t} \,\, G \left( \frac{t}{\hat{t}},\,\, k \hat{\xi} \right) \;.
%\ee
%in terms of some system-specific function $G$, and the scaling parameters $\hat{t}$ and $\hat{\xi}$ depending on the quench rate $\tau_Q$ via the scalings of Eqs.~(\ref{t-hat})-(\ref{xi-hat}) [for a homogeneous 3D system].

Setting this aside for now, I note that the Kibble-Zurek relation of Eq.~(\ref{eq:kz}) enables us to predict the scaling of the number of defects spontaneously generated within such system during the linearly-quenched phase transition crossing with the quench timescale $\tau_Q$.
Based on the typical size of independent domains $\hat{\xi}$ in the broken-symmetry phase (set by the equilibrium correlation length $\xi$ evaluated at the freeze-out time $-\hat{t}$), and information about the system dimensionality, $d$, and defect dimensionality, ${\cal D}$, 
%(e.g.~a soliton, or straight vortex has $d=1$, a vortex ring has .....), 
%one can recast such relation in terms of the number of defects spontaneously generated within such system during the linearly-quenched phase transition crossing. 
this takes the form~\cite{delcampo-kibble-zurek}
\be
n_{\rm defect} \sim 
\left( \frac{\hat{\xi}^{\cal D}}{\hat{\xi}^d} \right) \sim
\left( \tau_Q \right)^{-\alpha}\;, \label{eq:n-defect}
\ee
where the exponent $\alpha > 0 $ in general depends on the critical exponents ($\nu,\, z$) and the system/defect dimensionalities ($d,\, {\cal D})$. 
For example, $(d-{\cal D})=1$ for a 2D planar soliton in a 3D system or a point-like solitonic defect in a 1D system, while $(d-{\cal D})=2$ in both cases of a linear vortex filament in a 3D system, or a point-like vortex defect in a 2D system, because two dimensions are needed to accommodate the phase profile around such a vortex. 
As expected, one finds that -- compared to slow quenches -- faster quenches (i.e.~small $\tau_Q$) for which $|\epsilon/\dot{\epsilon}|$ has a smaller slope, and thus yields a smaller value of $\hat{\xi}$, lead to the generation of more defects: this is because more independent regions of smaller size can be accommodated in a given system size.
In the specific 3D homogeneous example,
$\alpha = (d-{\cal D}) \nu / (1 + \nu z)$, whereas this becomes modified in the presence of harmonic confinement (see later).
The power of the presented `cartoon' Kibble-Zurek mechanism is that it predicts the correct {\em scalings} with $\tau_Q$ for a linearly-quenched homogeneous system.
Nonetheless, an immediate shortcoming of such adiabatic-impulsive-adiabatic scenario which can be identified, is that it ignores the temporal growth of correlations within the critical region; moreover, realistic experiments (besides being performed in finite-size systems) are often conducted in spatially-varying potentials, thus requiring the extension of such mechanism to inhomogeneous settings. Both these issues are addressed in Secs.~\ref{sec:kz-spgpe}-\ref{sec:kz-sonic} below.

In the context of ultracold atom experiments, beyond the previously-mentioned ring-trap geometry~\cite{dalibard-ring-KZ-2014,dalibard-multiring-merging}, the Kibble-Zurek phenomenon across a thermal phase transition has also been  studied %in the context of ultracold quantum gases 
in different settings, including box-like geometries in 3D~\cite{Hadzibabic-KZ} and 2D~\cite{Dalibard-KZ-2D}, elongated 3D~\cite{lamporesi2013spontaneous,donadello2016creation,liu_dynamical_2018,Trento-KZ-2022} and oblate (pancake-like)~\cite{KZ-Shin-1-EarlyTime-Coarsening,KZ-Shin-2-Early-Coarsening-Beyond-KZ,KZ-Shin-3-Inhomogeneous} inhomogeneous traps, and even in the context of strongly-interacting fermionic superfluids~\cite{shin-KZ-Fermi-2019},~\footnote{Tuning the interactions across the BEC-BCS crossover through the unitarity limit in a strongly-interacting superfluid revealed a near-constant, i.e.~interaction-strength-independent, critical exponent $\alpha$.}.

The insightful experiment at Cambridge in a 3D box-like geometry~\cite{Hadzibabic-KZ} probed the system spatial correlation function at different times, by interfering the system with a displaced copy of itself: they used such measurements to explicitly demonstrate the existence of critical slowing down; by determining the scaling of the correlation length with $\tau_Q$ -- and using the expected (and previously experimentally extracted~\cite{Donner:2007} value of $\nu \approx 2/3$ discussed earlier) -- they were able to show that the dynamical critical exponent $z = 3/2$, consistent with the expectation for the 3D XY model universality class.
%and to obtain the expected correlation length scaling with $\tau_Q$, which can implicitly yield a measurement of the dynamical critical exponent $z$ (once the information about the expected value of $\nu$ is fed in).
%
Many of these early ultracold experiments also probed the number of generated vortices during a controlled quench, and its dependence on the quench timescale $\tau_Q$~\cite{dalibard-ring-KZ-2014,Dalibard-KZ-2D,donadello2016creation} (see subsequent discussions for more recent works).
On the whole, good overall agreement with the expected behaviour was obtained for sufficiently slow quenches as long as other appropriate factors, such as dimensionality and type of confinement (e.g.~harmonic, ring-trap), were taken into account: however a common potential limitation in such cases arises from the fact that experimental defect counting is typically only possible at times significantly longer than $+\hat{t}$ (which can generally be rather restrictive for most defect types due to their relatively short lifetimes), or after allowing the system to relax and/or expand for some time to make defect detection easier:
%-- except for the case of persistent currents): 
such aspects are further discussed below.

%-------------------------------------
\begin{figure*}[t!]
\centering
\includegraphics[width=0.8\linewidth]{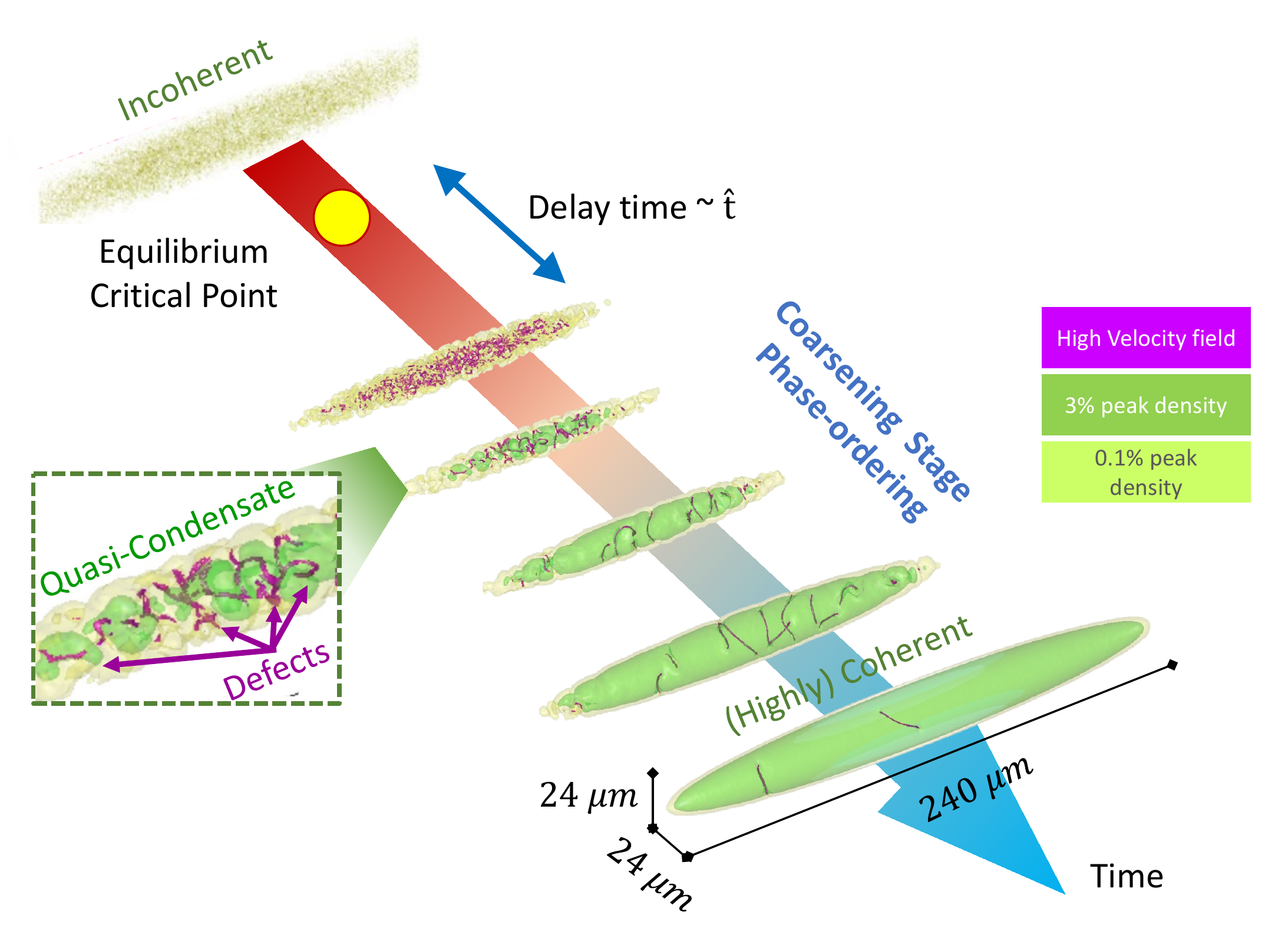}%{Fig-6a.png}\\
\caption{
Schematic of entire condensate growth process during a
 driven cooling quench from an initial equilibrium thermal configuration:
Shown are density isosurfaces of the most populated  (Penrose-Onsager) mode based on numerical simulations of Eq.~(\ref{spgpe}) 
for the parameters of a recent elongated 3D harmonic trap experiment of Refs.~\cite{lamporesi2013spontaneous,donadello2016creation,liu_dynamical_2018}.
 After the characteristic Kibble-Zurek delay time of $\sim \hat{t}$, one sees the spontaneous emergence of a `turbulent' quasi-condensate state (green) containing a large number of spontaneously-generated defects (shown in purple) whose density decreases as one moves away from the trap centre: these tangled defects gradually decay to a small number of well-formed vortices which are rather long-lived and become trapped in the growing macroscopically-occupied condensate mode.
 During such evolution, the relative weight of the dominant eigenvalue (compared to the next largest one) increases significantly, revealing clear evidence of the transition from a quasi-condensate exhibiting multiple equally-likely populated modes (around $\hat{t}$) to a pure BEC (with one dominant eigenvalue) in the latter density snapshots, in which the relative volume occupied by the defects has significantly decreased.
}
\label{fig:6a}
\end{figure*}
%-------------------------------------

\subsection{Analyzing Kibble-Zurek in an Elongated Ultracold Atomic System} \label{sec:kz-spgpe}

To gain further direct microscopic insight into the complex system dynamics over all timescales and physical observables, and probe the physics beyond the `cartoon' Kibble-Zurek picture presented so far, I briefly review here key results obtained in the context of a particular sequence of experiments conducted at Trento in an elongated 3D harmonic geometry~\cite{lamporesi2013spontaneous,donadello2016creation,liu_dynamical_2018,Trento-KZ-2022}. In such experiments, the system was cooled through the phase transition at different rates $\dot{T}$, with approximately fixed initial and final states.

A schematic of such process -- based on extensive numerical simulations~\cite{liu_dynamical_2018} via Eq.~(\ref{spgpe}) --  is shown in Fig.~\ref{fig:6a}. An initially incoherent gas is dynamically quenched through the critical point (which is itself shifted by the combination of fluctuations and mean-field effects). As predicted, the system reacts (`unfreezes' its dynamics) with a certain delay time $+\hat{t}$ after crossing the equilibrium critical point, at which point the highly-symmetry-broken state looks rather `chaotic': here the randomly distributed and tangled purple lines of variable lengths and orientations depict spontaneously generated defects in the form of highly-excited vortices in a strongly-turbulent state.
Notice that the defects only emerge within an elliptical volume -- consistent with the imposed anisotropic harmonic confinement -- and that the defect density is highest at the trap centre, where the particle density is also highest: I will comment later on the significance of such observations.

%-------------------------------------
\begin{figure*}
\centering
\includegraphics[width=1.0\linewidth]{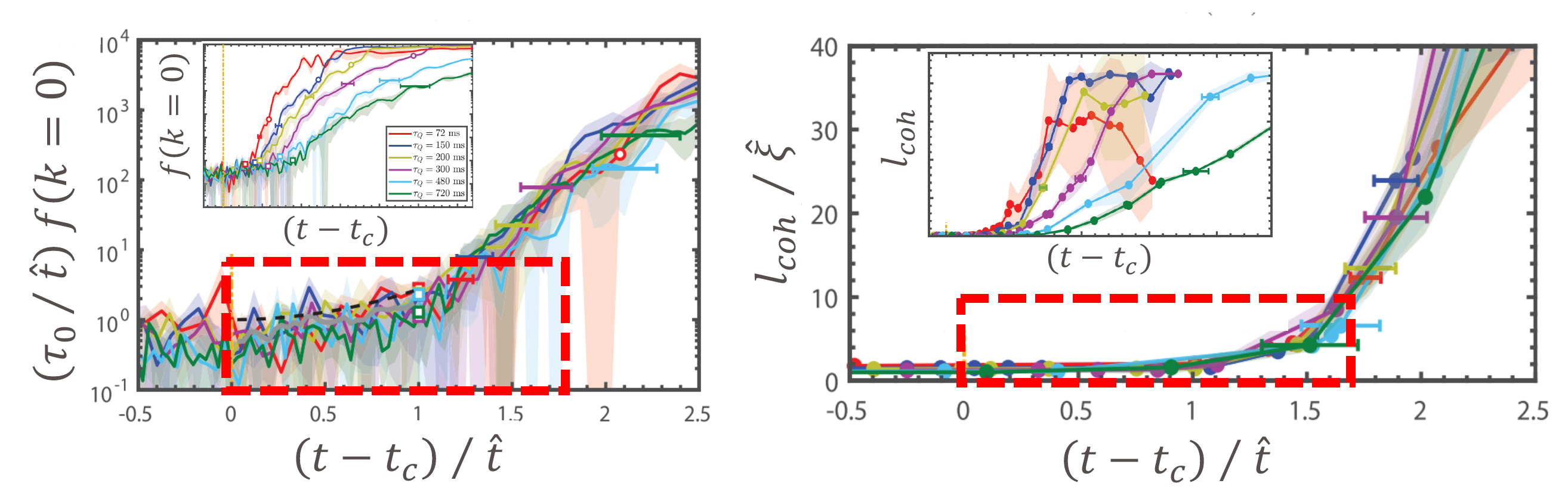}%{Fig-6b.png}
\caption{
%Effect of Quench Rate for the parameters of Fig.~\ref{fig:6a}
 Numerical demonstration of the Kibble-Zurek scaling hypothesis revealing self-similar evolution for the parameters of Fig.~\ref{fig:6a} in terms of appropriately rescaled (i) zero-momentum mode occupation $f(k=0)$, and (ii) coherence length $l_{coh}$ for a range of different $\tau_Q$. Main plots show rescaled evolution of insets, with time measured in terms of $(t-t_c)$ (where $t_c = t(\epsilon=0)$) and scaled to $\hat{t}$. Rescaled vertical axes respectively plot (i) $(\tau_0/\hat{t})f(k=0)$ (where $\tau_0$ is a non-universal constant time) and (ii) $l_{coh}/\hat{\xi}$.
 In the unscaled insets (plotted in terms of $(t-t_c$), the value of the quench timescale
$\tau_Q$ decreases from the lower (slower quench, green) to the upper (faster quench, red) curves.
Adapted with permission from I.~K.~Liu {\em et al.}, Kibble-Zurek dynamics in a trapped ultracold Bose gas, Phys. Rev. Research {\bf 2}, 033183. Copyright (2020) by the American Physical Society.
%(b) Characteristic single-trajectory temporal evolution from a fixed initial to a fixed final state for 3 different values of $\tau_Q$ (quenching speed increases from quasi-adiabatic (top) to fast (bottom), with middle trajectory showcasing  a typical case where Kibble-Zurek scaling is expected to be fully applicable. Remarkably, faster quenches (bottom) contain (on average) more defects at a very late evolution time that slower quenches.
}
\label{fig:6b}
\end{figure*}
%-------------------------------------

%
Based on the Kibble-Zurek scaling hypothesis, all system observables become universal during the time interval $t \in [-\hat{t},\hat{t}]$, 
thus uniquely specifying the temporal window during which one should be investigating universal Kibble-Zurek scaling effects for constant-rate cooling quenches $\tau_Q = T_c/|\dot{T}|$.

The emergence of such universal features can be seen in Fig.~\ref{fig:6b} which showcases the unscaled (insets) and appropriately-rescaled dynamical evolution of (i) the zero-momentum mode occupation $f(k=0)$ [left], and (ii) the evolving system correlation length $l_{coh}$ [right]. Unscaled images are plotted in terms of the time $(t-t_c)$ lapsed since crossing the equilibrium phase transition, whereas such shifted time has been additionally scaled by $\hat{t}$ in the rescaled plots [main plots]. Moreover, in the rescaled plots the vertical-axis variables have been respectively rescaled as $\tau_0/\hat{t} f(k=0)$ and $l_{coh}/\hat{\xi}$.
These simulations~\cite{Liu-KZ-PRR} reveal universal features up to a timescale of $\sim \hat{t}$ after the system crosses the equilibrium critical point. Although a delayed response was also found in more recent experiments with the same system, at such early times there was neither direct access to the spatially-resolved density distribution nor the vortex number, with such delay instead attributed to other factors masking the ability to extract $\hat{t}$ directly~\cite{Trento-KZ-2022}. 
%in these inhomogeneous experiments~\cite{Trento-KZ-2022}. 
%
%such timescale is not currently experimentally accessible, and so one has to rely on numerical simulations, revealing such scaling features very clearly in Fig.~6(b). 

%-------------------------------------
\begin{figure*}[t]
\centering
\includegraphics[width=1.0\linewidth]{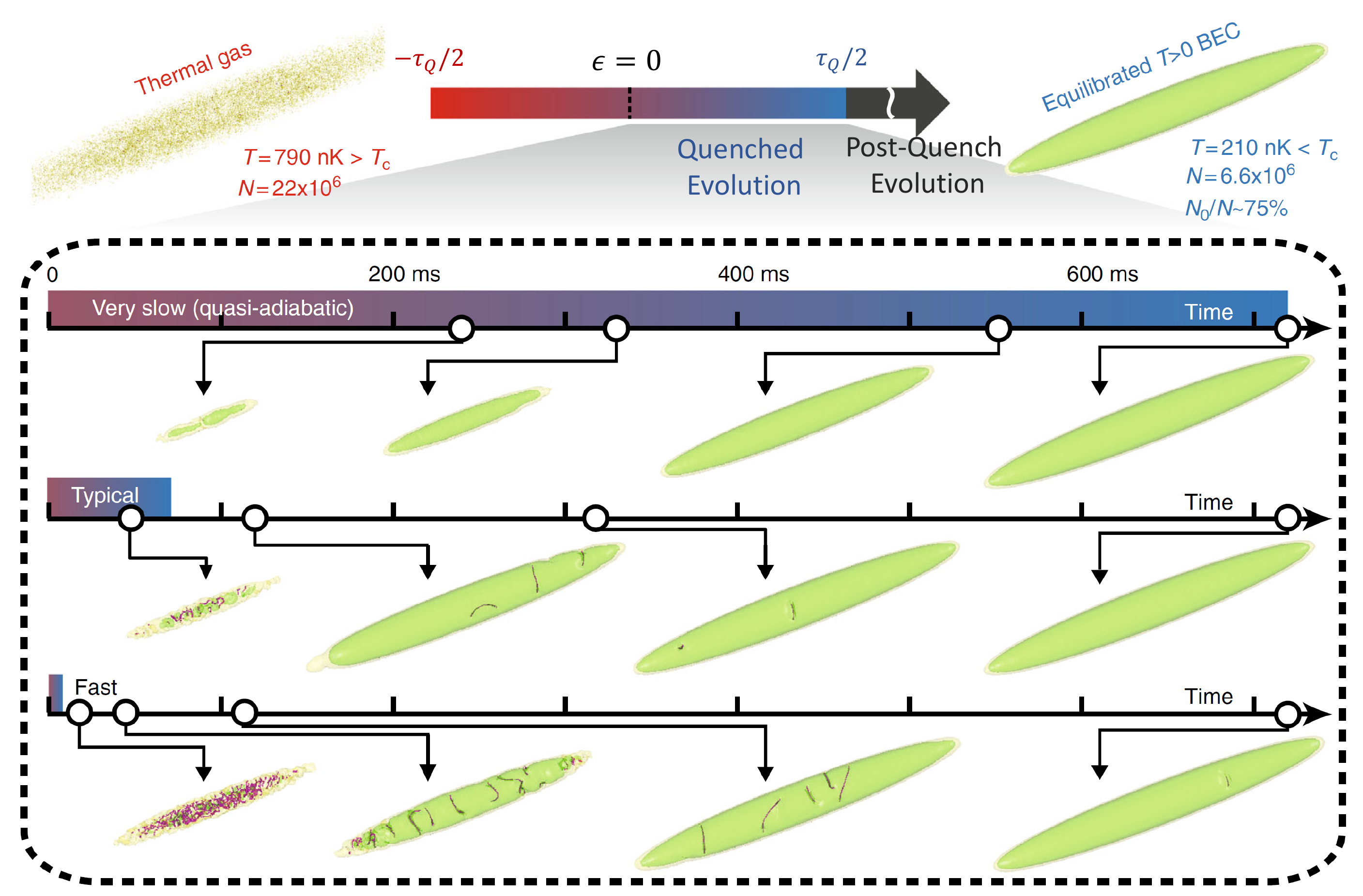}%{Fig-9.png}
\caption{
Dependence of dynamical growth on quench rate  for the parameters of Fig.~\ref{fig:6a}
%(a) Numerical demonstration of the Kibble-Zurek scaling hypothesis revealing self-similar evolution for appropriately rescaled (i) zero-momentum mode occupation $f(k=0)$, and (ii) coherence length $l_{coh}$ for a range of different $\tau_Q$. Main plots show rescaled evolution of insets, with time measured in terms of $(t-t_c)$ (where $t_c = t(\epsilon=0)$) scaled to $\hat{t}$. Rescaled vertical axes respectively plot (i) $(\tau_0/\hat{t})f(k=0)$ (where $\tau_0$ is a non-universal constant time) and (ii) $l_coh/\hat{\xi}$.
%$\tau_Q$ decreases from the lower to the upper unscaled curves in the inset.
based on characteristic single-trajectory temporal evolution from a fixed initial (thermal) to a fixed final state (with a $\sim 75\%$ BEC fraction) for 3 different values of $\tau_Q$: schematic at the top shows symmetric quench protocol from a $\mu <0$, $T>T_c$ initial thermal state to one with $\mu>0$, $T<T_c$ indicating corresponding temperatures and atom numbers characteristic of such controlled quench experiment. The quench rate increases from quasi-adiabatic (top) to fast (bottom), with middle trajectory showcasing  a typical case where Kibble-Zurek scaling is expected to be fully applicable. Remarkably, faster quenches (bottom) contain (on average) more defects at a very late evolution time that slower quenches.
Adapted with permission from I.~K.~Liu {\em et al.}, Dynamical equilibration across a quenched phase transition in a trapped quantum gas, Commun. Phys. {\bf 1}, 24. Copyright (2018) by the Nature Publishing Group.
}
\label{fig:6c}
\end{figure*}
%-------------------------------------

As the system is further cooled (i.e.~$\epsilon>0$ increases), one obtains a clear glimpse of its quasi-condensate characteristics:
in particular, one sees the emergence of scattered localized regions of randomly-selected near-constant phase (and so high coherence {\em locally}), separated by chaotically evolving and highly-excited defects (vortices): at this stage there are many highly-occupied modes competing against each other, and such information is revealed by the study of the largest, approximately equally-dominant, eigenvalues~\cite{liu_dynamical_2018} -- providing direct evidence of quasi-condensation.
The state of the system can also be thought of as `turbulent' (in the broad interpretation of the word).

As time evolves, and the system is further externally cooled (at the same constant rate), the random quasi-coherent regions grow in size, with the defects separating such regions interacting with other defects, thus dissipating their energy and shrinking in size. 
Naturally, the process of defect generation through the symmetry-breaking process and defect decay through coarsening coexist temporally, making it hard to isolate the relative importance of such competing mechanisms. However, at sufficiently late times
%This is the beginning of the second evolutionary stage of coarsening, or phase-ordering dynamics. At sufficiently late times 
after the dynamical phase transition crossing (i.e.~for $\epsilon(t) \gg 0$), and within a specific temporal window, such dynamics are generally expected to exhibit a new set of self-similar (universal) scaling behaviour associated with the phase-ordering process~\cite{bray_theory_1994,damle_phaseordering}.
The complicated inhomogeneous and anharmonic elongated 3D geometry of this system largely conceals such features, which are therefore discussed in more detail later on in  the `cleaner' context of a homogeneous 2D quantum gas in Sec.~\ref{sec:scalinghypothesis}.
The net effect of this lengthy process is the gradual establishment of domains of increased volume and coherence, and thus the gradual transition from a state of `quasi-condensation' to that of true condensation, with properties set by the external system parameters ($T$ and $\mu$).
%
%the density grows faster in the centre of the trap and the size of domains increases, with defects gradually becoming engulfed within the growing coherent atomic density. This is the second evolutionary stage of coarsening, or phase-ordering dynamics. During such phase the density grows and the engulfed defects start interacting with each other.

During such dynamics, the tight transverse direction rapidly constrains the system evolution, and dictates the specific type of defect preferentially forming in such a system: this corresponds to the macroscopic excitation with the lowest energy in the given geometry~\footnote{In this case, such defect is a so-called `solitonic vortex' which exhibits a $2 \pi$ phase winding characteristic of quantized vorticity, but with a non-uniform gradient, such that its phase profile resembles the characteristic planar jump of a dark soliton from larger distances~\cite{donadello_observation_2014}.}.
%The trapped vortices continue interacting, and 
As the defect density gradually decreases, one sees the emergence of a phase-coherent condensate within which are embedded few long-lived defects: this is in fact a very clear dynamical demonstration of the trapping of defects emerging as relics of the phase transition in the growing highly-coherent system. 
The particular geometrical set-up constraints their interactions, such that they only interact rather infrequently when few defects remain. 
However, when they do, the defects can reconnect and dissipate away further energy~\cite{Trento-PRX-VortexReconnection}. After significant such evolution, the system eventually expels all of its defects: at this point, the competition between the various quasi-condensate modes is complete, and one arrives at a fully phase-coherent Bose-Einstein condensate. Such a transition is further confirmed by a study of the ratio of the largest to second-largest eigenvalue of the system as a function of time~\cite{liu_dynamical_2018,Liu-KZ-PRR}.
Density profiles (green) in Figs.~\ref{fig:6a}, \ref{fig:6c} showcase the evolution of the spatial distribution of the most populated eigenstate, even though one only becomes genuinely dominant (the Penrose-Onsager mode) at late times.

Obviously, details of such process are heavily dependent on the quench protocol (e.g.~temperature quench, interaction quench, geometry/phase-space quench), system geometry and dimensionality and interaction strength (which respectively set the typical number, type and size of the defects).  A characterisitc illustration of how the quenching rate (which for given initial and final states is simply a function of the inverse quench timescale $\tau_Q$) affects the dynamics can be found in Fig.~\ref{fig:6c} which displays the system evolution over an extended fixed total evolution time for 3 different quench rates~\cite{liu_dynamical_2018}. As evident, more defects are generated {\em on average} early on for more rapid quenches (bottom);
moreover, because of this -- and despite rapid defect annihilation and expulsion -- typically more defects remain visible at a later absolute time when performing rapid cooling (i.e.~for systems which cross the equilibrium critical point $\epsilon=0$ faster), than in more gradual cooling schemes. 
For completeness, I note that while the example trajectories chosen to be displayed here are characteristic of the average behaviour, each single cooling realization (for a fixed value of $\tau_Q$) can lead to some variation in the number (and distribution) of generated defects; therefore, one needs to perform a number of distinct numerical/experimental realizations to appropriately extract relevant averaged parameters (e.g.~defect number evolution).

Performing such averaging over different realizations, both experiments~\cite{donadello2016creation} and simulations~\cite{liu_dynamical_2018} found a power law decay in the defect number with $\tau_Q$, with an exponent broadly consistent  with the analytically predicted exponent $\alpha$ for this setting~\cite{del_campo_universality_2014}: while such features/measurements were only experimentally accessible at rather late evolution times, much exceeding $\hat{t}$, numerical simulations demonstrated that the actual overall scaling behaviour was rather insensitive to evolution time: this is presumably because after defects have formed, they primarily propagate `freely' in the confining potentials for significant fractions of time, thus being rather long-lived: this is particularly true when the number of defects is not too large and they are spread over a large volume, as is the case for not-very-fast quenches.
Importantly -- and rather intuitively -- such Kibble-Zurek scaling was not however seen for very fast quenches ($\tau_Q \sim O(\hat{t})$): this arises when the maximum number of defects that can be accommodated in the system volume (or the maximum number that can be experimentally, or numerically, detected) has been reached in a given geometry, thus giving rise to a plateau in the defect number for small $\tau_Q$: this was clearly observed experimentally (at late times)~\cite{donadello2016creation}, and found to be consistent both with a simple model and with numerical simulations~\cite{liu_dynamical_2018}.
We also note that the decay of fluctuations was found to obey a universal power law, with a  distinct exponent to that corresponding to the growth of condensation~\cite{liu_dynamical_2018,Trento-KZ-2022}.

The quenching of defect number for very rapid quenches (i.e.~the deviation from the Kibble-Zurek scaling law $n_{\rm defect} \sim (\tau_Q)^{-\alpha}$) was also observed in strongly-interacting fermionic superfluids~\cite{shin-KZ-Fermi-2019} and in more recent bosonic experiments in an oblate geometry~\cite{KZ-Shin-1-EarlyTime-Coarsening,KZ-Shin-2-Early-Coarsening-Beyond-KZ,KZ-Shin-3-Inhomogeneous}.
The latter series of experiments also reported observational evidence for universal coarsening dynamics during the early stages of defect formation. 
Moreover, they were able to observe spatial variations in the generated defect number: both these features are in qualitative agreement with the presented numerical simulations~\cite{liu_dynamical_2018}.

All above features indicate that -- rather unsurprisingly --there is more to driven phase transition crossing than the simple (but still very elegant and powerful) adiabatic-impulsive-adiabatic Kibble-Zurek model of Fig.~\ref{fig:5}.
Below I address more systematic extensions of such a model, as highly relevant in light of such recent experimental developments.

\subsection{Beyond Homogeneous Kibble-Zurek Considerations} \label{sec:kz-sonic}

The preceeding discussion has highlighted two limitations of the adiabatic-impulsive-adiabatic Kibble-Zurek scenario of Fig.~\ref{fig:5} which I address here.

\begin{figure}[t]
\centering
\includegraphics[width=1.0\linewidth]{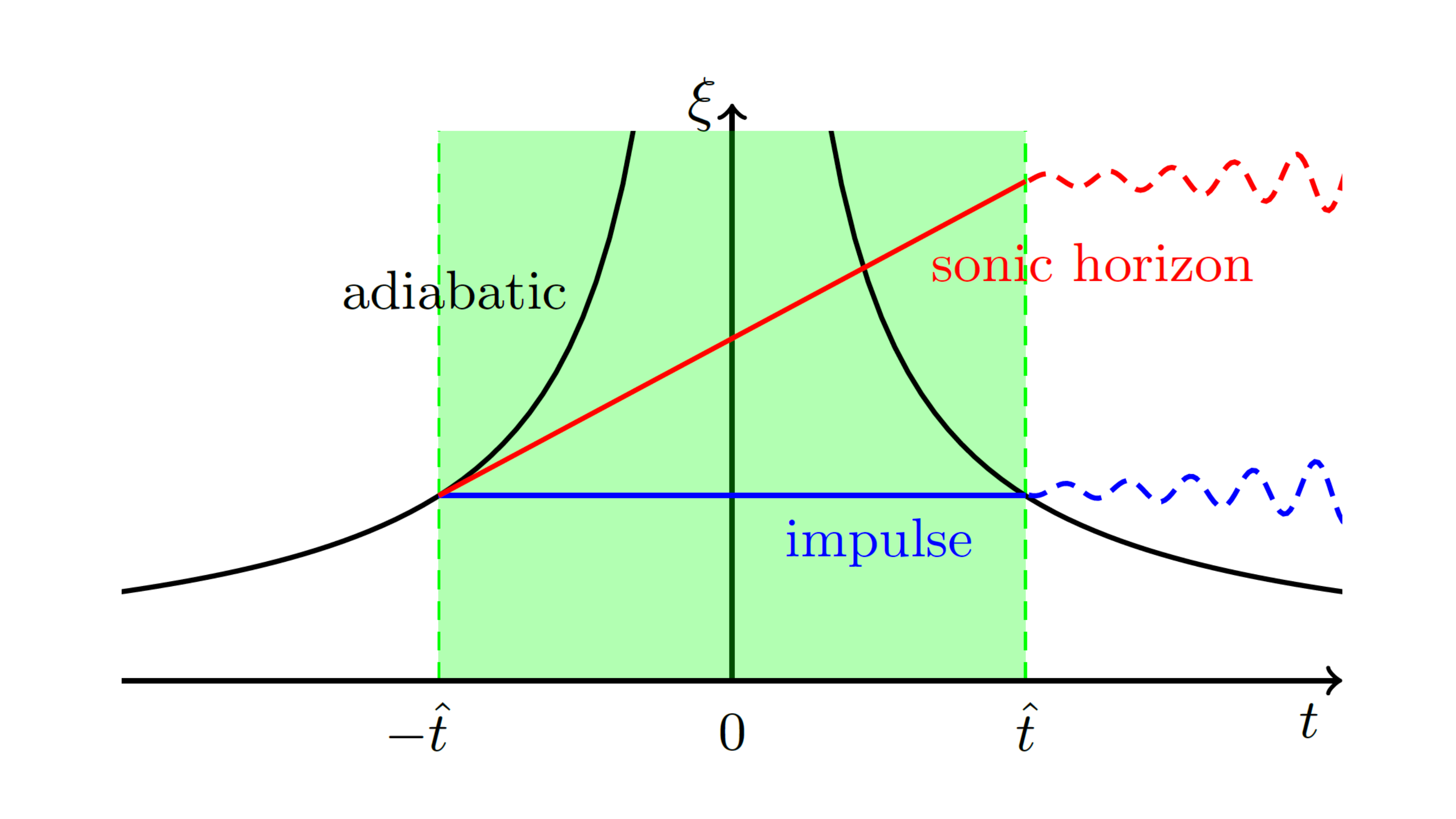}%{Fig - Sonic KZ.pdf}\\
\caption{
Extended view of the Kibble-Zurek mechanism explicitly accounting for a sonic horizon propagation with a critical speed of $\hat{\xi}/\hat{t}$.
Unlike the `cartoon' regime of impulsive dynamics for $t \in [-\hat{t},\, \hat{t}]$ shown in the Fig.~\ref{fig:5}, during which the correlation length is assumed to remain frozen to its value at $-\hat{t}$, this improved picture allows the `sonic cone' to grow during such time window by a constant rate set by the instantaneous slope of $\xi(-\hat{t})/\hat{t}$. As a result, the typical radius of the region over which broken symmetry emerges grows from $\hat{\xi}$ to few $\hat{\xi}$.
Reprinted with permission from D.~Sadhukhan {\em et al.}, Sonic horizons and causality in phase transition dynamics, Phys.~Rev.~B {\bf 101}, 144429~\cite{dziarmaga_sonic}. Copyright (2020) by the American Physical Society.
}
\label{fig:kz-more}
\end{figure}

The first one, concerns the `freezing' of the correlation length $\xi$ during the entire evolution $t \in [-\hat{t},\, \hat{t}\,]$, to a constant (time-independent) value corresponding to its instantaneous value at 
the time $-\hat{t}$ when symmetry-breaking occurs: in other words
$\xi \rightarrow \hat{\xi} = \xi\left(\epsilon \left(-\hat{t}\, \right)\right)$, which would imply a constant size of domains during $[-\hat{t},\, \hat{t}\,]$, which is somewhat counter-intuitive.
In fact, during such a temporal period, causality implies the existence of a characteristic velocity for the growth of correlations which -- in this context -- can be defined as~\cite{dziarmaga_IKZM_1999,zurek-soliton,dziarmaga_sonic}
\be
\hat{v} = \frac{\hat{\xi}}{\hat{t}} \;.
\ee
Given that the typical locally-coherent regions grow with such a velocity even during $t \in [-\hat{t},\, \hat{t}]$, the emerging effective size of domains at $\hat{t}$ is in fact larger than previously suggested, with $\xi\left( +\hat{t} \right)$ typically a few times its corresponding value at $-\hat{t}$~\cite{dziarmaga_sonic}: a schematic of such enhanced picture is shown in Fig.~\ref{fig:kz-more}.
Although such considerations evidently decrease the total number of spontaneously-generated defects, rather remarkably they do not actually
invalidate the previous arguments in terms of the existence of {\em scalings} of the defect number with quench timescale [Eq.~(\ref{eq:n-defect})], provided of course that the system inhomogeneity determining the shape of the region of increasing density is appropriately accounted for in the exponent $\alpha$.

In this context, it should be noted that the `cartoon' Kibble-Zurek scenario discussed in Sec.~\ref{sec:kz-hom} [Fig.~\ref{fig:5}] was formulated for a homogeneous system and thus assumed that
the system reacts, in a statistical sense, in the same way irrespective of which part of the system one looks at.
However, as clearly visible from Fig.~\ref{fig:6a} -- and commented upon earlier -- the density is highest at the centre of a harmonically-confined gas, implying that coherence grows from the trap centre radially outwards.
As a result, both the critical temperature of the system, and hence the distance to criticality $\epsilon$ become position-dependent~\cite{zurek-soliton,del_Campo_2011_IKZM,delcampo-kibble-zurek,del_campo_universality_2014}, i.e.~$T_c \rightarrow T_c({\bf r})$ and $\epsilon(t) \rightarrow \epsilon({\bf r},t)$.
A natural question then arises, as to the extent to which the previous `cartoon' picture holds under inhomogeneous trapping, and how to appropriately generalize this.

In the context of the homogeneous Gross-Pitaevskii equation (with repulsive interactions $g>0$), symmetry-breaking is induced simultaneously across the entire system when the chemical potential transitions from negative values (corresponding to a purely thermal cloud) to positive values. Due to the presence of the trap, one can instead define an effective spatially-dependent chemical potential $\mu({\bf r},t) = (\mu(t) - V({\bf r}))$ thus outlining at any given time the  region with $\mu({\bf r},t)>0$ where symmetry-breaking takes place. 
Such region grows radially outwards from the trap centre (here in an elliptical manner), thus defining a phase transition front which propagates in the system with a characteristic speed
\be
v_F({\bf r}) = \frac{T_c(0)}{\tau_Q} \left( \frac{dT_c({\bf r})}{d{\bf r}} \right) \;.
%\frac{\dot{T}}{(dT_c(r)/dr)} \;.
\ee
%where $T_c(r)$ labels the now position-dependent critical temperature.
As such, the simultaneous emergence of symmetry breaking in a homogeneous system can be described in terms of an infinite phase transition front velocity $v_F$, and this is also (locally) the case at the centre of a harmonic trap where local evolution mimics that of a homogeneous system.
The extent to which the emerging symmetry-breaking features in the phase transition of a system have a global or a local character  thus depends on the competition between the characteristic velocity $\hat{v}$ for the growth of correlations set by causality and the velocity $v_F$ for the growth of the critical front as determined by the system geometry, thus also accounting for the shape of the growing volume of the coherent density front. The generalization of the Kibble-Zurek effect which accounts for such spatial variations and local features is known as the inhomogeneous Kibble-Zurek mechanism, and is extensively discussed in Refs.~\cite{zurek-soliton,del_Campo_2011_IKZM,delcampo-kibble-zurek,del_campo_universality_2014,dziarmaga_sonic,dziarmaga_IKZM_1999}.

%-------------------------------------
%-------------------------------------

As a result, the previously discussed `cartoon' picture of Sec.~\ref{sec:kz-hom}, known as the {\em homogeneous} Kibble-Zurek picture remains {\em qualitatively} valid as long as (and within the regions where) 
\be
v_F > \hat{v} \;,
\ee
i.e.~the growth of the critical front happens on a faster rate than  such causality-induced coherence growth associated with the sonic horizon: in such a regime, the phase of each newly-formed domain is chosen locally.
As explained above, this is always the case in the centre of the harmonic trap, and in a region near the propagating front (due to critical slowing down)~\cite{delcampo-kibble-zurek}.
%thus this condition is broadly satisfied in some (potentially sizeable) fraction of the system. 
As a result, defects can in general only form spontaneously over a restricted volume in an inhomogeneous system, thus leading to a reduced number of emerging defects in comparison to the homogeneous predictions. 
In the opposite limit of $v_F < \hat{v}$, the growing condensate can rapidly communicate its choice of phase to the neighbouring domains, and so no further defect formation can occur.
%
%In other words, if the sonic horizon grows faster than the evolution of the critical front, the 
For an evolving system to enter a regime where it can probe its true inhomogeneous nature, the sonic horizon should be growing faster than the evolution of the critical front.

This, in turn, places constraints on the values of $\tau_Q$ required in a given system for a quench to allow the inhomogeneous nature of the phase transition to be fully probed: applying such simple arguments to the above elongated 3D experiment suggested the need for significantly slower quenches, thus also requiring longer evolution times and enhanced system stability.
Although such a regime was not reached in the experiments of Ref.~\cite{donadello2016creation}, it should nonetheless be noted that 
(i) the corresponding numerical simulations discussed in Sec.~\ref{sec:kz-spgpe} clearly demonstrated the spatial dependence of spontaneous defect generation over a restricted volume, and that
(ii) both observed and numerically-computed scalings of the defect density with $\tau_Q$ were found to be consistent with predictions based on critical exponents for a harmonic trap~\cite{donadello2016creation,liu_dynamical_2018} (leading to an amended power $\alpha = (d-{\cal D})(1+2 \nu)/(1+\nu z)$~\cite{del_Campo_2011_IKZM,del_campo_universality_2014} when such geometrical features of the growth were taken into account).
%and (ii) that the corresponding numerical simulations discussed in Sec.~\ref{sec:kz-spgpe} clearly demonstrated the spatial dependence of spontaneous defect generation over a restricted volume.

%reproducing the power-law scaling of Eq.~(\ref{eq:n-defect}) for the emerging defect number with $\tau_Q$, but with an amended power $\alpha = (D-d)(1+2 \nu)/(1+\nu z)$ relevant for a 3D harmonic trap -- consistent with (late-time) experimental observations~\cite{donadello2016creation} and (both early- and late-time) numerical simulations~\cite{liu_dynamical_2018}.

Furthermore, it has been noted that a quasi-2D geometry may be a more favourable setting for observing such spatial signatures of the inhomogeneous nature of the dynamical phase transition.
This was indeed shown in very recent experiments in a quasi-2D geometry: specifically, following experiments highlighting the anticipated importance of coarsening dynamics during the early stages of spontaneous defect formation~\cite{KZ-Shin-1-EarlyTime-Coarsening,KZ-Shin-2-Early-Coarsening-Beyond-KZ} observations of the spatial distribution of defects in such a geometry found (after allowing the system to relax further and expand) a local suppression of spontaneous defect generation with $\tau_Q$ which was more pronounced in the outer region~\cite{KZ-Shin-3-Inhomogeneous}: in other words, the local value of the power law exponent $\alpha(r)$ dictating the defect number through the power law $n_{\rm defect} \sim \tau_Q^{-\alpha(r)}$ was inferred after expansion to increase locally as one moves away from the trap centre.
While such an observation is consistent with the interplay of causality and inhomogeneity, they consistently found higher exponents than might be anticipated: although various factors could be at play here, associated with the fact that observations were made after relaxation and expansion (and so not at $\hat{t}$), the ongoing defect dynamics and coarsening, and intricate details of the trapping potential, this does also pose interesting questions meriting further investigation regarding the applicability of the inhomogeneous Kibble-Zurek effect.
%
%for such observation. At the time of writing this chapter, this remains an active topic of investigation in ultracold quantum gases, although we note interesting very recent observations in oblate atomic condensates which have observed the spatial suppression of defect generation in outer, lower-density, inhomogeneous regions}\cite{KZ-Shin-3-Inhomogeneous}.
%
%Notwithstanding the above analysis, there is an important comment to be made: The extended Kibble-Zurek picture presented here does in fact both confirm (up to a multiplicative constant) the significance of $\hat{\xi}$ as a key relevant lengthscale, while simultaneously reproducing the power-law scaling of Eq.~(\ref{eq:n-defect}) for the emerging defect number with $\tau_Q$, but with an amended power $\alpha = (D-d)(1+2 \nu)/(1+\nu z)$ relevant for a 3D harmonic trap -- consistent with (late-time) experimental observations~\cite{donadello2016creation} and (both early- and late-time) numerical simulations~\cite{liu_dynamical_2018}.

Irrespective of the precise details just discussed of corrections to the basic Kibble-Zurek scheme which allow for evolution of the characteristic lengthscale and the interplay between inhomogeneity and causality, as the system exits the critical region in the $\epsilon >0$ side of the phase transition at $t = +\hat{t}$, it will be left in a rather non-equilibrium state (the faster the quench, the more non-equilibrium the state is likely to be -- see Fig.~\ref{fig:6c}): such a non-equilibrium state features a random mosaic of domains of different, near-constant, phase, separated by defects of variable length, orientation and curvature, in a way which accounts for the differences in phase between the establishing micro-domains. 
The relaxation of such a state through the gradual decay of the number/length of defects will ultimately lead to the final coherent state for the given system parameters. Here one can distinguish two cases:
Either one continues driving the system to the final desired equilibrium and studies its relaxation in an `open' (driven) environment, or one leaves the  system in such a highly chaotic/turbulent state and observes its dynamical evolution towards relaxation in a `closed' manner: the fundamental dynamics governing these two cases are different. Nonetheless, in both cases, universal laws can (and do, in general) emerge in certain temporal (and momentum) windows. Below I discuss the general concept of self-ordering by means of a scaling hypothesis, paying attention to non-equilibrium features, particularly in terms of universal late-time dynamics and other more general non-equilibrium features facilitating a bidirectional dynamic scaling, both of which can leave a significant imprint in the decay of superfluid turbulence.

Although very distinct dynamical regimes with different scaling laws, the dynamically evolving  $\hat{\xi}$ discussed above does bear some notional analogy to the  dynamically-growing lengthscale $L_c(t)$ emerging in the scaling regime discussed below.
%which bears some notional analogy to the dynamically evolving  $\hat{\xi} discussed above. %Sec.~\ref{sec:phaseordering-main}.

\section{Universality during Phase-Ordering Dynamics} \label{sec:phaseordering-main}

So far I have discussed the role of finite-duration quenches in a system, focusing on an incoherent to coherent thermal phase transition. 
We have discussed a broad dynamical range from quasi-adiabatic to near-instantaneous transitions, and explained how universal Kibble-Zurek dynamics emerges in the critical region provided the quench is not too fast.
As discussed, %(in the context of a specific experiment)
the relaxation process in the system %-- far from the critical region -- 
is governed by the decay of defects giving rise to growing regions of constant phase, with the quasi-condensate giving way to a macroscopic condensate~\cite{witkowska-deuar,liu_dynamical_2018}, a process known as phase-ordering.

In order to present the key features of such process in its most pure form,  it is convenient to consider the idealized limit of instantaneous quenches. Moreover, based on the above findings on the role of inhomogeneous trapping, and in order to suppress transversal (e.g.~Kelvin-wave-like) excitations on the defects, I initially restrict the discussion to two-dimensional homogeneous systems, before returning to more general non-equilibrium settings.
%providing some more general remarks.

\subsection{Scaling Hypothesis and Late-Time Coarsening} \label{sec:scalinghypothesis}

The dynamics through which the system expels all defects and orders its phase is known as phase-ordering kinetics, and has a long history.
Here I do not attempt to give a full account -- referring the reader to excellent reviews~\cite{bray_theory_1994,rutenberg1995phase,rutenberg_energy-scaling_1995} -- but rather to discuss its features in the specific context of quenched quantum gases.

Naturally, even when quenched instantaneously, the system requires some time to order itself: as argued before, defects -- which emerge at the interfaces across regions of different phases -- gradually disappear, thus allowing adjacent regions of particular (randomly-chosen) phases to merge, leading to the temporal growth of the typical domain size. A key point to note here is that this is also a scaling phenomenon: from a statistical point of view, the distribution of such domains -- each with a constant phase, yet no phase coherence between them -- looks at late times remarkably similar to the one at earlier times, with the growth of the typical domain size fully accommodated through a (simple) global change of scale.

%-------------------------------------
\begin{figure*}[t!]
\centering
\includegraphics[width=0.8\linewidth]{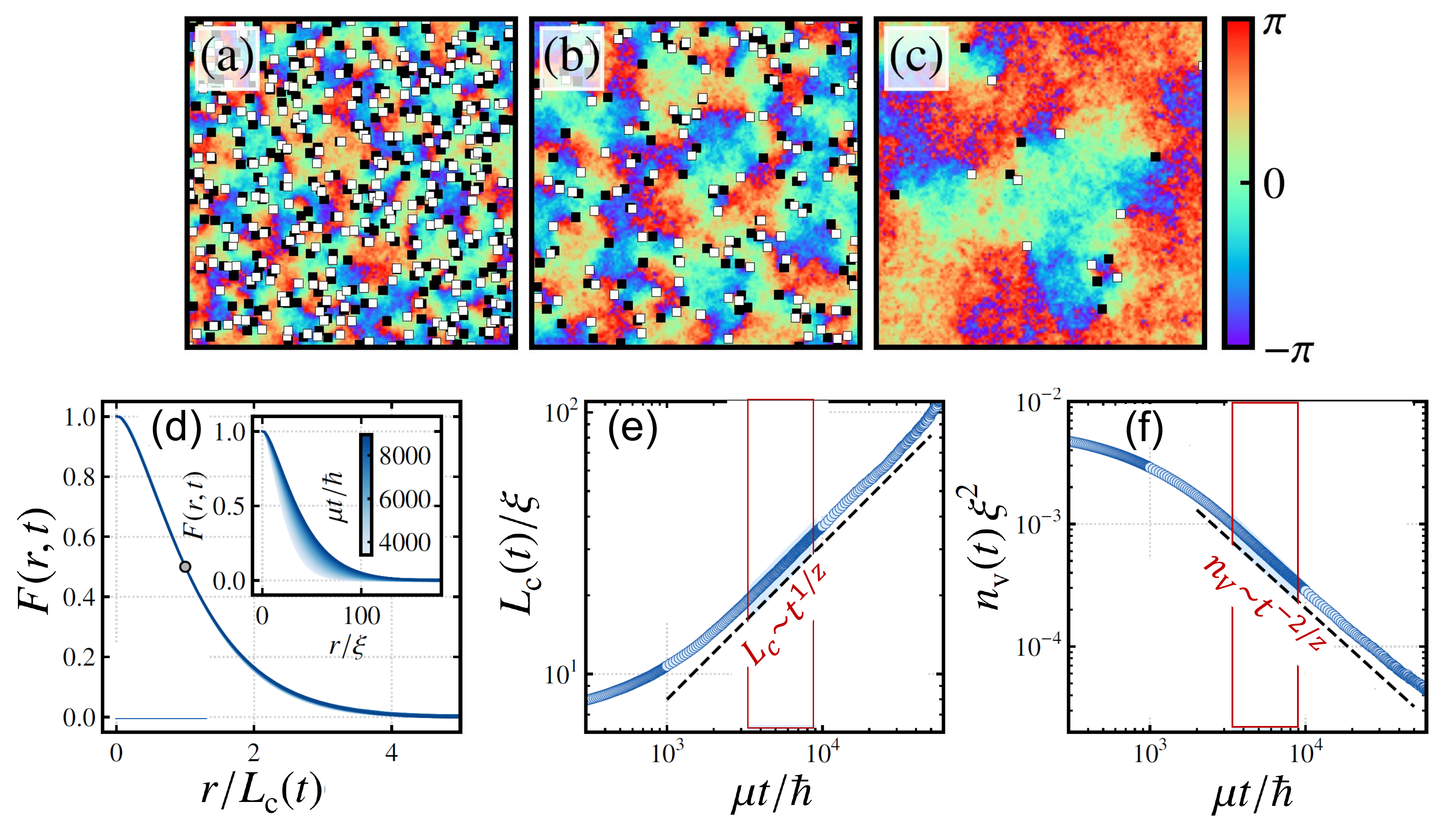}%{Fig - 2D.pdf}
\caption{
Phase-ordering kinetics of a 2D homogeneous Bose gas following an instantaneous quench across the BKT phase transition:
(a)-(c) Evolutionary snapshots of the phase, showing the reduction in the number of vortices/antivortices (white/black squares respectively).
(d) Self-similarity through the collapse (main plot) of the spatial correlation functions evaluated at different times $t$ within the `collapse window', scaled to $\hbar/\mu$ (inset) onto a universal scaling function $F$ in terms of the scaled variable $r/L_c(t)$ according to Eq.~(\ref{eq:C-scaling}).
(e) Post-quench growth of the characteristic lengthscale $L_c(t)$, clearly highlighting the time window when all self-similar correlation functions perfectly collapse onto one another.
(f) Corresponding decreasing evolution of the vortex number density $n_V$.
Although both (e)-(f) plot the variation across the entire temporal domain, note that the very late-time evolution is somewhat sensitive to numerical boundary effects (as can be seen from the slight deviation of both behaviours at very late times).
Adapted with permission from A. Groszek {\em et al.}, Crossover in the dynamical critical exponent of a quenched two-dimensional Bose gas, Phys. Rev. Research {\bf 3}, 013212~\cite{groszek_comaron_PRR}. Copyright (2021) by the American Physical Society.
}
\label{fig:phaseordering}
\end{figure*}
%-------------------------------------

We recall at this stage that a dynamical scaling hypothesis was previously considered in the context of the Kibble-Zurek scenario [Eq.~(\ref{eq:kz-scaling-hyp}), Sec.~\ref{sec:kz-hom}].
The scaling function was generally argued to change in time both in terms of its shape (through its explicit dependence on $t/\hat{t}$) and its spatial extent.
%with the characteristic length scale, $\xi$, assumed to remain constant, frozen to its value at $\hat{\xi} = \xi(-\epsilon)$.
In stark contrast to this, I note here that during the late-time self-similar dynamics and in the scaling limit  $r \gg \xi$, the shape of the correlation function remains unchanged (i.e.~there is no explicit dependence on time), with its spatial range changing in time only through a dynamically-growing domain size $L_c(t)$, i.e.~through a dependence of the form $r/L_c(t)$; such an increase in the characteristic length scale 
%we highlight here that the typical domain size -- denoted here by $L_c$ -- is itself growing in time: such an effect 
is consistent with the number of independent domains and defect number decreasing.

%While such a dynamical scaling hypothesis was previously considered in the case of the `cartoon' Kibble-Zurek scenario -- in which $\xi$ was assumed to remain frozen to its value at $\hat{\xi} = \xi(-\epsilon)$ -- we highlight here that the typical domain size -- denoted here by $L_c$ -- is itself growing in time: such an effect is consistent with the number of independent domains and defect number decreasing, i.e.~$L_c \rightarrow L_c(t)$.
%%(see also the preceeding Sec......).
The existence of a single (dynamical) dominant lengthscale in the system allows us to formulate the correlation function in the form~\cite{bray_theory_1994}
%and corresponding momentum distributions as
\be
C({\bf r},t) \sim F \left( \frac{ r}{L_c(t)} \right) 
\label{eq:C-scaling}
\ee
%\hspace{0.15cm} {\rm and} \hspace{0.15cm}
with the corresponding momentum distributions taking the form
\be
n(k,t) \sim \left[L_c(t)\right]^d G\left(k L_c(t)\right) \;.
\label{eq:n-scaling}
\ee
Here $F$ is an (unspecified) \npp{universal} function, and $G$ is its Fourier transform,
%In other words, there is no {\em explicit} dependence on time, 
with temporal variation fully accounted for by the change in scale.
%~\footnote{Note that this remains true even when looking at the different-time correlation functions, for which all evolution can in fact be directly parametrized in terms of $r/L_c(t)$ and $r/L_c(t')$.}.

To calculate how such a characteristic lengthscale varies for a non-conserved field (i.e.~in the presence of external dissipation), one can utilize energetic considerations, taking into account the system symmetries and the fact that the defects are not isolated (thus introducing a lengthscale beyond which a defect becomes screened by other defects).
By estimating a critical velocity for the growth of such domains $u = dL_c(t)/dt$ and integrating, one can obtain the functional form of $L_c(t)$, as explicitly discussed in Ref.~\cite{bray_theory_1994}. This is generally found to exhibit a power law scaling of the form $L_c(t) \sim t^\beta$, with the critical exponent $\beta$ related to the dynamical critical exponent $z$ through $\beta=1/z$.
We note that in the special case of point vortices in a purely 2D geometry the presence of external dissipation (e.g.~through coupling the system to an external bath) leads to well-known logarithmic corrections such that $L_c(t) \sim (t/{\rm log}(t))^{1/z}$.

The topic of phase-ordering has been extensively discussed in the literature (see, e.g.~\cite{Huse_Coarsening_PhysRevE.47.1525,rutenberg_energy-scaling_1995,damle_phaseordering,jelic2011quench,Nam_Coarsening_2012,Gasenzer_2_-NTFP,groszek_comaron_PRR}).
To make this point clear, I briefly discuss here recent extended numerical simulations~\cite{groszek_comaron_PRR} in a 2D box with periodic boundary conditions 
%The generic scaling of the correlation function 
in the context of both the conservative and the dissipative Gross-Pitaevskii equations, 
%was recently confirmed numerically in 2D (for which $g^{(1)}(r) \sim r^{-\alpha}$) 
 respectively via Eqs.~(\ref{gpe}), (\ref{spgpe}).
 
 In both cases, phase-ordering during the post-quench evolution takes place through the annihilation of vortex-antivortex pairs (white/black squares in temporal evolution snapshots shown in Fig.~\ref{fig:phaseordering}(a)-(c)).
 Spatial correlations at a general time $t$ are probed in terms of the equilibrium correlation function $g^{(1)}_{eq}( r,t)$ -- the latter obtained after very lengthy numerical propagation --  through the generic relation
%
%We have already seen that 2D systems are different to 3D, giving rise to equilibrium quasi-condensates exhibiting an algebraically-decaying correlation function $g^{(1)}(r) \sim r^{-\alpha}$, due to the crucial role played by vortices.
%
%In the late-time dynamics, the system enters a self-similar regime, such that the (non-equilibrium) system correlation function, scaled to the equilibrium correlation function $g^{(1)}_{eq}(r) \sim r^{-\alpha}$ takes the general dynamical scaling form
\be
\frac{g^{(1)}({\bf r},t)}{g^{(1)}_{eq}({\bf r},t)} =  F\left( \frac{{\bf r}}{L_c(t)} \right) \;. \label{eq:scaled-g1}
\ee
Correlation functions in terms of the scaled time-dependent variable ${\bf r}/L_c(t)$ are found to collapse perfectly onto a single function [Fig.~\ref{fig:phaseordering}(d)] within a certain time window, which thus identifies both the evolutionary stage when the scaling hypothesis [Eq.~\ref{eq:C-scaling}] holds and the manner in which $L_c(t)$ grows, directly confirming an evolution of the form $L_c(t) \sim t^\beta$.
Correspondingly, directly counting the decay in the vortex number, one confirms a scaling for the vortex density of the form $n_V \sim 1/L_c^2 \sim t^{-2 \beta}$.
Identifying $\beta = 1/z$ allows the extraction of the dynamical critical exponent $z$ through these fits during the scaling interval. The behaviour shown in Fig.~\ref{fig:phaseordering} is generic, broadly applicable to both the conservative and the dissipative Gross-Pitaevskii equations.

%Here $L_c(t)$ denotes the time-dependent correlation length, % scaling as $L_c(t) \sim t^{1/z}$;  with the corresponding vortex density scaling as $n_V \sim 1/L_c^2(t)$ %$ \sim t^{-2/z}$ -- see Fig.~\ref{fig:phaseordering}.
%. The existence of such scalings, and thus of universal features, has been numerically confirmed via Eqs.~(\ref{gpe})-(\ref{spgpe}) over an extended late-time evolution window (see Fig.~\ref{fig:7}) [ ]. 

Nonetheless, a detailed numerical comparison of the predictions between the conserved and dissipative regimes leads to  slightly different behaviour and modifications in the precise value of the critical exponent: this can be traced back to the fundamental question of `isolated' (conservative) vs.~`open' (dissipative) quantum systems~\cite{Hohenberg-Halperin-RevModPhys.49.435}: in the purely dissipative limit (i.e.~Eq.~(\ref{spgpe}) without the 1 in $(1-i\gamma)$), the existence of vortices leads to logarithmic corrections to the system dynamics, such that $L_c(t) \sim (t/{\rm log}(t))^{1/z}$ with $z=2$ exactly, and such behaviour is explicitly reproduced in the simulations [see also subsequent discussion in the context of
instantaneously-quenched driven-dissipative exciton-polariton condensate systems (Sec.\ref{sec:excitonpolariton})]. However, in the purely conservative limit (and without including such logarithmic corrections), the value of $z$ was found to lie approximately 15-20\% below the $z=2$ value~\cite{groszek_comaron_PRR}. While a possible explanation for such a change in $z$ is likely to arise from a power-law vortex mobility due to vortex-sound interactions, a detailed understanding of such subtleties (as well as its extension to inhomogeneous systems) is still lacking.
%-- but see also
%but is beyond the scope of this chapter.
%We note that the existence of such self-similar dynamics (and the specific value of $z=2$ in the presence of logarithmic corrections) has also been explicitly confirmed in 

\subsection{Evolution around Non-Thermal Fixed Points} \label{sec:ntfp}

At this point, %To better understand the highly non-equilibrium features discussed in this section, 
the reader is briefly reminded (in rather basic terms) of the notion of `fixed points' in renormalization group theory -- for more details please refer to, e.g.~Refs.~\cite{nishimori-phasetransition-book,binney1992theory,RG-newphysics} {\em ... and ...... Chapter by Fradkin .....} .
In such a language one typically 
scales out microscopic details in favour of global features of macroscopic observables; as such, one deals with
 a parameter space of available many-body configurations of the system expressed in terms of a limiting set of relevant `coupling' parameters: long-time evolution between possible many-body system configurations is then modelled through flowing trajectories within such space, in a manner governed by the system Hamiltonian. Such a picture can be characterized by a small number of fixed points which define the nature of the system evolution for different configurations. Specifically,  flow around (towards/away from) such points qualitatively indicates the system evolution in parameter space. 
Naively, one can think of a fixed point as arising when a change of the scaling parameter, e.g.~$L_c(t)$, leaves the correlation function unchanged~\cite{binney1992theory}.
For example, for a system undergoing a single phase transition, one can directly identify certain fixed points: in the common case of a thermal phase transition, one identifies a low-temperature fixed point (denoting the system equilibrium at $T=0$, e.g.~in the fully-coherent state, or more generally in the $\epsilon \rightarrow +\infty$ limit) and a high-temperature fixed point (denoting the system in the $T \rightarrow \infty$, or $\epsilon \rightarrow - \infty$ limit): ultimately
%as $t \rightarrow \infty$, 
the system flows to either of those, with all equilibria below/above the critical point formally mapping onto the former (condensed), or latter (thermal) by an appropriate rescaling of relevant parameters. Such trajectories are separated by a critical line/surface, which contains a `mixed' critical fixed point, such that it is attractive along all the directions within the critical surface, and repulsive along those pointing away from the critical surface~\cite{binney1992theory}. Note that exactly at the critical point, the system exhibits fluctuations at all lengthscales, making it impossible to eliminate (as argued above) short-scale behaviour through a sequence of transformations, thus implying that a critical point is a fixed point.
More generally, a fixed point is characterized by a pure scaling form and a coupling which no longer changes under a rescaling.
The previously discussed phase-ordering dynamics is a simple example of a scaling phenomenon controlled by a (strong coupling) fixed point~\cite{bray_theory_1994}.

In general, various different classes of fixed points may exist. In particular, some attention is currently being devoted to the context of so-called %one can also identify another class of fixed points, called 
`non-thermal' fixed points, which 
arise as a fixed point in a typical far-from-equilibrium scenario: as such, they 
can be thought of as non-equilibrium configurations exhibiting scaling in both space and time, in terms of a limited number of properties.
These were originally introduced by Berges and coworkers in the context of reheating after early-universe inflation~\npp{\cite{Berges:2008wm,Berges:2008sr,Orioli:2015dxa,10.1093/acprof:oso/9780198768166.003.0002}}, and have been increasingly investigated across different physical systems, including heavy-ion collisions~\npp{\cite{Berges_Ion_PhysRevLett.114.061601,RevModPhys.93.035003}} and quantum gases~\cite{NTFP_2011_PhysRevB.84.020506,NTFP_2012_PhysRevA.85.043627,Nowak_2014,Gasenzer_2_-NTFP,chantesana_kinetic_2019,schmied_non-thermal_2019}, also finding applications in a range of experiments~\cite{NTFP_Oberthaler_2018,NTFP_Schmiedmayer_2018,NTFP_Hadzibabic_2021,Bagnato_Universal_2022,NTFP_Bagnato_Review}. The emergence of such points builds upon the concepts of self-similarity and critical behaviour, generalizing them to far from equilibrium dynamics. Under appropriate conditions, a system may spend a significant amount of time around such a non-thermal fixed point. 
The discussion below is largely based on recent reviews~\cite{chantesana_kinetic_2019,schmied_non-thermal_2019,NTFP_Bagnato_Review}.

%-------------------------------------
\begin{figure}[t]
\centering
\includegraphics[width=1.0\linewidth]{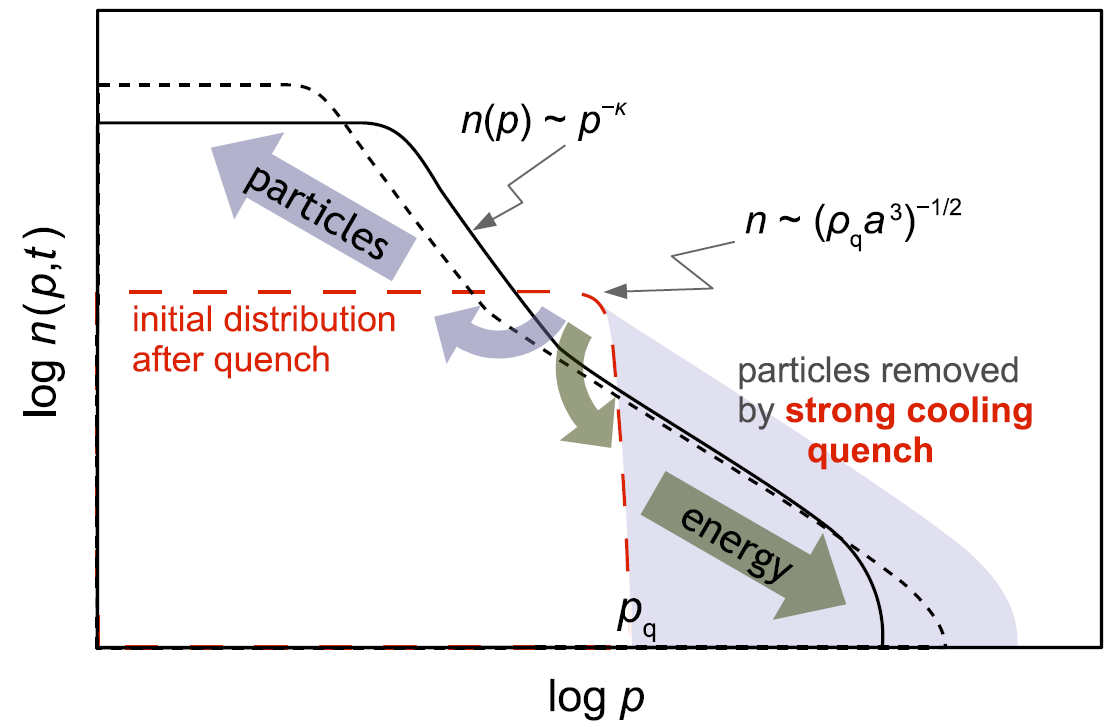}%{Fig-12-FINAL.png}
\caption{
Schematic of the bidirectional flow exhibited by the spectrum $n(k,t)$ vs.~$k$ [shown here in terms of $p = \hbar k$] around a non-thermal fixed point: A highly non-equilibrium initial distribution with equal population across all modes up to a cutoff, and random phases -- shown here by the dashed red lines -- evolves in a manner such that particles flow in an inverse manner towards lower $k$ (through conservation of particle number: $\alpha, \, \beta > 0$), while energy flows towards higher $k$ ($\alpha, \, \beta < 0$), carried by few particles in a manner which conserves total system energy.
Reprinted with permission from I.~Chantesana {\em et al.}, Kinetic theory of non-thermal fixed points in a Bose gas, Phys.~Rev.~A, {\bf 99}, 043620~\cite{chantesana_kinetic_2019}. Copyright (2019) by the American Physical Society.
}
\label{fig:7}
\end{figure}
%-------------------------------------

In a nutshell, the dynamics of a system in the vicinity of a non-thermal fixed point acquires a generalized evolution according to
\be
n(k,t) = \left( \frac{t}{t_0} \right)^\alpha \,\, n \left( \left( \frac{t}{t_0}\right)^\beta k, \,\, t_0 \right) \;,
\label{ntfp}
\ee
where $\alpha$ and $\beta$ are two scaling parameters which are, in general, distinct and which, combined, identify the universality class of the system.
The relation between $\alpha$ and $\beta$ is in fact set by the relevant conservation laws of the system. 
Essentially, these define -- within a broad momentum range -- how particles are redistributed in the self-similar regimes, and relate the scaling exponents $\alpha$ and $\beta$.
In the simplest case of particle number conservation (and for a particle with quadratic dispersion relation), this leads to $\alpha = \beta d$ (in the long-wavelength, or infrared, regime), where $d$ is the system dimensionality. Further noting that the characteristic lengthscale $L_c(t) \approx t^{\beta}$, and $\beta = 1/z$ practically reduces such expression to that of Eq.~(\ref{eq:n-scaling}) governing the self-similar phase-ordering kinetics previously discussed: as such the preceeding discussion of Sec.~\ref{sec:scalinghypothesis} can be seen as a special case of non-thermal fixed points under total particle number conservation.

However, the concept of non-thermal fixed points is more general. In particular, I note that the scaling function [Eq.~(\ref{ntfp})] can behave differently in different momentum limits: namely the previously-discussed `inverse' particle transfer towards lower momenta arises from the conservation of particle number in the infrared (low-momentum) limit.
Scalings due to energy conservation instead affect the ultraviolet (high-momentum) limit, leading to energy transport towards higher momenta and $\alpha = (d+z) \beta$.
As a result, evolution around a non-thermal fixed point leads to a bi-directional flow (see Fig.~\ref{fig:7}).
%Such a picture was recently confirmed in experiments in an isolated quench-cooled atomic gas~\cite{NTFP_Hadzibabic_2021} (see Fig.~\ref{fig:bidirectional}).

As a highly non-equilibrium feature, it is not surprising that
the potential emergence of a non-thermal fixed point following a thermal quench would require a very specific engineered initial set of conditions.
In this context I note the seminal numerical work of Berloff and Svistunov~\cite{Berloff2002a}. They considered a quench in a 3D box, starting with a strongly non-equilibrium initial condition\npp{~\cite{svis91,kagan-qc-growth,Kagan:1994,kaga97}} $\Phi({\bf r},t=0) = \sum_{k} a_k e^{i {\bf k} \cdot {\bf r}}$, for which the magnitudes of the complex amplitude $a_k$ were obtained from the self-similar solution of the (wave/semiclassical) Boltzmann equation describing the system dynamics sufficiently before the critical phase transition region, and their phases were uniformly distributed between $0$ and $2 \pi$.
Evolving by means of the classical field Eq.~(\ref{gpe}), they observed, as would be expected, particle build-up towards lower momenta, with most particles restricted to modes below a certain cutoff: this is a direct indication of the gradual build-up of a quasicondensate, arising from the macroscopic occupation of, and competition between, numerous low-momentum modes. Significantly however, filtering out higher momentum modes revealed a tangle of vortices, a characteristic of a `strongly' turbulent state.

As the governing equation [Eq.~(\ref{gpe})] is conservative, this tangle was found to gradually decay, following the relaxation process of superfluid turbulence in a non-driven system through the generation of sound waves and shrinking/annihilation of defects via a typical energy cascade.
Similar features, revealing also information about the vortex structure and its decay products, were found when doing a limited time averaging, thus connecting to the previous discussion on the application of short-time averaging during evolutionary dynamics [Sec.~\ref{sec:PO}].
We also note here that probing the state of an initially non-equilibrium system after very long classical field evolution in a separate numerical work, revealed the existence of vortex tangles, with the total vortex linelength increasing significantly as the system energy, and thus temperature, were increased~\cite{davis_burnett_2002}.
Considering a similar scenario, Nowak {\em et al.}~\cite{Nowak_2014} studied the system evolution following a sudden thermal quench by removing high-momentum particles from the distribution:
When a relatively small amount of energy was removed through the quench, they also found the system to evolve towards a coherent condensate, with clear evidence of a tangle of vortices after filtering short-wavelength fluctuations.
However, when a stronger quench was performed by instantaneously removing all the particles above some momentum cutoff value, they found a long-standing strong wave turbulence inverse cascade towards lower energies, with a  rather distinct power law power spectrum $n(k) \sim k^{-5}$: such behaviour, which is consistent with so-called `Porod' high-momentum tails $n(k) \sim 1/k^{d+2}$~\cite{schmied_non-thermal_2019,bray_theory_1994} corresponds to the case of randomly-distributed vortices at momenta exceeding the inverse lengthscale corresponding to the distance between defects, a feature which is a direct consequence of the 
\npp{ velocity field (${\bf v} \sim (1/r) \, \hat{{\bf \phi}}$) }
around a vortex line.
%dominated by incompressible excitations.
Nonetheless, excitations in the \npp{high-momentum} part of the spectrum still revealed $n(k) \sim k^{-2}$, leading to a bimodal spectrum.
Importantly, \npp{the above-discussed} features were now found to become observable without the need to filter out short-wavelength fluctuations.
Ultimately, the overall momentum distribution takes the form $n(k) \sim k^{-2}$, leading to the relaxation of the system -- but this now happens on a much longer timescale.
This was considered as evidence of the system spending significant time around a non-thermal fixed point.

This example clearly shows the distinct features that a thermal phase transition can exhibit depending on the nature and details of the strength of the cooling quench, i.e.~in this case on the initial non-equilibrium conditions for the classical field evolution.
Further evidence of the emergence and decay of the arising vortex tangle and properties of the order parameter were also discussed in Refs.~\cite{Cugliandolo_Kobayashi_EPL,Cugliandolo_Kobayashi_Long,parker_barenghi_quenched}.

%-------------------------------------
\begin{figure}[t!]
\centering
\includegraphics[width=0.8\linewidth]{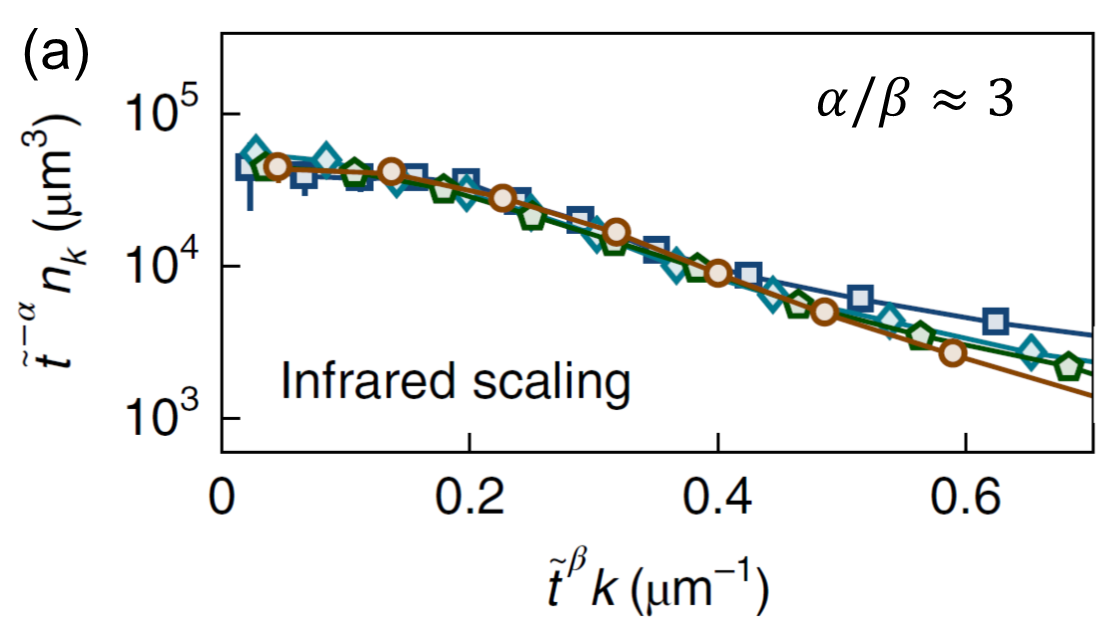}\\%{Fig-13a-neww.png}\\
\includegraphics[width=0.8\linewidth]{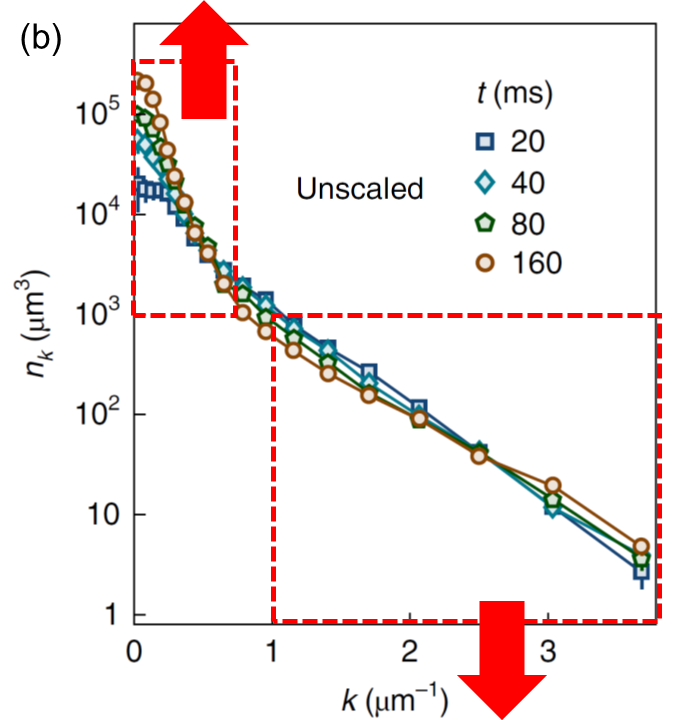}\\%{Fig-13b-new.png}\\
\includegraphics[width=0.8\linewidth]{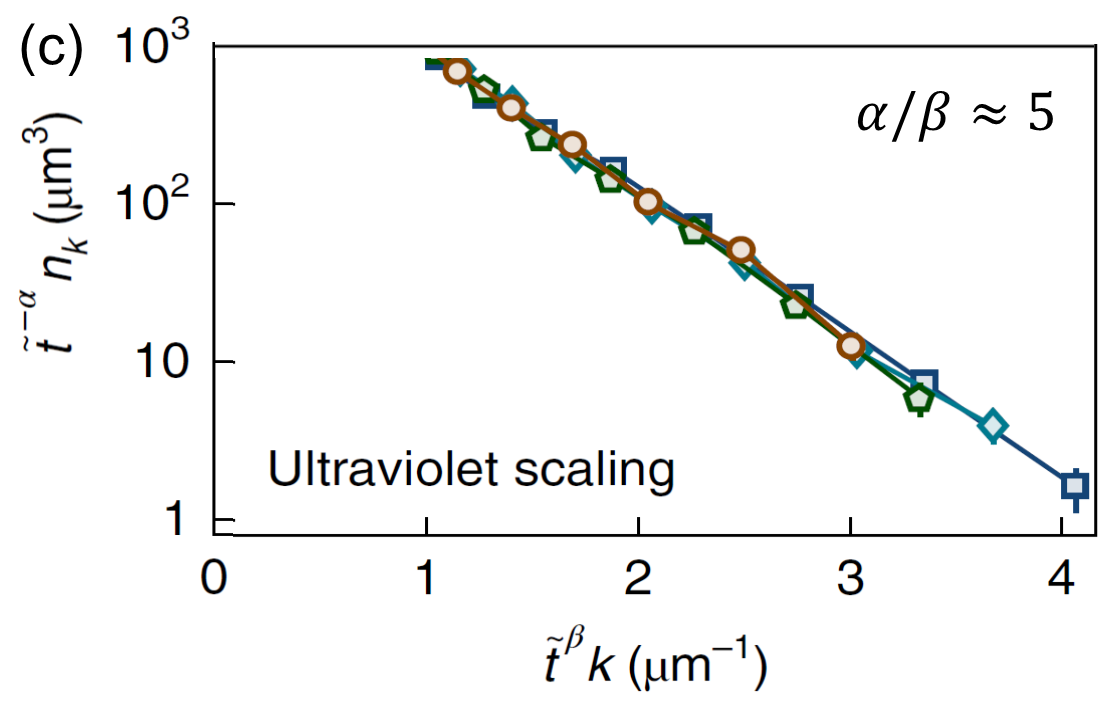}%{Fig-13c-FINAL.png}
\caption{
Experimental demonstration of bidirectional flow according to Eq.~(\ref{ntfp}) in an isolated $d=3$ ultracold atomic gas in a box quenched from just above the critical temperature to a highly non-equilibrium situation by removing significant fractions of atoms and energy, such that the system ultimately relaxes to a state with an $\approx 40\%$ BEC fraction: The unscaled momentum distribution shown in (b) at different evolution times following the non-equilibrium quench is scaled differently in the limiting (a) infrared case (for which one finds $\alpha,\,\beta>0$, with $\alpha/\beta \approx 3$, consistent with the condition $\alpha/\beta=d$ for particle-conserving transport), and (c) ultraviolet case (for which one finds $\alpha /\beta \approx 5$ with $\alpha,\,\beta <0$ consistent with the prediction $\alpha/\beta=d+2$ for energy-conserving transport).
Adapted with permission from 
J.~A.~P.~Glidden {\em et al.}, Bidirectional dynamic scaling in an isolated Bose gas far from equilibrium, Nat.~Phys.~{\bf 17}, 457~\cite{NTFP_Hadzibabic_2021}. Copyright (2021) by the Nature Publishing Group.
}
\label{fig:bidirectional}
\end{figure}
%-------------------------------------

Dynamical properties of evolution consistent with Eq.~(\ref{ntfp})
was recently observed in experiments in an isolated quench-cooled atomic gas~\cite{NTFP_Hadzibabic_2021}.
By removing a significant fraction of the atoms and energy of a system initially just above the critical temperature and doing this process with interactions switched off, the experiment was able to generate a highly-non-equilibrium state, and observe its relaxational dynamics, thus verifying the expected distinct scaling relations across the ultraviolet and infrared limits, corresponding to different conservation laws~\cite{NTFP_Hadzibabic_2021} (see Fig.~\ref{fig:bidirectional}).
Moreover, this experiment was also able to confirm the emergence of a quasicondensate in the early evolution stages, with the coherence across the system gradually growing to span the whole system, in a manner consistent with the system equilibrium for the given atom number and total energy.

For completeness, I note here that
%such properties have in fact been recently experimentally observed in ultracold atoms in the context of .... 
manifestations of evolution consistent with Eq.~(\ref{ntfp}) have also been reported (prior to the above experiment) in 
a quasi-one-dimensional spin-1 BEC~\cite{NTFP_Oberthaler_2018}, and a single-component BEC~\cite{NTFP_Schmiedmayer_2018}, while also bearing close analogies to the inverse cascade evolutions observed in 2D experiments~\cite{2D_Onsager_Vortex_Monash,2D_Onsager_Vortex_UQ}.
%... and ... turbulence ... 
We refer the reader to these works and excellent recent reviews on non-thermal fixed points -- which also address in some detail the relation of non-thermal fixed points to weak-wave turbulence and strong turbulence~\cite{schmied_non-thermal_2019,chantesana_kinetic_2019} -- and their experimental study in ultracold quantum matter~\cite{NTFP_Bagnato_Review}.

\subsubsection{Relation of Quenched Turbulence Dynamics to `Controlled' Turbulence Experiments}

%For a more in-depth discussion of the relation between non-thermal fixed points, weak-wave turbulence and strong turbulence, and the relevance of such fixed points to other dynamical aspects, such as prethermalization, the reader is referred to the review articles~\cite{schmied_non-thermal_2019,chantesana_kinetic_2019}.
%
%Given the above references to and the broad interest in the topic of turbulence, we make some further brief remarks here:
In this chapter, I have focused on the emergence of `turbulent' features through the instantaneous (Sec.~\ref{sec:phaseordering-main}) or driven (Sec.~\ref{sec:kz-main}) cooling across a thermal phase transition, starting from an incoherent state, i.e.~quenching from $\epsilon<0$ to $\epsilon>0$, and studying the relaxation of the emerging non-equilibrium states generated, which typically include some type of vortex tangle.
As well-known~\cite{Barenghi_PhysRep_2016,Turbulence_PNAS_2014,Tsubota_Review_2017}, an open, dynamically-driven, system can reach a steady-state under continuous driving, subject to a dissipation mechanism at a different scale from that where energy is being injected, thus acquiring a momentum spectrum exhibiting universal features in a certain `inertial' range. On the other hand, 
a closed turbulent non-equilibrium system will gradually see its vorticity decay in a manner which can also display universal scaling laws: for completeness, I briefly consider both these cases below.

Firstly, if the system is left isolated shortly after its quench into
 such a turbulent state, one could also compare its subsequent relaxation evolution
  to studies characterizing the decay of superfluid turbulence.
The topic of superfluid turbulence has a rich history~\cite{Barenghi_Turbulence_Book_2001,Vinen_Review_2002}, with an extensive body of literature on liquid helium supplemented by recent work on ultracold quantum gases which gives access to different lengthscales~\cite{Turbulence_PNAS_2014,Barenghi_PhysRep_2016,Tsubota_Review_2017}.
The  relaxation dynamics can be different, depending on whether the system is primarily a superfluid, or has a significant (potentially dominant) thermal fraction.

Current studies, based on experimental observations~\cite{walmsley_golov,bradley_helium} and detailed numerical simulations~\cite{baggaley_barenghi_2012,Baggaley_other} make a distinction between two types of superfluid turbulence at relatively low temperatures when a well-formed condensate exists (see also recent reviews~\cite{Turbulence_PNAS_2014,Barenghi_PhysRep_2016}).
The most common form of turbulence discussed is that of Kolmogorov, or `quasi-classical' turbulence, which can exhibit (under continuous injection of energy) a cascade of kinetic energy from large to small eddies, thus giving rise to the well-known classical Kolmogorov spectrum $E_k \sim k^{-5/3}$ for $k \ll 1/l$ (where $l$ is the typical intervortex spacing): this form of turbulence -- which is thus a property of the motions at lengthscales larger than the intervortex spacing -- is now understood to be associated with the existence of turbulent bundles of vortices. Once the injection of energy is discontinued, this leads to a total vortex linelength decay (characterizing the intensity of turbulence) which scales as $L_V \sim t^{-3/2}$.

On the other hand, some systems can exhibit a (potentially transient) turbulent state, somewhat similar to random flow, which is dominated by the presence of individual randomly-oriented tangled vortices (with most of the energy contained in the intermediate scales): in such a regime, known as `ultraquantum', or `Vinen' turbulence,  the vortex linelength decays instead according to $L_V \sim t^{-1}$.
%Both regimes have been observed in liquid helium experiments [ ], and in fact the ultraquantum regime can also emerge as a precursor state to the Kolmogorov one [ ].

Early controlled low-temperature turbulence experiments with ultracold atoms focussed on the generation of tangled vorticity through either shaking across two distinct axes~\cite{Bagnato_Turbulence_2009}, or intense periodic driving~\cite{Hadzibabic_Turbulence}, in both cases starting from a $T \approx 0$ pure condensate. 
As such dynamics are generated by initiating the dynamical quenching from the coherent side, it is clearly a very distinct excitation process to the thermal quenches starting from an incoherent system discussed in this Chapter.
Among other findings, this has led to the observation of the expected power-law spectrum $n(k) \sim k^{-\gamma}$ in a box geometry, 
%of the form $n(k) \sim k^{-\gamma}$, with a $\gamma$ value close the the predicted close to value for the direct cascade exponent of 
consistent with (Kolmogorov-Zakharov) weak-wave turbulence of compressible superfluids~\cite{Hadzibabic_Turbulence}; further characterization of the flux of such a cascade through dissipation, has enabled the probing of the propagation of the cascade front in momentum space at both
 short- and long-time evolution~\cite{Hadzibabic_Turbulence_2}; very recent experiments from the same group have also facilited a direct measurement of the equation of state in experiments.
The concept or reversibility was also probed in such quenches:
Experiments in a box revealed that when such a turbulent system was left undriven for a significant period, it would reform into a near-pure condensate consistent with the initial state, i.e.~the process was found to  be fully reversible~\cite{Hadzibabic_Turbulence}.

Nonetheless, I also note here a recent related experiment~\cite{Bagnato_Universal_2022} studying the evolution of $n(k,t)$ from a non-equilibrium initial state generated by the application of a misaligned sinusoidally-varying magnetic field on an initial highly-coherent harmonically-confined BEC: the excitation was performed in such a manner so as to perturb the system through dynamical rotations and distortions.
Rather interestingly, such an experiment observed a behaviour broadly consistent with the scaling relation of Eq.~(\ref{ntfp}), but in this case both $\alpha$ and $\beta$ exponents were found to be negative (beyond the expected $\alpha,\,\beta<0$ values in the ultraviolet regime) also in the infrared regime: this should be contrasted to the previously-discussed dynamics observed following a strong non-equilibrium quench in a box-like geometry~\cite{NTFP_Hadzibabic_2021} for which in the infrared regime both $\alpha,\,\beta >0$.
%(and related, as expected~\cite{schmied_non-thermal_2019}, through $\alpha = \beta D$).

Although understanding such issues is still an open matter, the ongoing experiments evidence the important role that an inhomogeneous geometry can play, and the richness of the dynamical features it can generate, which may have a non-negligible impact on the types of turbulence generated and their subsequent relaxation. Moreover, differences in the dynamical properties of turbulent states generated starting from a well-formed superfluid, or an incoherent thermal cloud also highlight the important role that the thermal cloud can play on such dynamics.

For a more in-depth discussion of the current understanding of the relation between non-thermal fixed points, weak-wave turbulence and strong turbulence, and the relevance of such fixed points to other dynamical aspects, such as prethermalization~\cite{Prethermalization-Berges:2004ce,Prethermalization-Gring2011a}, the reader is referred to the review articles~\cite{schmied_non-thermal_2019,chantesana_kinetic_2019}.

%(e.g.~equal mode populations up to some high momentum .... consistent with $\mu$ .... but random phases being imposer on the classical field) and an instantaneous cooling quench. This is somewhat reminiscent of the scenario studied in the pioneering numerical work of Berloff and Svistunov.... Gasenzer revisited .... open questions ... and there are significant links between the concepts of NTFP, weak-wave and strong turbulence and quasi-condensation.

\section{Selected Further Considerations} \label{sec:other}

\subsection{Beyond Continuous Single-Component Atomic Gases} \label{sec:other-atomic}

The properties of ultracold atomic gases can be controlled experimentally by a range of different tuneable parameters. In this Chapter, I have focused on the role of temperature in inducing a transition across criticality between two single-component thermalized states with very different characteristic properties.
Relevant comments have also been made regarding the role of geometry and dimensionality in single atomic species confined in continuous potentials, such as harmonic traps and box-like potentials.
The exquisite control offered in experiments with ultracold atoms has opened up the way for a plethora of other types of quenches -- with, no doubt, new physics yet to be explored in them, although significant progress towards various probes of universality  have already been performed.
While this Section cannot do justice to such exciting developments, for completeness I note in passing a few such areas of relevance below.

A transition across criticality can also be engineered at fixed temperature by changing the interaction energy, 
%which controls the location of the critical point through the dependence of the critical temperature on interactions.
%Moreover, a change in the interaction energy 
which can also control the type of emerging equilibrium phase transition: in this vein, Ref.~\cite{Smith:2011-prl-interactions} demonstrated the smooth crossover from an interaction-driven BKT transition to a BEC transition in the limit of vanishing interactions. Such issues are further discussed in Ref.~\cite{UBEC-Smith} {\em (... see also Chapter by Smith ....)}.
For example, the interaction energy can be tuned by changing either the atom number/density, or the effective strength $g$ of contact interactions appearing in the Gross-Pitaevskii equation -- which provides another exciting knob for probing universal physics.

Moreover, one can also control the dominant type and range of interactions in dipolar atomic condensates, which offer tuneability in the relative strength of local and long-range interactions~\cite{Dipolar-Review-2023}.
Furthermore, degenerate Fermi gas settings provide an added `dimension' to such topics through control of the interaction strength, as they encompass different physical regimes depending on the value of $(k_F a)^{-1}$, where $k_F$ is the Fermi wavevector and $a$ the s-wave scattering length. In particular, such systems facilitate a study of the BEC limit (in the sense of atoms pairing up to form closely-bound molecules which then Bose-condense), the BCS limit (where the formation of Cooper-pairs of atoms in momentum space leads to familiar BCS-like effects), and -- even more interestingly -- the unitarity regime  between these two, exhibiting infinite-range interactions. The physics of the unitarity limit is however also accessible in Bose gases.
Moreover, the co-existence of mixtures of quantum-degenerate gases and components with a spin degree of freedom (or, so-called, spinor gases) enable such studies to be generalized to quantum mixtures.

In the vein of the main topics covered in this Chapter, I note a significant (but not exhaustive) body of literature on Kibble-Zurek-type quenches, including, for example, through interaction-induced quenches in mixtures~\cite{KZ-Mixture-Sabbatini2011} and spinor condensates~\cite{KZ-Spinor-StamperKurn,KZ-Spinor-saito2007kibble,KZ-Spinor-Swistock2013,KZ-Spinor-Witkowska2013,KZ-Spinor-williamson_universal_2016,QKZ-Spinor-Chapman-PhysRevLett.116.155301,KZ-Spinor-Gerbier}, prethermalization quenches in 1D~\cite{UBEC-Schmiedmayer,Prethermalization-Gring2011a}, superfluid quenches across unitarity~\cite{Hadzibabic-Bose-Unitary-Quench-Nature,Hadzibabic-Bose-Unitarity-Quench-PhysRevLett.119.250404}, and quantum quenches~\cite{dziarmaga2010dynamics,QKZ-RevModPhys.83.863}  in periodic potentials (optical lattices)~\cite{Greiner:2002a,KZ-deMarco-1,QKZ-Schneider,QKZ-lattice-chin-doi:10.1126/science.aaf9657} which could facilitate applications in quantum simulations~\cite{bloch_quantum_2012,dziarmaga-quantum-computer-defects,QKZ-Dwave-PhysRevResearch.2.033369}, or in the presence of disorder~\cite{quench-deMarco2}, with open realms for studying Kibble-Zurek physics in systems such as dipolar gases and photon condensates.
An important feature of quantum quenches is that they are performed  under conditions of negligible thermal dissipation, thus potentially facilitating  better control and cleaner scalings -- while allowing for the possibility of the emergence of new interesting physical regimes.

Clearly these, and other settings not mentioned here, demonstrate the power of controlled quantum gas experiments with ultracold atomic gases. Nonetheless, new physics corresponding to the dynamics in an open quantum system can be probed in another experimentally-well-controlled quantum gas platform, namely that of exciton-polariton systems, which I discuss next.

\subsection{Driven-Dissipative Condensation in Exciton-Polariton Systems} \label{sec:excitonpolariton}

A somewhat distinct physical system, that of exciton-polariton condensates {\em (... see also Chapter on quantum fluids of light ....)} can be studied by 
embedding a 2D quantum well in a semiconductor microcavity  driven by external pumping (laser)~\cite{deng2010exciton}. The presence of the cavity allows the emergence of a new type of quasiparticle which is a mixture of a photon and a matter excitation: under sufficiently strong pumping, this leads to  `hybridization' and a new dressed-state bosonic quasiparticle whose spectrum facilitates the emergence of a quasi-ordered state with a high degree of coherence~\cite{kasprzak2006bose,carusotto2013quantum,PolaritonReview-Nature2022}.
%macroscopic quantum coherence [ ]. 
As this is a strictly 2D system, technically one should not speak of `condensation', but rather of a `BKT'-type phase transition to a `quasi-condensate' state exhibiting quasi-long-range order, as discussed in more detail below. Experimentally, there are various distinct pumping schemes which can be used to generate such an exciton-polariton system: one of these used rather commonly is that of incoherent pumping, which allows the quasi-ordered state to form spontaneously, through relaxation in the hybridized energy spectrum (relaxation in the so-called `lower polariton' branch).

%-------------------------------------
\begin{figure*}[t!]
\centering
\includegraphics[width=0.515\linewidth]{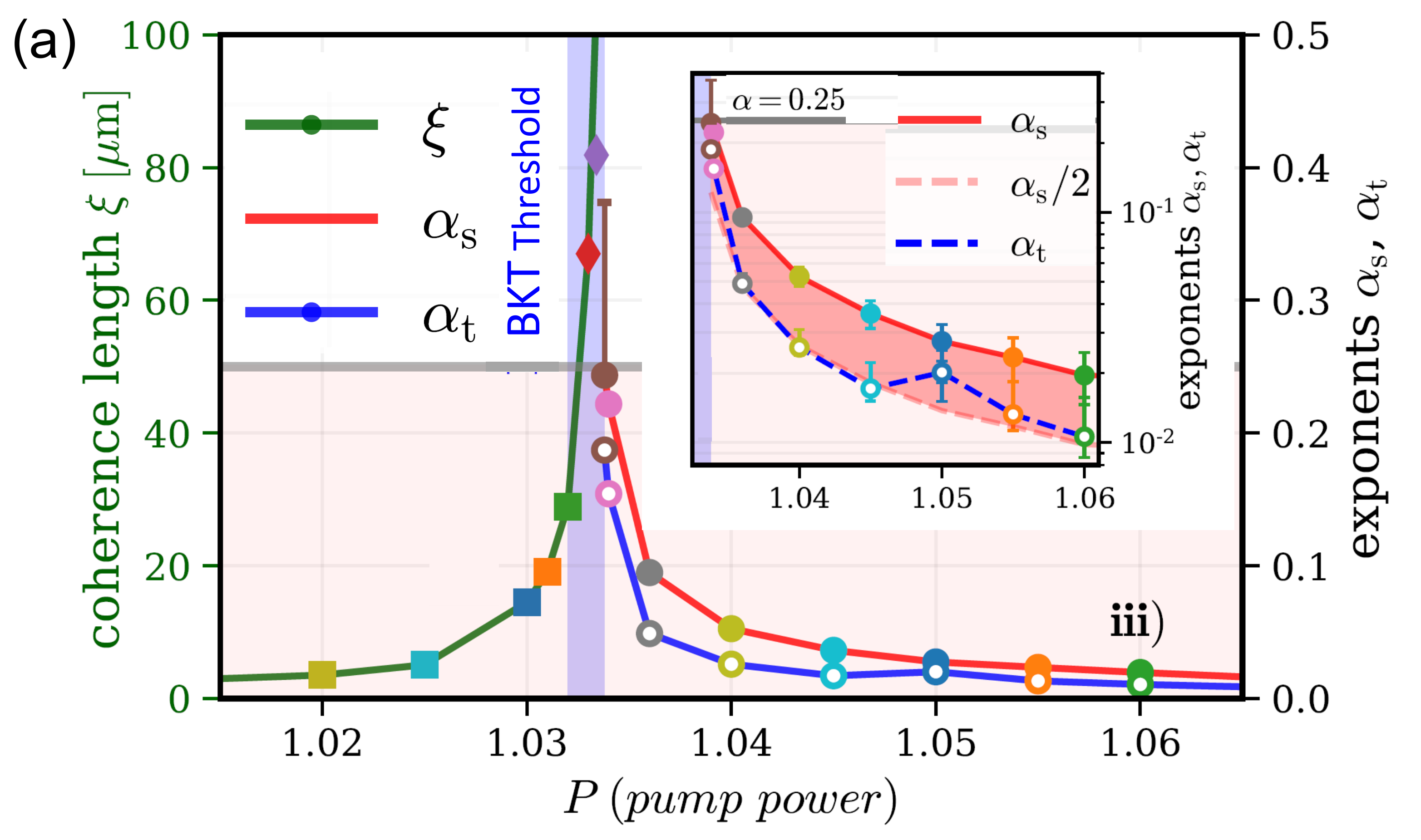}%{Fig-polaritons-a.pdf}
\includegraphics[width=0.495\linewidth]{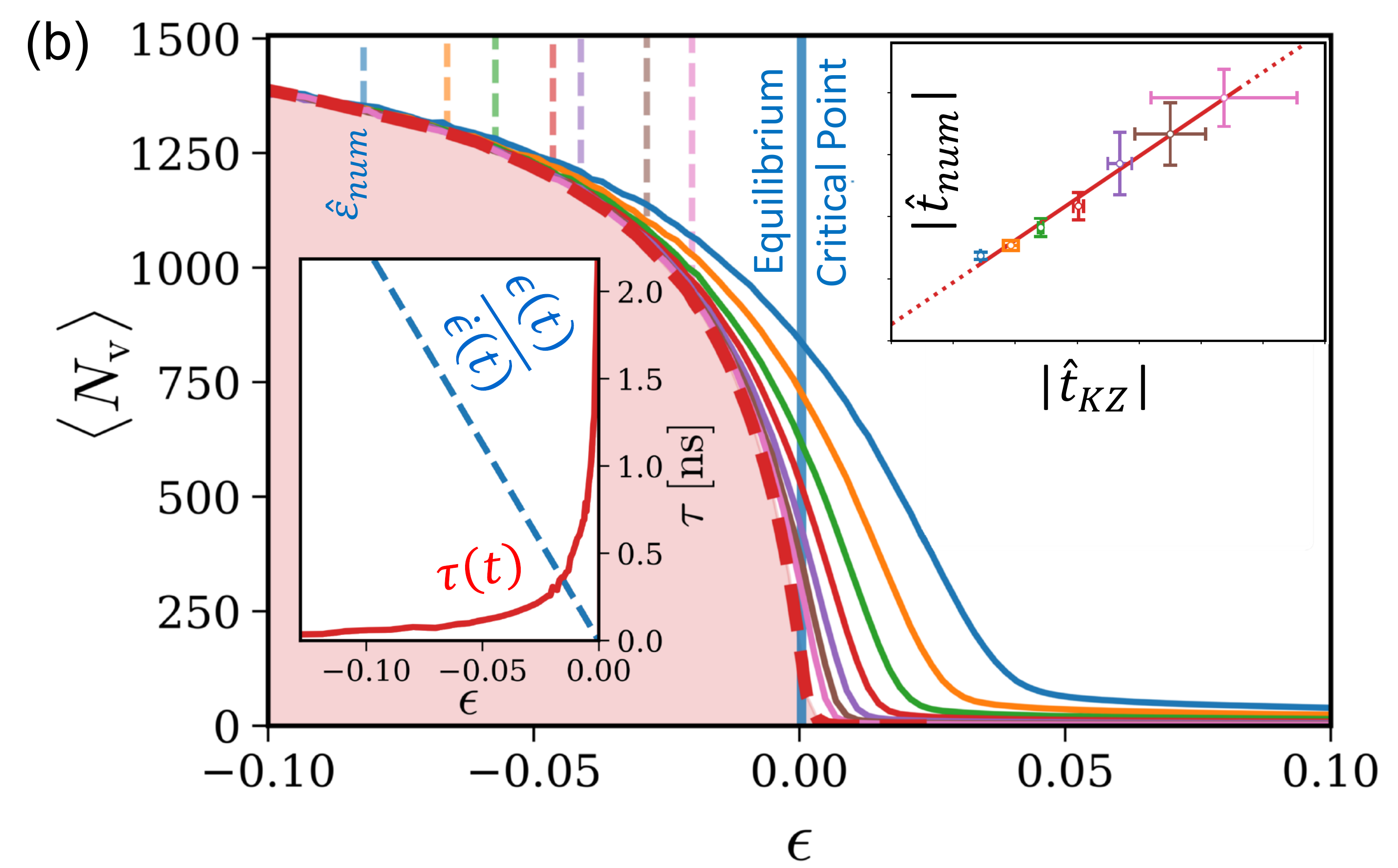}\\%{Fig-polaritons-b.pdf}\\
\includegraphics[width=0.55\linewidth]{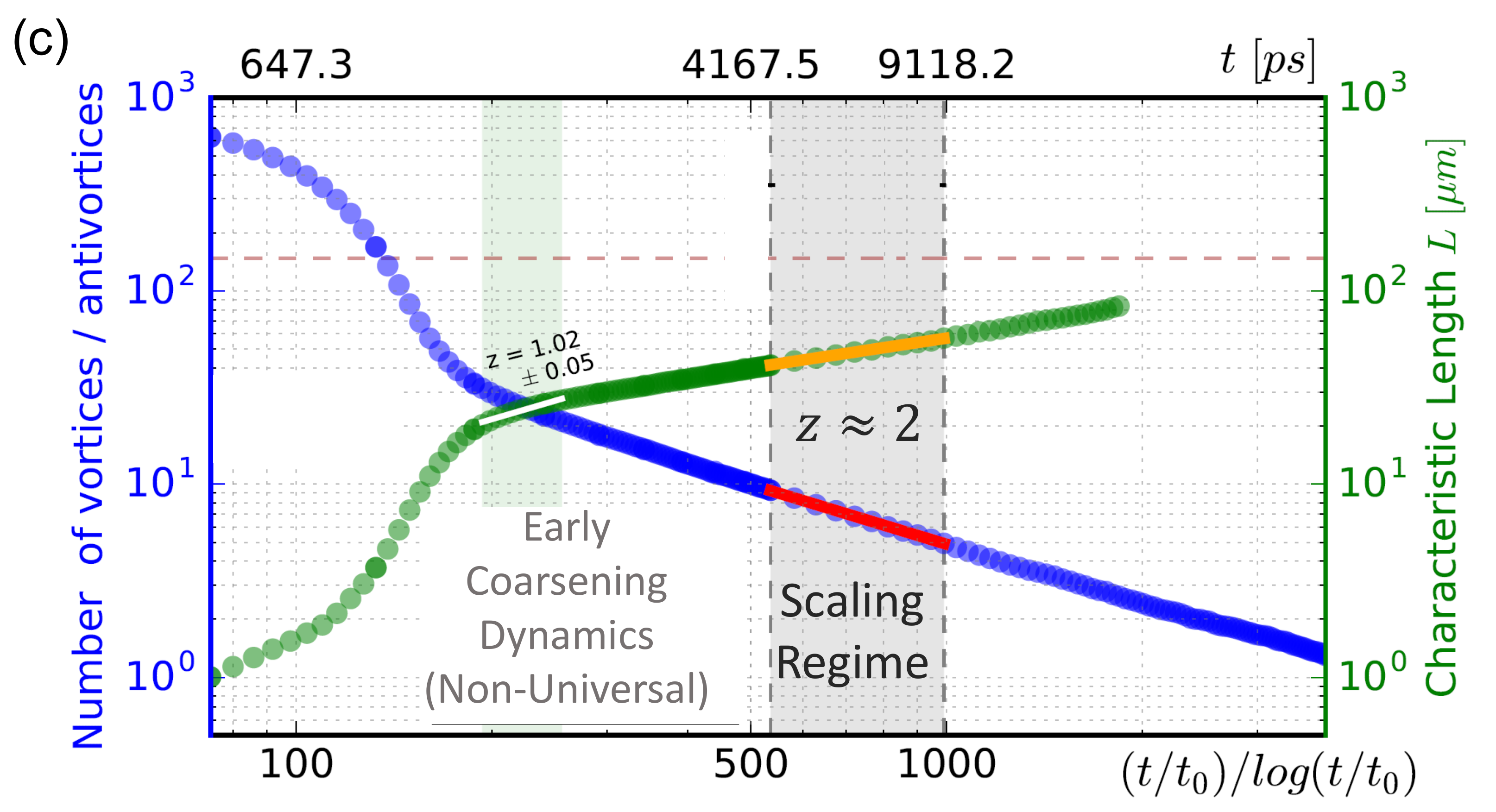}\\%{Fig-polaritons-c.pdf}\\
\caption{
Universal  features of a non-equilibrium driven-dissipative exciton-polariton condensate for typical experimental parameters.
(a) Dependence of the coherence length, $\xi$, defined by $g^{(1)}({\bf r}) \sim e^{-r/\xi}$ on the incoherent side of the phase transition, and of the spatial (temporal) exponents $\alpha_s$ ($\alpha_t$) defined on the quasi-ordered side by $g^{(1)}({\bf r}) = r ^{-\alpha_s}$ ($g^{(1)}(t) = t ^{-\alpha_t}$).
Horizontal axis depicts the external pump power in units of the critical value predicted by mean-field theory, with the transition occurring above the critical pump threshold (here $P_{\rm cr}\sim 1.033$): fluctuations clearly shift this by (up to) a few \%, towards a higher pump value.
Inset: zoomed-in version clearly revealing $\alpha_s = 2 \alpha_t$, as direct evidence of the non-equilibrium nature of the phase transition in such parameter regime.
Adapted with permission from P.~Comaron {\em et al.}, Non-equilibrium Berezisnkii-Kosterlitz-Thouless transition in driven-dissipative condensates, Europhys.~Lett.~{\bf 133}, 17002~\cite{Comaron_2021}. Copyright (2021) by the Institute of Physics.
(b) Demonstration of validity of the Kibble-Zurek relation [Eq.~(\ref{eq:kz})]. Main panel shows the dependence of vortex number on $\epsilon = (P-P_{c})/P_{c}$ for the equilibrium case (dashed red line) and the corresponding  dependence for $\langle N_v(t) \rangle$ in terms of $\epsilon(t)$ for a linear external quench in the pump via $\epsilon(t) = t/\tau_Q$, with $\tau_Q$ decreasing between curves as one moves from left to right.
For each $\tau_Q$, there is a corresponding value of $\epsilon=\epsilon(\hat{t}_{num})$ at which the (dynamical) curve determining the post-quench number of vortices (solid) as a function of $\epsilon$ deviates from the corresponding equilibrium one (dashed red) -- these values are shown by the dashed vertical lines, and yield the numerical estimate $\hat{t}_{num}$.
Insets: (bottom left) Relation between diverging relaxation time $\tau$ and characteristic timescale $\epsilon(t)/\dot{\epsilon}(t)$ in the vicinity of the critical point, with their intersection defining the freeze-out time $-\hat{t}_{KZ}$.
(top right) Demonstration of $|\hat{t}_{num}| \sim |\hat{t}_{KZ}|$ validating the expected Kibble-Zurek proportionality relation (up to some microscopic, non-universal prefactor).
Adapted with permission from A. Zamora {\em et al.}, Kibble-Zurek mechanism in driven-dissipative systems crossing a non-equilibrium phase transition, Phys.~Rev.~Lett. {\bf 125}, 095301~\cite{Comaron-KZ-2020}. Copyright (2020) by the American Physical Society.
(c) Evolution of decreasing number of vortices/antivortices (blue curve) and growing characteristic lengthscale (green curve) highlighting the dramatic early post-quench coarsening evolution, and their corresponding dependence on time in the scaling regime, allowing us to extract the dynamical critical exponent $z=2$. Note that all axes are plotted on logarithmic scale, with the temporal axis also including the anticipated logarithmic correction to the system evolution.
Adapted with permission from P.~Comaron {\em et al.}, Dynamical critical exponents in driven-dissipative quantum systems, Phys.~Rev.~Lett. {\bf 121}, 095302~\cite{comaron2018dynamical}. Copyright (2018) by the American Physical Society.
}
\label{fig:polariton-nick}
\end{figure*}
%-------------------------------------

Under the assumption that the exciton bath can be adiabatically eliminated -- which is valid in many experimental set-ups -- the fundamental equation describing such a scenario can take the form of a generalized Gross-Pitaevskii equation already alluded to in Sec.~\ref{sec:modelling}~\cite{polariton-gpe-wouters,polariton-gpe-berloff}.
When further adding stochastic noise to this
%%%%%%%%%%%%%%%%%%%%%%%%%%%%%%%%%%%
(and in the usual convention of setting $\hbar=1$), this takes the form of a stochastic complex Ginzburg-Landau equation~\cite{chiocchetta2013non,carusotto2013quantum,wouters-kpz-1-PhysRevA.90.023615,comaron2018dynamical}:
\be
i \frac{\partial \Phi}{\partial t} = - \frac{\nabla^2}{2 m } +
g|\Phi|^2 + \frac{i }{2}
\left( \frac{P}{1+|\Phi|^2/n_\text{s}} -\gamma \right)   
%%%%%%%%%%5%+\frac{1}{2}\frac{P}{\Omega}\frac{\partial}{\partial t} \bigg]
\Phi +  dW
\label{eq:pol}
\ee
where $m$ is the polariton mass (which is many orders of magnitude smaller than the electron mass), $P$ is the strength of the homogeneous external drive, $g$ is the polariton-polariton interaction strength, $n_\text{s}$ is a (phenomenological) saturation density, and $dW$ a complex-valued noise term satisfying $\langle
dW^{*}({\bf r},t) dW ({\bf r'},t) \rangle
=[(P+\gamma)/2] \delta_{{\bf r},{\bf r'}}$, where $\gamma$ here is the inverse of the polariton lifetime. I note here that, despite the decaying nature of the quasiparticle, the system does nonetheless enter a non-equilibrium steady-state (provided external pumping is maintained).
Different variants of this equation exist in the literature, including extensions which account for a frequency-selective pumping mechanism favouring relaxation to low-energy modes~\cite{chiocchetta2013non,woutersliew2010,comaron2018dynamical}.
%
%In the present study, we use typical experimental parameters {\cite{nitsche2014algebraic}}: lifetime $\tau = 1/\gamma= 4.5 \mathrm{ps}$, $m = 6.2 \ 10^{-5} \ m_\text{e}$, $g = 6.82 \ 10^{-3} \ \mathrm{meV \mu m^2} $, {$\Omega = 11.09 \mathrm{ps^{-1}}$} and $n_\text{s}= 1500 \mathrm{\mu m^{-2}}$. 

For such systems, %(which are in fact at room temperature), 
the control parameter is not temperature, but pumping: in fact the quasi-coherent regime sets in when pumping {\rm exceeds} a threshold `critical pumping' value: in other words $\epsilon = (P-P_{c})/P_{c}$ such that $\epsilon>0$ occurs in the high pumping limit $P>P_{c}$ (recall that in the case of cold quantum matter $\epsilon>0$ labelling the coherent side is a low-temperature case $T<T_c$).
In the mean-field limit, when $\epsilon > 0$, Eq.~(\ref{eq:pol}) supports a non-trivial steady-state solution  in the form of $|\Phi|^2 = n_s (P/\gamma -1)$. 

Based on the previous discussion, one would expect a BKT transition associated with vortex-antivortex binding with increasing pumping strength; however the driven-dissipative nature of the system can modify such picture of `equilibrium condensation' previously discussed in the context of ultracold atomic gases~\footnote{For the sake of clarity, I note here that while ultracold atomic condensates can be thought of as equilibrium, i.e.~fully-relaxed, during their probed long timescale, the true thermodynamic equilibrium of such systems is in fact a solid, facilitated by three-body loss processes which are not usually dominant during experimentally-probed timescales in most cold atom settings, thus enabling `long' lifetimes of up to a few minutes.}, and this has led to significant discussion in the community about the nature of the phase transition~\cite{polariton-bkt-altman,polariton-bkt-szymanska,UBEC-Keeling}.

In effect, the extent that the phase transition from disordered to (quasi-)ordered state is an `equilibrium' phase transition or not -- and the nature of the arising correlation functions on the quasi-ordered state -- depend on the system parameters, which can be controlled by the quality of the sample on which experiments are performed (i.e.~the polariton lifetime), but also by the spatial pump profile, details of the pumping scheme, and any engineered anisotropy in the effective polariton mass. Depending on such parameters, these systems can also exhibit a broader range of universal behaviour, as discussed below.

We start with revisiting the universal dynamics during a gradual driven transition (Kibble-Zurek) and the late-time phase-ordering for parameters corresponding to recent exciton-polariton experiments~\cite{nitsche2014algebraic}. The equilibrium criticality phase diagram can be characterized by looking at the divergence of the characteristic coherence length, $\xi$ on the incoherent (here below threshold) pumping side $\epsilon <0$ and the corresponding behaviour of the exponent $\alpha_s$ of the spatial coherence power law decay on the $\epsilon >0$ (quasi)ordered side [Eq.~(\ref{eq:g1D}) with $\alpha(T) \rightarrow \alpha_s$]. Such behaviour is shown in Fig.~\ref{fig:polariton-nick}(a) as a function of the ratio of the pumping power to the corresponding critical value predicted by mean-field theory (respectively by green and red lines)~\cite{Comaron_2021}. 
Firstly, it should be noted here that due to the inclusion of fluctuations (via the noise term) the transition is in fact shifted from the predicted mean-field value by a few \%.
It is instructive to also consider the decay of the temporal correlations as one approaches the critical region from the $\epsilon >0$ (quasi-)ordered side: these are shown by the red line. The diverging behavior on the (quasi)ordered side is the same qualitatively for spatial and temporal correlations. However, it clearly becomes evident (see also inset) that $\alpha_s = 2 \alpha_t$~\cite{Comaron_2021}, with such a relation between their respective exponents giving the transition a genuine {\em non-equilibrium} flavour~\cite{Szymanska-noneqm-PhysRevLett.96.230602,Szymanska-noneqm-PhysRevB.75.195331}. Moreover, various experiments observed the universal scaling exponent characteristic of the spatial correlation function decay on the (quasi)ordered side to exceed the expected {\em equilibrium} value of (1/4) at the critical point~\cite{roumpos2012power,nitsche2014algebraic}: although this has also been seen in numerical simulations~\cite{polariton-bkt-szymanska},
the results presented here -- while broadly consistent with such a picture -- cannot provide a conclusive answer, because the precise value is also expected to be somewhat sensitive to finite-size effects~\cite{Comaron_2021}.
%nonetheless we stress that other simulations have conclusively shown $\alpha \gg 1/4$ in the critical region......
%

%

Having identified the precise location of the critical point, one can now
consider the dependence of the system properties as a function of the distance to criticality $\epsilon$, the first step towards performing and characterizing
 linear quenches across the phase transition. As expected the number of vortices in this 2D system at equilibrium decreases very dramatically in the vicinity of $\epsilon=0$ (dashed red line in Fig.~\ref{fig:polariton-nick}(b)). Quenching across the transition at different rates leads to such vortices surviving for significant amounts of time after the system parameters have been quenched to the (quasi-)ordered side~\cite{Comaron-KZ-2020}: this is here visible by the slower decay of vortices with $\epsilon$, where $\epsilon = \epsilon(t) = t / \tau_Q$ following the quenched system evolution through the externally imposed parameter. The slower decay of vortices in time as the system adapts to the external quench 
 across the phase transition leads to the expected divergence in the relaxation time (red line, $\tau(\epsilon)$, in bottom left inset to Fig.~\ref{fig:polariton-nick}(b)). The faster one quenches across the phase transition (i.e.~the smaller that $\tau_Q$ is), the deeper into $\epsilon(t)>0$ values that the vortices can be observed on the (quasi)ordered side.
 Of course, ultimately such vortices will decay to the corresponding equilibrium value at such value of $\epsilon$, marked by the dashed red line.
 %consistent with the absence of ODLRO and condensation in a homogeneous 2D system). 
 I recall here that the quench rate affects the slope of $\epsilon(t)/\dot{\epsilon}(t)$ (dashed blue line in bottom left inset to Fig.~\ref{fig:polariton-nick}(b)), and so the exact location of the intersection point of these two curves which, according to the fundamental Kibble-Zurek equation, defines the timescale $-\hat{t}$. Numerical simulations~\cite{Comaron-KZ-2020} have confirmed the validity of the Kibble-Zurek scenario, by revealing the expected proportionality between the observed relaxation time of vortex dynamics (essentially the time when the quenched solid lines in Fig.~\ref{fig:polariton-nick}(b) deviate from the equilibrium dashed red line) and the time predicted by matching $\tau(-\hat{t})$ to $\epsilon(-\hat{t})/\dot{\epsilon}(-\hat{t})$ (up to some irrelevant microscopic non-universal prefactor): such proportionality can be seen in the top right inset.

In the long-time limit, the vortices decay. As discussed in Sec.~\ref{sec:scalinghypothesis}, this decay (following instantaneous quenching into the (quasi-)ordered state) is expected to enter another scaling window, associated with phase-ordering~\cite{jelic2011quench}.
Already from Fig.~\ref{fig:polariton-nick}(b) one can see that the vortex decay slows down as their number decreases for a given value of $\tau_Q$, i.e.~along a particular curve shown in Fig.~\ref{fig:polariton-nick}(b).
This is because vortices move (on average) further and further apart, thus the frequency of their annihilations slows down. 
Then, at some late time, when there are only relatively few vortices left in the system, the behaviour changes drastically. Such behaviour (following now an instantaneous quench) has been mapped out in Fig.~~\ref{fig:polariton-nick}(c): after a very rapid decay in the number of vortices (blue) and corresponding increase in the characteristic lengthscale describing the typical size of the emerging quasi-ordered regions (green), the system enters the scaling regime~\cite{comaron2018dynamical}: in this regime, properties of the system  can be characterized by a single growing lengthscale and thus all correlation functions collapse onto a single function, in terms of such an appropriately rescaled variable $L_c(t)$ [see Eq.~(\ref{eq:scaled-g1})]. Interestingly, one now sees very clearly both the existence of the expected logarithmic corrections in the presence of external coupling (dissipation) [previously discussed in Sec.~\ref{sec:scalinghypothesis}] , and the anticipated value of the dynamical critical exponent ($z=2$)~\cite{jelic2011quench}. While this behaviour is expected to continue at later times, finite size effects make it harder to obtain concrete predictions of $L_c(t)$ from the spatial correlation function.

The above results have painted the picture of a `non-equilibrium' BKT phase transition, and an anticipated late-time phase-ordering law, with 
significant discussion in the literature addressing the nature of this phase transition~\cite{polariton-bkt-szymanska,polariton-bkt-altman,UBEC-Keeling}.
Parallel to this, there has been quite some interest in searching for a different kind of universal behaviour on the quasi-ordered ($\epsilon>0$) side of the transition,
%with some emphasis on the potential emergence of a rather different transition, 
associated with a so-called Kardar-Parisi-Zhang (KPZ) universality class~\cite{kpz-paper}.
%instead of the BKT one. 

The KPZ equation describes  the stochastic growth of the height of an interface and is established in diverse (classical) physical settings, including, e.g.~polymer growth, and spreading of wildfires.
In the present context, I note that the KPZ equation arises as an equation modelling the phase of the superfluid $\theta$, obtained from $\Phi({\bf r},t) = \sqrt{n({\bf r},t)} e^{i \theta({\bf r},t)}$ in the presence of a momentum-dependent damping rate under the assumption that the density fluctuations ($\delta n({\bf r},t) =n({\bf r},t)-\langle n({\bf r},t) \rangle$) vary slowly both in space ($\nabla \delta n \approx 0$) and in time ($\partial  n/\partial t \approx 0$).
Such an assumption makes the KPZ physics more relevant in regimes characterised by weak phase fluctuations, namely deep in the quasi-ordered phase $\epsilon>0$.
In that case, one obtains the KPZ equation in the form~\cite{polariton-bkt-altman,wouters-kpz-1-PhysRevA.90.023615,wouters-kpz-2,kpz-diehl-PhysRevB.94.104520,kpz-diehl-2-PhysRevB.92.155,kpz-minguzzi-PhysRevB.97.195453,kpz-Deligiannis_2020}
\be
\frac{\partial \theta}{\partial t} = \nu \nabla^2 \theta + \frac{\lambda}{2} (\nabla \theta)^2 + \sqrt{D} \, \eta \;.
\ee
The terms on the right-hand side respectively describe the diffusion leading to smoothening, a nonlinear contribution leading to critical roughening and a stochastic noise term, with the constants $\nu$, $\lambda$ and $D$ depending on microscopic system parameters.
This equation maps the evolution of the phase of the polariton mean field to a growing interface with the same internal dimension.

Let us consider the general normalized first-order correlation functions in space and time in the form $g^{(1)}({\bf r},{\bf r}+{\bf \Delta} {\bf r},\, t, t+\Delta t)$ and denote this by the short-hand notation $g^{(1)}({\bf \Delta r}, \Delta t)$. In general (upon neglecting density-density and density-phase correlations), one can show that~\cite{KPZ-1D-Experiment}
\be
g^{(1)}({\bf \Delta r},\Delta t) = \langle {\rm exp}\left(i \Delta \theta ({\bf \Delta r}, \Delta t) \right) \rangle \;
\ee
where $\Delta \theta$ denotes the change in phase from some reference value.
To lowest order in the fluctuations, this would thus give the dominant behaviour that
\bea
{\rm Fixed} \,\, \Delta t_0: \hspace{0.1cm} |g^{(1)}({\bf \Delta r}, \, \Delta t_0)| \sim {\rm exp} \left( -\frac{1}{2} \left( \frac{\Delta r}{\lambda} \right)^{2 \chi} \right) \nonumber \\
{\rm Fixed} \,\, \Delta r_0: \hspace{0.1cm} |g^{(1)}({\bf \Delta r_0}, \, \Delta t)| \sim {\rm exp} \left( -\frac{1}{2} \left( \frac{\Delta t}{\tau} \right)^{2 \beta} \right)
\label{eq:kpz-scaling}
\eea
thus introducing the two KPZ critical exponents $\chi$ and $\beta$ for space and time, respectively denoting the roughness and growth critical exponents (with $\lambda$ and $\tau$ the corresponding non-universal parameters).
As a result, KPZ scaling would show up as `stretched exponential' decay in the correlation functions in the limit of a relatively well-formed mean field, i.e.~for `sufficiently large' $\epsilon>0$ values (such behaviour is termed `stretched exponential' here because $2 \chi,\, 2 \beta\, \neq 1$ as would be the case in a `normal' exponential) : in order to unequivocally observe such behaviour one would generally need to observe very large systems (and for very long times),
 because of the existence of a characteristic minimum lengthscale for the onset of KPZ physics~\cite{polariton-bkt-altman}.
 %which should be larger than the system size. 
%
%with the conditions for the emergence of KPZ arising as a competition between such a KPZ length,  the average vortex lengthscale and the overall system size.

%-------------------------------------
\begin{figure}[t!]
\centering
\includegraphics[width=0.685\linewidth]{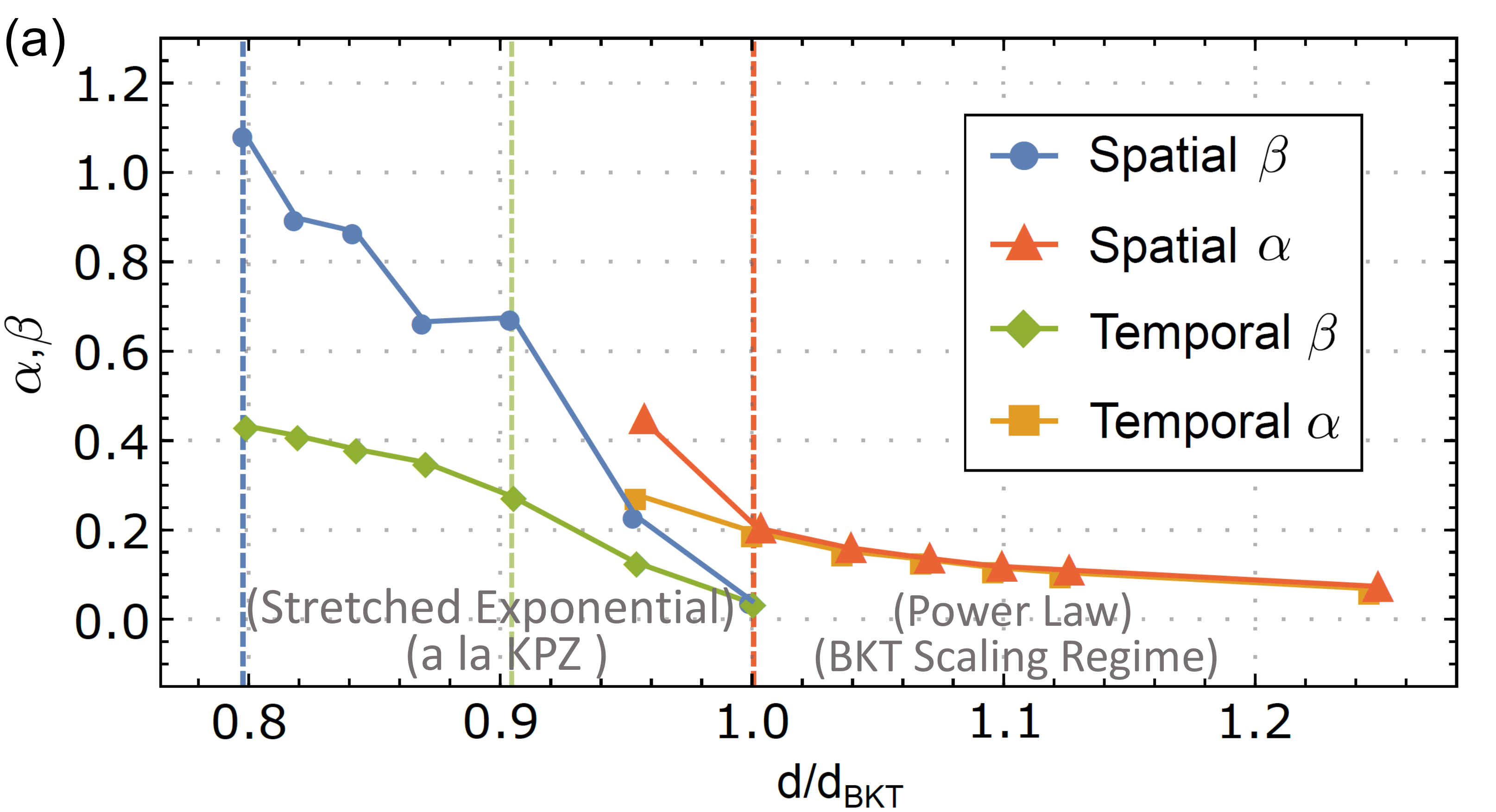}\\%{Fig-14a-new.png}\\
\includegraphics[width=0.693\linewidth]{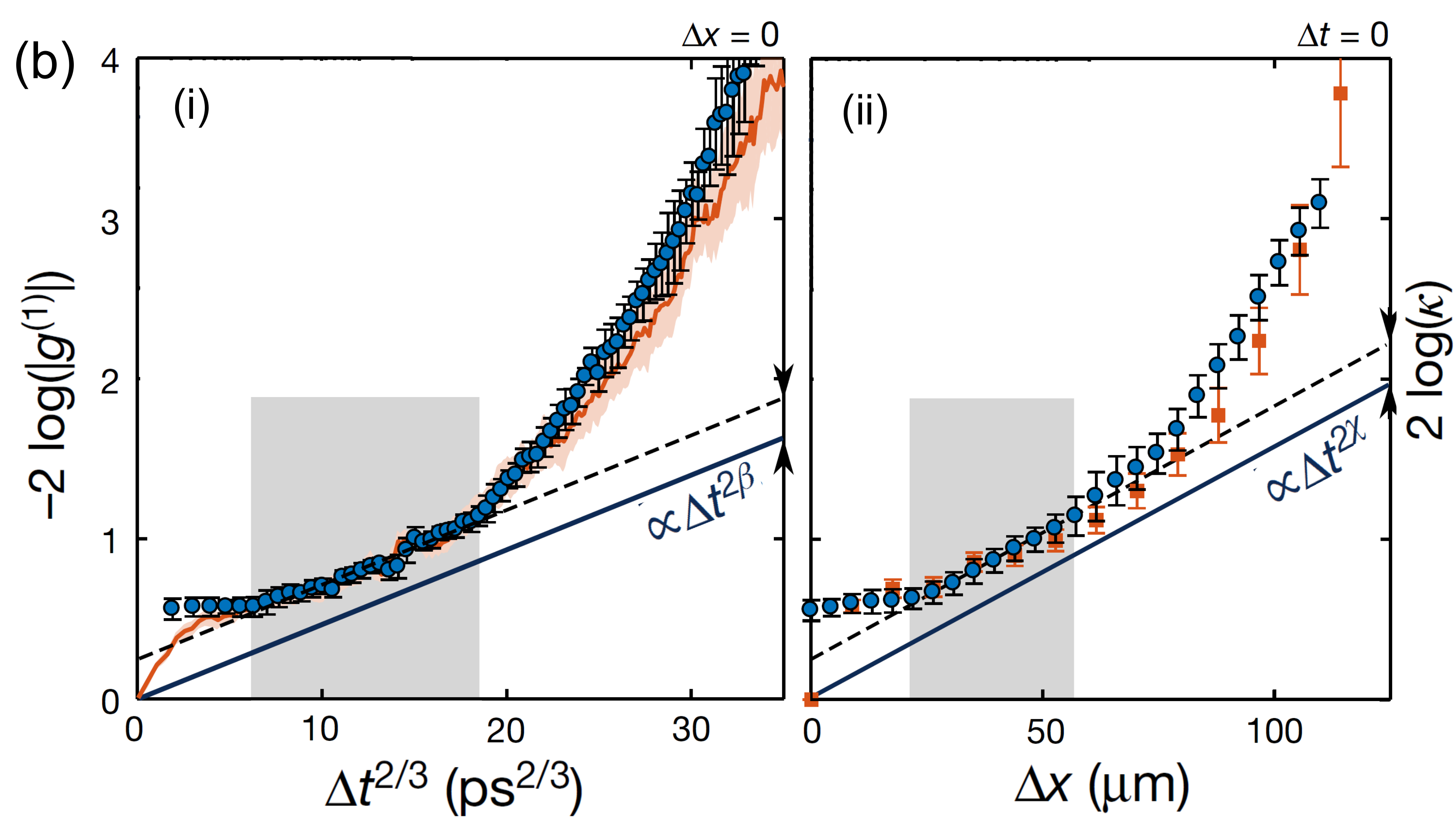}\\%{Fig-14a-new.png}\\
\includegraphics[width=0.718\linewidth]{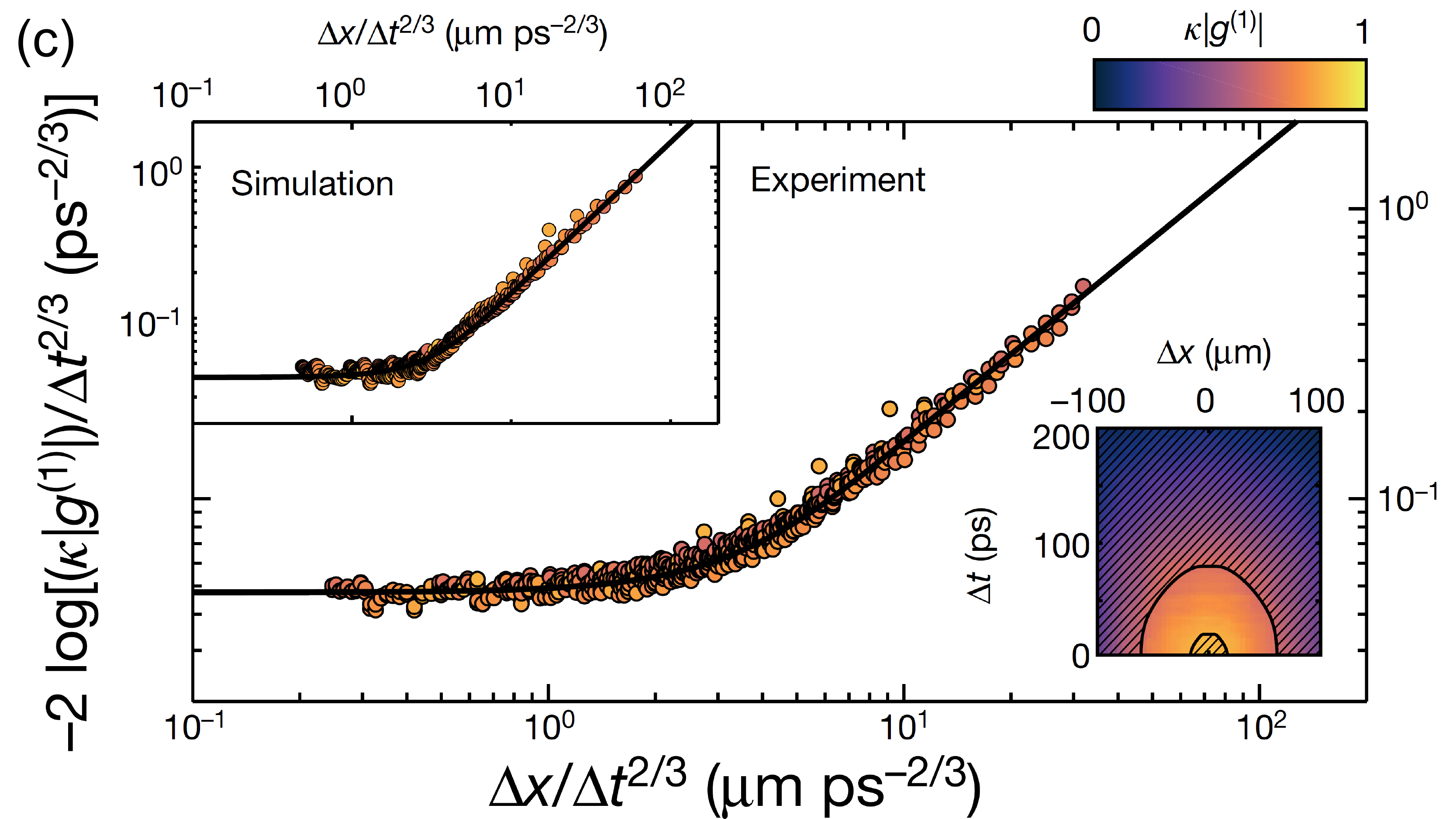}\\%{Fig-14bc-new.png}
\caption{
(a) Regime of 2D equilibrium BKT: Dependence of the spatial and temporal correlation function exponents as a function of the pumping (here $d$) scaled to the critical value ($d_{BKT}$) for the experimental parameters of Ref.~\cite{Sanvitto-BKT-KPZ}. Depicted values correspond to exponents labelling correlation decay either in the form of %the KPZ values 
 stretched exponentials (left part of plot), or algebraic decay (right part of plot) according to which one of these fits works best in each regime. Here one sees clearly that %BKT scaling 
 algebraic decay (characteristic of the 2D XY model) dominates (over stretched exponential decay) for all points above threshold, and that, importantly, the emerging spatial ($\alpha$) and temporal ($\beta$) exponents are equal and do not exceed the value of $0.25$ at the critical point.
 Adapted with permission from D.~Caputo {\em et al.}, Topological order and thermal equilibrium in polariton condensates, Nat.~Mat. {\bf 17}, 145~\cite{Sanvitto-BKT-KPZ}. Copyright (2018) by the Nature Publishing Group.
(b)-(c) Experimental demonstration of non-equilibrium KPZ scaling in 1D:
(b)(i)-(ii) Identification of spatiotemporal regime where both spatial and temporal correlations exhibit their expected stretched exponential scalings of Eqs.~(\ref{eq:kpz-scaling}) on the quasi-ordered side.
(c) Appropriately rescaled plot confirming the KPZ picture of Eq.~(\ref{eq:kpz-mixed}) for all data points in (b)(i)-(ii) within the corresponding scaling regimes for which $|g^{(1)}|$ takes the values shown within the annulus in the bottom right inset.
Note that such scaling function shows a plateau at small $y = y_0 \Delta r/(\Delta t )^{2/3}$ and linear growth at large $y$.
Insets: (top left) Corresponding analysis by means of numerical simulations; (bottom right) Coherence map showing value of $|g^{(1)}|$ (see colourbar) in terms of $\Delta r$ (denoted here by $\Delta x$) and $\Delta t$. 
%[Note that in subplots (b)-(c) the spatial separation $\Delta r$ of Eqs.~(\ref{eq:kpz-scaling})-(\ref{eq:kpz-mixed}) is denoted by $\Delta x$ here.]
Adapted with permission from Q.~Fontaine {\em et al.}, Kardar-Parisi-Zhang universality in a one-dimensional polariton condensate, Nature {\bf 608}, 687~\cite{KPZ-1D-Experiment}. Copyright (2022) by the Nature Publishing Group.
}
\label{fig:polariton-more}
\end{figure}
%-------------------------------------

Notwithstanding the earlier comments about a possible {\em non-equilibrium} BKT phase transition, an experiment conducted with high-quality samples (and so longer polariton lifetimes, thus facilitating a less non-equilibrium regime) found two interesting results~\cite{Sanvitto-BKT-KPZ} (see Fig.~\ref{fig:polariton-more}(a)): firstly, although spatial and temporal correlation functions could be fitted by both stretched exponential and power-law decays in the vicinity of the critical region, the power-law decay always fit better on the quasi-ordered side of the phase transition, thus explicitly verifying BKT-type correlations -- consistent with BKT being the transition mechanism.
%with observed correlations belonging to the 2D XY universality class. 
Secondly, unlike the previous discussion, the power-law exponents for the spatial and temporal decay on the quasi-ordered side were found to exactly match each other, both having a value clearly below 0.25: all this corroborated to an {\em equilibrium} BKT phase transition in this particular experimental setting, demonstrating that exciton-polariton condensates can control the extent of the `equilibrium' or `non-equilibrium' nature of the phase transition through the sample quality characterizing the polariton lifetime.

%The KPZ is associated with a strong-coupling fixed point in dimensions $D \ge 2$.
From a theoretical point of view, the universal scaling function for KPZ takes the form~\cite{kpz-Deligiannis_2020}
\begin{eqnarray}
C({\bf \Delta r},\, \Delta t) &=& \langle (\Delta \theta({\bf \Delta r}, \Delta t))^2 \rangle - \langle \Delta \theta ({\bf \Delta r}, \Delta t) \rangle^2  \nonumber \\ 
&=& - 2 {\rm ln} \left( \left| g^{(1)}({\bf \Delta r},\, \Delta t) \right| \right)  \nonumber \\
&=& C_0 (\Delta t)^{2 \beta} \,\, F \left( y_0 \frac{|{\bf \Delta r}|}{\Delta t^{1/z}}\right) \;,
\label{eq:kpz-mixed}
\end{eqnarray}
where  $F(y)$ is a universal scaling function (whose asymptotics are known), $z=\chi/\beta$ and $C_0$ and $y_0$ are non-universal constants.

%Nonetheless, the quest for KPZ-type physics has continued: Working with ........ samples ...... a recent experiment in a 1D setting managed to conclusively demonstrate the 

Such emergence of KPZ scaling was conclusively demonstrated in an appropriately-engineered (discrete) 1D setting~\cite{KPZ-1D-Experiment}, constituting the first realization of KPZ scaling in a quantum system.
%at the phase transition. 
In addition to identifying a spatial/temporal region where Eqs.~(\ref{eq:kpz-scaling}) hold [Fig.~\ref{fig:polariton-more}(b)(i)-(ii)], they were able to convincingly demonstrate the emergence of a universal spatiotemporal KPZ scaling function by plotting $C(\Delta r,\, \Delta t) / (\Delta t)^{2 \beta}$ as a function of $\Delta r/\Delta t^{1/z}$, for the particular exponent values of  $\beta=1/3$ and $z=3/2$, thus corresponding to $\chi=1/2$.

Extending such analysis to the 2D limit, the same team also managed to numerically demonstrate -- sufficiently above threshold so that vortices do not overwhelm KPZ effects~\cite{KPZ-2D-Deligiannis} -- the emergence of scaling according to Eq.~(\ref{eq:kpz-scaling}), with critical exponents consistent with those anticipated in the 2D KPZ universality class. This opens up the possibility of observing -- for the first time -- 2D KPZ physics in an experiment, a long-standing goal in non-equilibrium statistical physics.
The above brief discussion highlights driven-dissipative exciton-polaritons as ideal systems for tuning both the extent of equilibrium (or non-equilibrium) nature of the BKT phase transition, and the prevailing universality class through carefully engineered experiments~\cite{Szymanska-TuningUniversalities-PhysRevX.7.041006,New-KPZ-Diehl-PhysRevLett.128.070401}.

%For a system in the KPZ universality class, one expects the spatial and temporal correlation functions to scale with a so-called stretched exponential, i.e.~exhibit scaling of the form....
%
%the polariton lifetime, and so the strength of dissipation in the system, is directly controlled by the quality of the samples. Experiments with very high-quality samples have recently probed

\subsection{Bose-Einstein Condensation on Astrophysical/Cosmological Scales?} \label{sec:cosmo-bec}

Due to its universal nature, Bose-Einstein condensation has even been hypothesized to be at play across very different astrophysical/cosmological settings. Perhaps the most broadly known example is that of neutron stars~\cite{UBEC-Pethick}, whose core is expected to exhibit both superfluidity and superconductivity. However, various other forms of condensation have been discussed in the literature (see, e.g. early discussions in Ref.~\cite{BEC-Green-book}), including rather interestingly condensation of gravitons~\cite{UBEC-Dvali}.
Below I briefly outline an alternative idea gaining increasing attention currently in the literature in the context of dark matter modelling.

%-------------------------------------
\begin{figure*}[t!]
\centering
\includegraphics[width=0.7\linewidth]{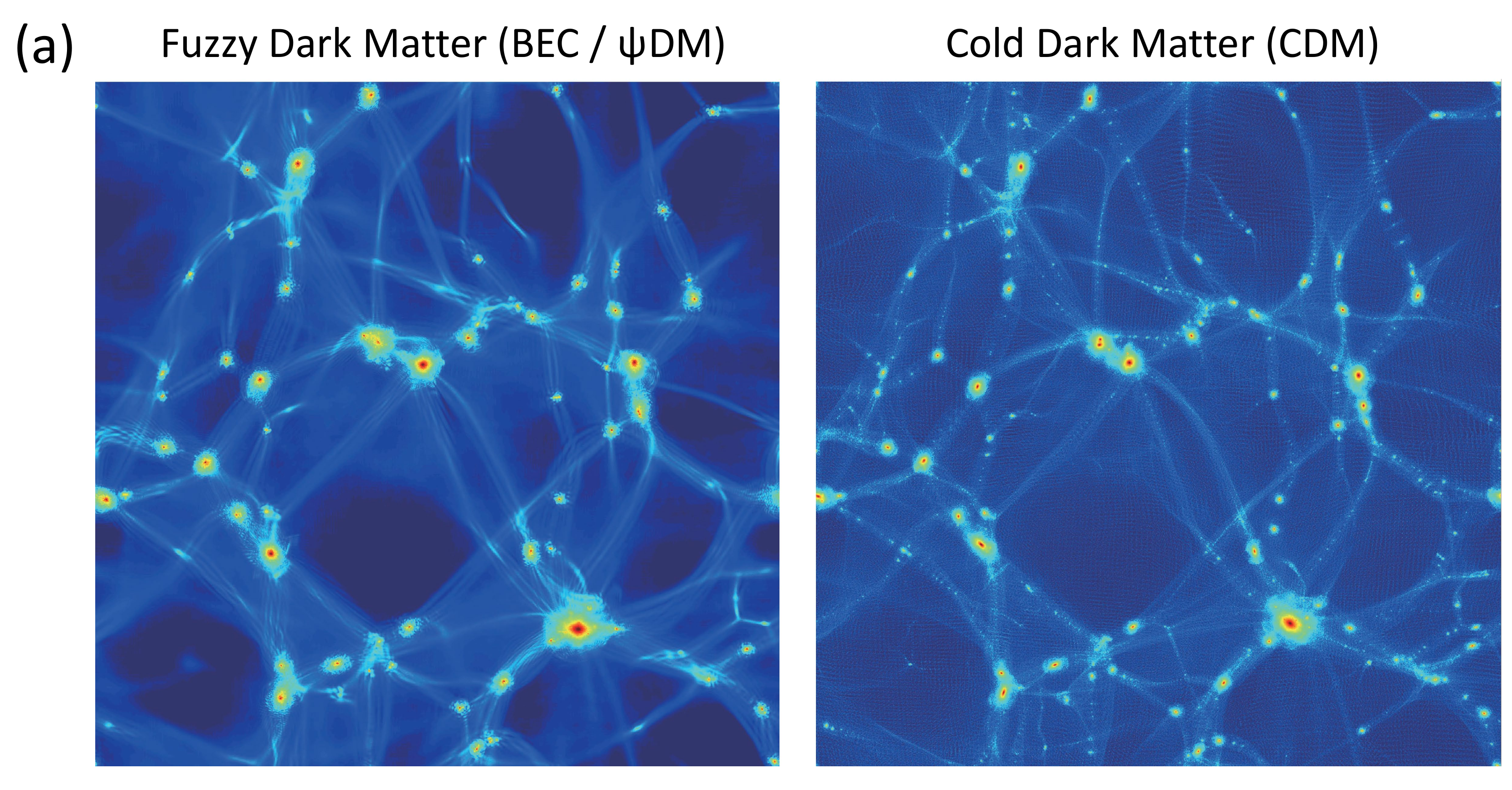}\\
\includegraphics[width=0.7\linewidth]{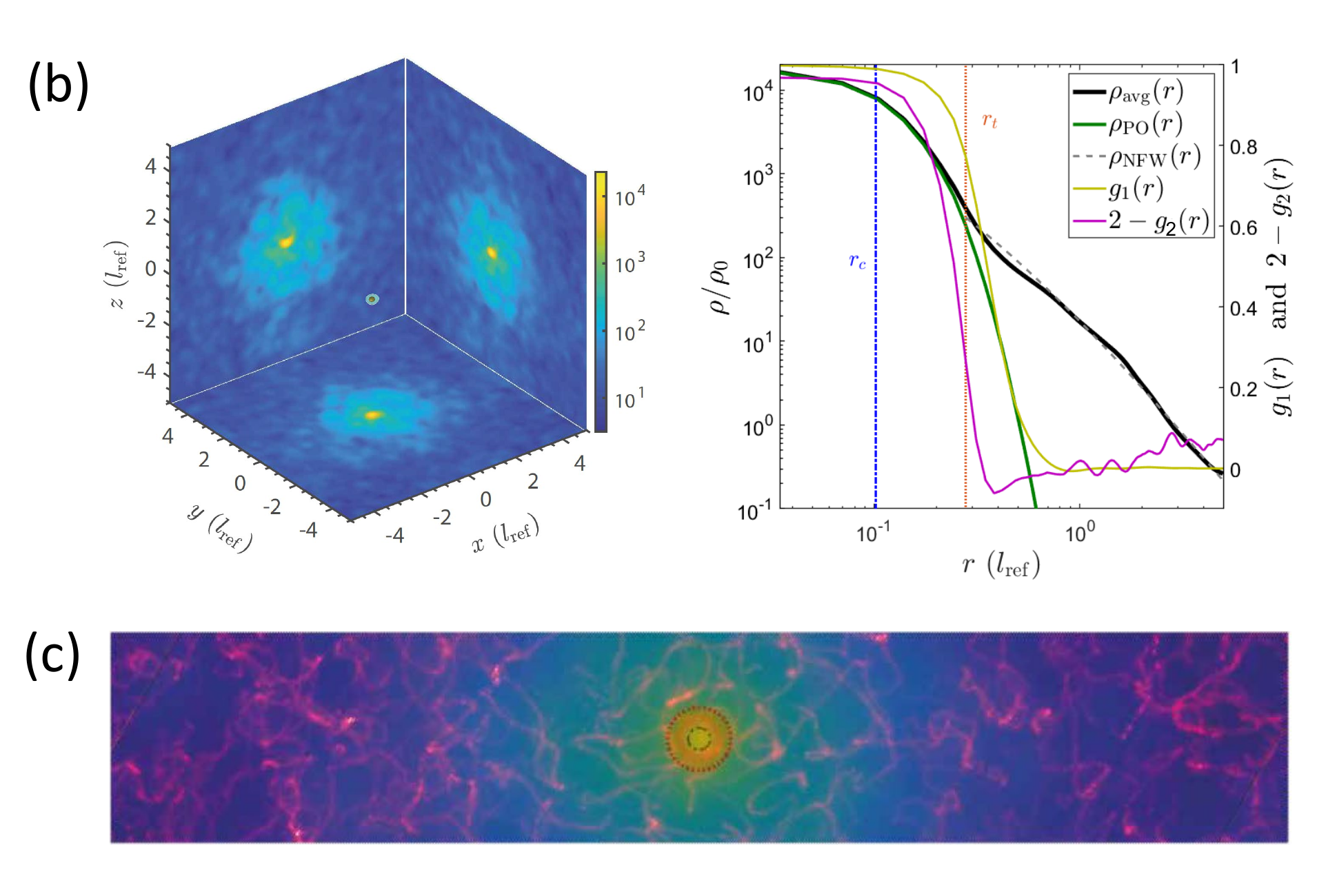}
\caption{
Cosmological condensation in the Fuzzy Dark Matter model (also known as $\psi$DM in Ref.~\cite{Schive-2014NatPh..10..496S}):
(a) Large-scale comparison of fuzzy dark matter predictions (left) to the established cold dark matter model based on N-body simulations (right) from a given set of initial cosmological conditions: large-scale features are excellently reproduced.
Reprinted with permission from H.-Y. Schive {\em et al.}, Cosmic structure as the quantum interference of a coherent dark wave. Nat.~Phys.~{\bf 10}, 496~\cite{Schive-2014NatPh..10..496S}. Copyright (2014) by the Nature Publishing Group.
(b) Fuzzy dark matter simulations focussing on an 
%single core-halo structure, consistent with a single galaxy 
isolated (single) galactic halo, which in the fuzzy dark matter model contains a characteristic solitonic core at its centre
(left image), reveal a number of interesting features shown in the right image: firstly, the radial density profile (black line) both exhibits a core at the centre (unlike the CDM model predictions which lead to an `unphysical' $1/r$ divergence as $r \rightarrow 0$ which is not shown here) and reproduces the expected NFW profiles (dashed brown line) at larger distances ($r > r_t$); importantly, the solitonic core exhibits near perfect coherence $g^{(1)}({\bf r}) \approx g^{(2)}({\bf r}) \approx 1$ for $r \le r_c$, with coherence only gradually lost with increasing radial distance up to $r_t$, at which point the NFW density profile sets in; in fact, the solitonic core can be fully described as a Penrose-Onsager condensate mode (green line) all the way up to $r_t$, with the remaining (incoherent) density concentrated primarily outside the solitonic region, but with a non-zero (albeit small) component still present at smaller radii: this is directly analogous to the condensate-thermal cloud bimodal distribution of Hartree-Fock theories for ultracold atomic gases (see, e.g.~Ref.~\cite{proukakis_finite-temperature_2008}) and the corresponding experimentally-obtained density distributions.
(c) Example schematic of the density distribution of the solitonic core (yellow core -- with both $r_c$ and $r_t$ radii shown to give a sense of scale) surrounded by regions of decreasing density filled with random vortices (purple lines) which scramble the phase between different domains of locally similar phase -- constituting a galactic-size example of a quasi-condensate state.
(b)-(c) \npp{ Adapted with permission from I.~K.~Liu {\em et al.}, Coherent and incoherent structures in fuzzy dark matter halos, Mon.~Notices R.~Astron.~Soc.~521 (3), 3625~\cite{Gary-MNRAS}. Copyright (2023) by Oxford University Press. }
}
\label{fig:cdm}
\end{figure*}
%-------------------------------------

The possibility of the existence of a cosmological-scale gravitating Bose-Einstein condensate of axionic particles has been proposed as a plausible  alternative picture of the nature of dark matter~\cite{Hu2000} (see also the related scenario of thermalizing QCD axions~\cite{UBEC-Sikivie}). More recently, the focus of such theoretical ideas appears to be in the context of ultralight axionic particles of mass $10^{-22}$ eV/$c^2$: due to their low mass, these can give rise to galactic-size de Broglie wavelengths (recall that $\lambda \sim 1/\sqrt{m}$), thus facilitating condensation on such unprecedented scales. While such systems cannot be controlled in any way (and so the question of phase transition crossing becomes less relevant), 
%-- unlike the previously discussed laboratory settings -- 
nonetheless, predictions of such models in the current (late-time, virialized) state could ultimately be related to cosmological observations.

As well-known, there is currently an established model of cold dark matter, which explains the large-scale structures of the universe very well through N-body numerical simulations~\cite{Frenk-White_DM}.
 Nonetheless, some discrepancies have been recently noted in relation to the small-scale structure~\cite{CDM-Challenges-2015PNAS..11212249W,CDM-Challenges,CDM-Challenges-2017Galax...5...17D} (most notably the inferred distribution of the dark matter density at the very centre of galaxies). One very elegant way (but admittedly not the only one) to resolve such discrepancies is to describe such features through Bose-Einstein condensation in the context of the so-called Fuzzy Dark Matter (FDM) model~\cite{Schive-2014NatPh..10..496S,Mocz-2017MNRAS.471.4559M,Marsh:2016,Hui:2021,Ferreira}: The underlying idea here is that dark matter consists of ultralight axionic particles, of mass $10^{-22}$eV/$c^{2}$. Due to their high density and low mass, such particles would have a very high phase-space density allowing the emergence of BEC on galactic scales. 
The power of such an approach is that it describes `short-range' (on cosmological scales) features by the emergence of a so-called gravitationally self-bound soliton of dark matter, while still reproducing the very successful large-scale predictions of the usual cold dark matter model (CDM) -- most notably the Navarro-Frenk-White (NFW) radial density profiles at large distances from the galactic core~\cite{NFW_1996}.

The equation governing such cosmological condensates is a Schr\"{o}edinger equation, embedded within the gravitational field~\cite{Schrodinger-Poisson-Widrow-1993ApJ...416L..71W,SPE-Seidel-1990PhRvD..42..384S,Schive-2014NatPh..10..496S,Mocz-2017MNRAS.471.4559M}. In general, one could also envisage the existence of interactions between such axionic particles \npp{(although some constraints on their maximum strength do exist)~\cite{Kehagias-PhysRevD.97.023529,Mocz:2023,Munoz-Delgado-10.1093/mnras/stac3386}}, giving rise to the following system of coupled equations, loosely termed the Gross-Pitaevskii-Poisson equations:
\begin{equation}
    i\hbar\frac{\partial \Phi(\mathbf{r},t)}{\partial t} =\left[ -\frac{\hbar^2\nabla^2}{2m} + g\rho(\mathbf{r},t)+mV(\mathbf{r},t) \right]\Phi(\mathbf{r},t)  \label{eq:GPPE}
\end{equation}
where the Newtonian potential $V(\textbf{r},t)$ follows the Poisson equation,
\begin{equation}
    \nabla^2 V(\mathbf{r},t)  = 4\pi G \left( \rho(\mathbf{r},t) -\bar{\rho} \right) \label{eq:Poisson}
\end{equation}
with the mass density $\rho({\bf r},t)=|\Phi({\bf r},t)|^2$, and $\bar{\rho}$ its spatially-averaged constant value (and $G$ denotes the gravitational constant).
(Interestingly this same set of equations has also been recently used to model glitches in pulsars~\cite{brachet-pulsars-PhysRevResearch.4.013026}.)

%Significant work has already been done demonstrating that such a model reproduces the large-scale structure of the Universe very effectively [ ], in addition to giving some encouraging results on smaller scales. 
While the validity of the fuzzy dark matter model has yet to be tested, even in the absence of interactions (and we may still be a long way away before some of the emerging predictions can be directly tested against observational data), 
there is a rapidly-growing body of work (see, e.g.~recent reviews~\cite{Marsh:2016,Hui:2021,Ferreira}) addressing the model's predictions both on the cosmological large scale (Fig.~\ref{fig:cdm}(a))~\cite{Schive-2014NatPh..10..496S,Mocz-2017MNRAS.471.4559M,Fuzzy_May-Springel}, and on the level of a single (isolated) cored-halo regime corresponding to the distribution of dark matter on the level of a typical galaxy (Fig.~\ref{fig:cdm}(b)-(c))~\cite{Mocz-2017MNRAS.471.4559M,Schwabe-Niemeyer-2016PhRvD..94d3513S,Chan-CoredHalo-10.1093/mnras/stac063,Gary-MNRAS}.
The important new contribution from the fuzzy dark matter picture discussed in the above works (and many references therein) is that the density at the centre of the galaxy behaves very differently to CDM predictions, with the emergence of a well-formed soliton whose density remains finite as ${\bf r} \rightarrow 0$.
Essentially this leads to a bimodal profile of a central soliton surrounded by fluctuating dark matter density satisfying the NFW profile:
%In the latter scenario -- where a much more detailed characterization can be facilitated relatively easily -- 
%Essentially this leads to a bimodal profile, 
this is somewhat reminiscent of the density profiles of ultracold atomic gases in harmonic traps, with gravitational attraction in the galactic case playing a role analogous to the harmonic confinement of ultracold atomic gases -- such an important and interesting analogy has been explicitly discussed in Ref.~\cite{Gary-MNRAS}. 

Addressing the coherent properties of such a (virialized) cored-halo system, Ref.~\cite{Gary-MNRAS}
demonstrated that the discussed solitonic core is in fact fully coherent in the sense of a dominant Penrose-Onsager mode, with a nearly constant spatial correlation function in the central solitonic region exhibiting $g^{(1)}({\bf r}) \approx g^{(2)}({\bf r}) \approx 1$ for $r \lesssim r_c$, where $r_c$ labels a typical soliton radius. Beyond this, there is an intermediate region of gradually decreasing  coherence spatially, with global phase coherence across the soliton core lost between the soliton core and an outer `transition' region $r_t$, at which point $g^{(1)}({\bf r}) \approx 0 $ and $g^{(2)}({\bf r}) \approx 2$, consistent with a chaotic field. Nonetheless, closer inspection demonstrates that the outer halo region ($r > r_t$) contains randomly-distributed regions of near constant density, which locally feature enhanced phase coherence. 
Such domains are separated by randomly-oriented vortices which scramble the phase between them: in the cosmological context, such regions are termed `granules'.
Such a picture closely resembles a cosmological-size quasi-condensate, as shown schematically -- based on actual numerical data -- in Fig.~\ref{fig:cdm}(c)~\cite{Gary-MNRAS}; as such vortices are randomly oriented/distributed and move around (albeit on rather cosmologically-slow timescales), there is no overall phase coherence after radial, or temporal averaging. Moreover -- at least in the absence of self-interactions -- this system  exhibits familiar scaling laws of (ultraquantum) turbulence, with such states in fact rather similar qualitatively to the early dynamical phases of condensate formation discussed in Sec.~\ref{sec:kz-spgpe}~\cite{Gary-MNRAS}; interestingly recent works have shown  such a structure to be very long-lived, unlike the rapidly decaying turbulent states generated in non-driven ultracold atomic systems.

The predictions of the fuzzy dark matter model may end up not having a direct bearing on real-world observables. Nonetheless, the mere fact that a cosmological model of a BEC can actually generate features seemingly consistent with our current understanding of large-scale structure in the universe is truly remarkable in its own right, and demonstrates the power and universality of such a physical phenomenon.

\section{Concluding Remarks} \label{sec:conclusions}

In this Chapter I have given a broad overview of manifestations of macroscopic coherence across a diverse range of physical systems, spanning from the very tiny (exciton-polaritons, atoms) to the very large (cosmological).
%addressing both quasi-ordered systems, and those exhibiting off-diagonal long-range order.
While the specific physical manifestation across such systems might be different, with Bose-Einstein condensation associated with a single macroscopic observable and off-diagonal long-range order giving way -- in some limits -- to related behaviour in terms of a Berezinskii-Kosterlitz-Thouless transition to a quasi-ordered state, or even to a state exhibiting stretched exponential correlation functions consistent with those of the Kardar-Parisi-Zhang model over a restricted spatiotemporal regime, such systems nonetheless exhibit 
evidence of universal behaviour in appropriate temporal domains and/or spatial regions.

some common observable features. This facilitates an exciting cross-fertilization of ideas, which have borne out of seemingly distinct fields such as low-temperature physics, statistical physics, condensed matter, atomic molecular and optical physics. Interestingly, many of the ideas discussed in this Chapter have their origins in cosmological or theoretical physics studies, even if perhaps some of those features are now becoming more the playground of controlled quantum matter experiments.

For example, the concept of defects arising during a dynamical phase transition becoming embedded into, and thus observable within, the long-time evolution of the system -- first proposed in the context of a hot Big Bang model -- has not only found significant validation in a diverse range of coherent quantum systems, but also led to exciting studies pushing new frontiers in quantum gas experiments conducted with an unprecedented flexibility and accuracy -- such as experiments with ultracold atoms. In this context, one is now investigating the modifications to this Kibble-Zurek scenario of driven quenching across the phase transitions imposed by  inhomogeneities and the interplay with causality. 

Moreover, the concept of highly-non-equilibrium scenarios, associated with the dynamics around non-thermal fixed points emerged initially from studies of reheating after early universe inflation, but recent years have seen a number of distinct experiments with ultracold atoms revealing evidence of dynamics consistent with predicted scalings around such non-thermal-fixed points.
Such non-equilibrium scenarios have a close connection with the emergence of (strong) turbulence, in a generalized manner, i.e.~with the state of a system filled with random tangled defects which can also exhibit various types of universal scaling laws in appropriate regimes. Beyond the familiar and already observed direct energy cascade Kolmogorov turbulence of a driven system and associated decay of bundles of vortices once driving is removed, experiments with liquid helium and ultracold atomic systems have revealed a new `type' of turbulence, known as `ultraquantum' (or `Vinen') turbulence, which is dominated by the tangling and relaxation of individual vortices. The rapid quenching of a phase transition creates a strongly `turbulent' state with broadly similar features qualitatively, whose subsequent evolution ultimately follows universal scaling laws -- whether in the context of spatially self-similar solutions during phase-ordering, the stronger spatiotemporal evolution around non-thermal-fixed points, or even evolution related to the Kardar-Parisi-Zhang spatiotemporal self-similarity. Interestingly, such features also connect the  driven or instantaneous breaking of a continuous symmetry with phenomena investigated in fluids, including classical fluids and growth of interfaces.
%
%Moreover, in the context of driven-dissipative (exciton-polariton) superfluids, we have seen evidence of the rather different Kardar-Parisi-Zhang universality, connecting the stochastic growth of the height of interfaces applicable to a broad range of classical problems such as ...... to the establishment of a common phase in quantum gases.

As such, systems exhibiting macroscopic quantum coherence in controlled quantum matter experiments give access to a range of different universal types of behaviours, and the emergence of a range of scaling laws having wide applicability to very different physical systems across both the classical and the quantum realm. Combined with the enhanced quantum control in these systems which facilitates the on-demand construction of system hamiltonians -- thus making them ideal systems for quantum simulations -- such systems collectively open up a very exciting, highly-interdisciplinary realm for a deeper understanding of the non-equilibrium physical world.

\acknowledgments
The unified presentation given here has been made possible through collaborations (or extended discussions) on these topics with many excellent colleagues and friends over a period of years, including
%and I am grateful to Tom Billam, 
Tom Billam, Tom Bland, Jerome Beugnon, Iacopo Carusotto, Paolo Comaron, Leticia Cugliandolo, 
%Galaa Dagdavorj, 
Franco Dalfovo, Jacek Dziarmaga, Jean Dalibard, Piotr Deuar, Gabriele Ferrari, 
%Andrew Groszek, 
Andrew Groszek, Giacomo Lamporesi, Fabrizio Larcher,
%Gary Liu, Anna Minguzzi, 
Rob Smith, Henk Stoof, Marzena Szymanska, and Alex Zamora, to whom I am grateful.
In such a context, I would particularly like to highlight Dr I-Kang (Gary) Liu and Dr Paolo Comaron who have been pivotal to my respective understanding of Kibble-Zurek physics and polariton condensation, as well as Dr Gerasimos Rigopoulos and Dr I-Kang Liu for our recent collaboration into coherent cosmological endeavours.
During the preparation of this chapter I have benefited enormously from discussions with, and direct feedback on this manuscript from, Vanderlei Bagnato, Carlo Barenghi, Paolo Comaron, Jacek Dziarmaga, Thomas Gasenzer, Gary Liu, Gerasimos Rigopoulos, and Marzena Szymanska.
I also thank Jacek Dziarmaga, Thomas Gasenzer, Anna Minguzzi, Hsi-Yu Schive, Rob Smith, and Wojciech Zurek for explicit permissions to reprint key figures from their earlier works, and the UK EPSRC, Leverhulme Trust, and the Horizon-2020 framework for funding -- with particular note to the Quantera grant `Non-equilibrium dynamics in Atomic systems
for QUAntum Simulation' (NAQUAS) which refocused my interest on such topics.

\bibliography{Proukakis}

\end{document}